\documentclass[fleqn,usenatbib]{mnras}
 \usepackage{xspace}
\usepackage{aas_macros}
\usepackage{bm}
\usepackage{gensymb}
\usepackage{epsfig}
\usepackage{url}
\usepackage{hyperref}

\usepackage[normalem]{ulem} 
\usepackage{natbib}
\usepackage{latexsym}
\usepackage{epsfig}
\usepackage{amsmath}
\usepackage{amssymb}
\usepackage{wasysym}
\usepackage{graphicx}
\usepackage{dcolumn}
\usepackage{verbatim}
\usepackage{enumitem}
\usepackage[titletoc]{appendix}
\usepackage{amsfonts}
\usepackage{fontawesome}
\usepackage{fancyvrb} 
\usepackage{tikz} 
\usetikzlibrary{calc}
\usepackage[export]{adjustbox}

\usepackage{booktabs}

\usepackage{multirow}

\usepackage{pifont}  

\usepackage[normalem]{ulem}

\newcommand{\ba}{\begin{eqnarray}}
\newcommand{\ea}{\end{eqnarray}}
\newcommand{\be}{\begin{equation}}
\newcommand{\ee}{\end{equation}}

\newcommand{\Msun}{\mathrm{M}_{\odot}} 

\definecolor{grey}{rgb}{0.4,0.4,0.4}
\definecolor{dullmagenta}{rgb}{0.4,0,0.4}
\definecolor{darkblue}{rgb}{0,0,0.4}
\definecolor{midblue}{rgb}{0,0,0.5}
\definecolor{midred}{rgb}{0.5,0,0}
\definecolor{orange}{rgb}{1,0.5,0}
\definecolor{lightbrown}{rgb}{0.75,0.5,0.25}
\definecolor{tan}{cmyk}{0.14,0.42,0.56,0}
\definecolor{djunglegreen}{cmyk}{0.99,0,0.52,0}
\definecolor{lightgreen}{rgb}{0,1,0}
\definecolor{olivegreen}{cmyk}{0.64,0,0.95,0.40}
\definecolor{midgreen}{rgb}{0.0,0.675,0.0}
\definecolor{darkgreen}{rgb}{0,0.5,0}
\definecolor{ceruleanblue}{rgb}{0.0, 0.2, 0.7}
\definecolor{burgunday}{rgb}{0.5, 0.0, 0.13}
\definecolor{hvred}{RGB}{186,12,47}

\hypersetup{
    colorlinks=true,
    linkcolor=ceruleanblue,
    filecolor=ceruleanblue,      
    urlcolor=midblue,
    citecolor=burgundy,
}

\usepackage[all]{xy} 
\usepackage{amsfonts}

\makeatletter
\def\l@subsubsection#1#2{}
\makeatother

\newcommand{\qpls}{{QPLS}}

\usepackage{xcolor}
\definecolor{burgundy}{rgb}{0.5, 0.0, 0.13}

\defcitealias{QPLS2025}{Paper~I}
\newcommand{\PaperI}{\citetalias{QPLS2025}\xspace}

\title[Binary-lens amplification of starlight]{Binary-lens amplification of starlight: \\
from magnification flares to quasi-periodic tracking of orbital evolution}

\author[H. Wang et al.]{
Hanxi Wang,$^{1,2}$\thanks{E-mail: hanxi.wang@st-hughs.ox.ac.uk }
Bence Kocsis$^{3,1,4}$\thanks{E-mail: bence.kocsis@physics.ox.ac.uk }
and Miguel Zumalac\'arregui$^{5,6}$\thanks{E-mail: miguel.zumalacarregui@csic.es }
\\
$^{1}$St Hugh's College, University of Oxford,
St Margaret's Rd, Oxford, OX2 6LE, UK\\
$^{2}$ Department of Physics, Astrophysics, University of Oxford,
Denys Wilkinson Building,\\
Keble Road, Oxford, OX1 3RH, UK\\
$^{3}$Rudolf Peierls Centre for Theoretical Physics, University of Oxford,
Clarendon Laboratory,\\
Parks Road, Oxford, OX1 3PU, UK\\
$^{4}$Institute of Physics, E\"otv\"os University,
P\'azm\'any P. s. 1/A, Budapest 1117, Hungary\\
$^{5}$Instituto de F\'isica Te\'orica UAM/CSIC, \\
C/ Nicol\'as Cabrera 13--15, Universidad Aut\'onoma de Madrid,
Cantoblanco, 28049 Madrid, Spain\\
$^{6}$Max Planck Institute for Gravitational Physics
(Albert Einstein Institute), \\
Am M\"uhlenberg 1, D-14476 Potsdam-Golm, Germany
}

\date{Accepted XXX. Received YYY; in original form ZZZ}

\pubyear{\the\year{}}

\begin{document}
\label{firstpage}
\pagerange{\pageref{firstpage}--\pageref{lastpage}}
\maketitle

\begin{abstract}
Supermassive black hole binaries (SMBHB) are expected at the center of galaxies as a result of galaxy mergers. The dynamics and evolution of SMBHBs shed light on the evolution and formation of galaxies, the gravitational wave background and the nature of gravity. Many methods have been proposed to identify sub-parsec SMBHBs via the detection of gravitational waves or the modulation of the emission from the surrounding gaseous environment. In a previous study, we proposed a new method of identifying SMBHBs in non-active galactic nuclei via gravitational lensing. The caustics, i.e. the source positions with a divergent magnification, exhibit a complex extended geometry which rotates with the binary period. Each caustic-crossing produces an extreme magnification by up to a factor of $10^{6}$. In this work, we generalize the previous study and describe the rich phenomenology of \textit{binary-lens amplification of starlight} (BLAST). We explore the electromagnetic transients produced when accounting for the relative proper motion of the source, either a single or a binary star, the eccentricity and precession of the SMBHB, and the relative inclination. Systems where a star rapidly drifts across the caustic produce a single- or double-peaked flare. In the opposite limit, a star with no relative proper motion produces \textit{quasi-periodic lensed starlight} (QPLS), i.e., repeated peaks that track the system's orbital evolution. We determine the expected magnification and peak duration in different circumstances, and calculate the magnification curves and their Fourier transforms. This study may help construct search pipelines to identify SMBHBs in electromagnetic time-domain surveys. 
\end{abstract}

\begin{keywords}
gravitational lensing: strong -- transients -- black hole mergers-- black hole physics
\end{keywords}





\section{Introduction}

Supermassive black holes (SMBHs) are ubiquitously found in galactic nuclei \citep{Kormendy_Ho2013}. When galaxies merge, their central black holes drift towards each other and form bound binaries. The study of the formation and evolution of SMBHBs will improve our understanding of their interaction with the surrounding environment~\citep{Begelman:1980vb,Escala2004,Volonteri_2010_review,Milosavljevic:2002bn,LISA:2022yao}, the formation and evolution of galaxies~\citep{Volonteri2003,Kormendy2013}, the gravitational wave (GW) background~\citep{NANOGrav:2023gor,EPTA:2023fyk,Reardon:2023gzh,Xu:2023wog}, fundamental physics, and the nature of gravity~\citep{Alonso-Alvarez:2024gdz,Kocsis+2008}. Many SMBHBs with wide separation (parsec to kilo parsec) have been identified~\citep{DeRosa:2019myq}, while binaries with closer separation have not yet been confirmed~\citep{DOrazio:2023rvl}. Sub-parsec binaries of mass $10^5-10^8\Msun$ can be identified with future GW detectors in the mHz band, e.g. LISA~\citep{LISA:2017pwj,LISA:2024hlh} and TianQin~\citep{TianQin:2015yph,Luo:2025sos}. Several methods have also been proposed to identify closer binaries with electromagnetic (EM) observations. Most methods are based on observing the gas emission modulated by the orbital motion of the SMBHB, due to e.g. gravitational torques, accretion variability, relativistic Doppler boost or self-lensing~\citep{Haiman_Kocsis_Menou2009,DOrazio:2015jcb,Charisi:2018qyp,Xin:2019dkw,Song:2019lxb,DOrazio:2017ssb,Davelaar:2021eoi,Kelley+2021,Krauth+2024}.

We recently proposed a new method for discovering SMBHBs through repeated gravitational lensing magnification~\citep{schneider:1986,Schneider:1992}, presented in \PaperI~\citep{QPLS2025}, which we named quasi-periodic lensed starlight (\qpls{}):
SMBHBs act as natural telescopes, leading to extremely high amplification of stars in the same galaxy closely aligned with them. The orbital and inspiral motion of the massive binary induces the rotation and evolution of the caustics, where magnification due to lensing is divergent. If a source lies in the vicinity of the caustics, it will undergo quasi-periodic cycles of high-magnification, producing a lightcurve modulation. The high magnification of the caustic crossings can potentially amplify intrinsically bright stars to a comparable brightness as the host galaxy or active galactic nuclei (AGN). Compared to previous methods of identifying SMBHBs using EM observation, QPLS does not rely on the existence of AGNs, which are estimated to be absent in 90\% of the galaxies~\citep{Kauffmann+2003}. 

Here we introduce binary-lens amplification of starlight (BLAST) as an overarching phenomenological framework that generalizes QPLS. In addition to slowly-moving stars, we consider the caustic crossings induced by an arbitrary relative proper motion between the source and binary lenses. During the survey duration, BLAST might be observed as a number of magnification flares, producing, on a given system, a smaller number of amplification peaks than the previously studied quasi-periodic magnification curves. For example, during a `caustic transit', the star travels across the entire caustic during the observational period, producing 2-4 magnification peaks, corresponding to fold crossings. 

Magnification peaks caused by rapidly moving stars do not trace the orbital motion, but the corresponding magnification curves still contain valuable information about the geometry and the size of the caustics. The study of the QPLS magnification curves in \PaperI pointed out that the redshifted SMBH binary orbital period may be measured as a function time, possibly detecting its inspiral. That paper focused on stars with no proper motion, face-on and equal-mass lensing binaries for simplicity, and presented the magnification curves produced by both quasi-circular and eccentric binaries observed for a large number of orbital periods. Quasi-circular and eccentric systems exhibit different morphologies of the magnification curves, e.g. in terms of peak multiplicity per period. Other parameters of the binary lensing system (e.g. relative inclination of the source and lens) could also transform the morphologies of the magnification curves, producing potential degeneracies. Understanding these morphologies is essential for inferring the properties of the SMBHB from an observed lightcurve. 

In this paper we extend the discussion of BLAST phenomena. The paper is organized as follows: Sec.~\ref{sec:overview} gives an overview of binary lenses and the BLAST regimes. In Sec.~\ref{sec:caus}--\ref{sec:mag}, we work out the binary lens system analytically, giving estimates of the size of the caustics and magnification during caustic crossings for different topologies. In Sec.~\ref{sec:static_caustic}, we present the magnification curves of caustic transits. In Sec.~\ref{sec:different_lightcurves}, we explored the parameter space of the binary lensing system and discussed the different morphologies of the \qpls{} magnification curves. In sec.~\ref{sec:spectrum}, we analyze the frequency spectra of several \qpls{} magnification curves.

\begin{figure*}
    \centering
    \includegraphics[width=1\linewidth]{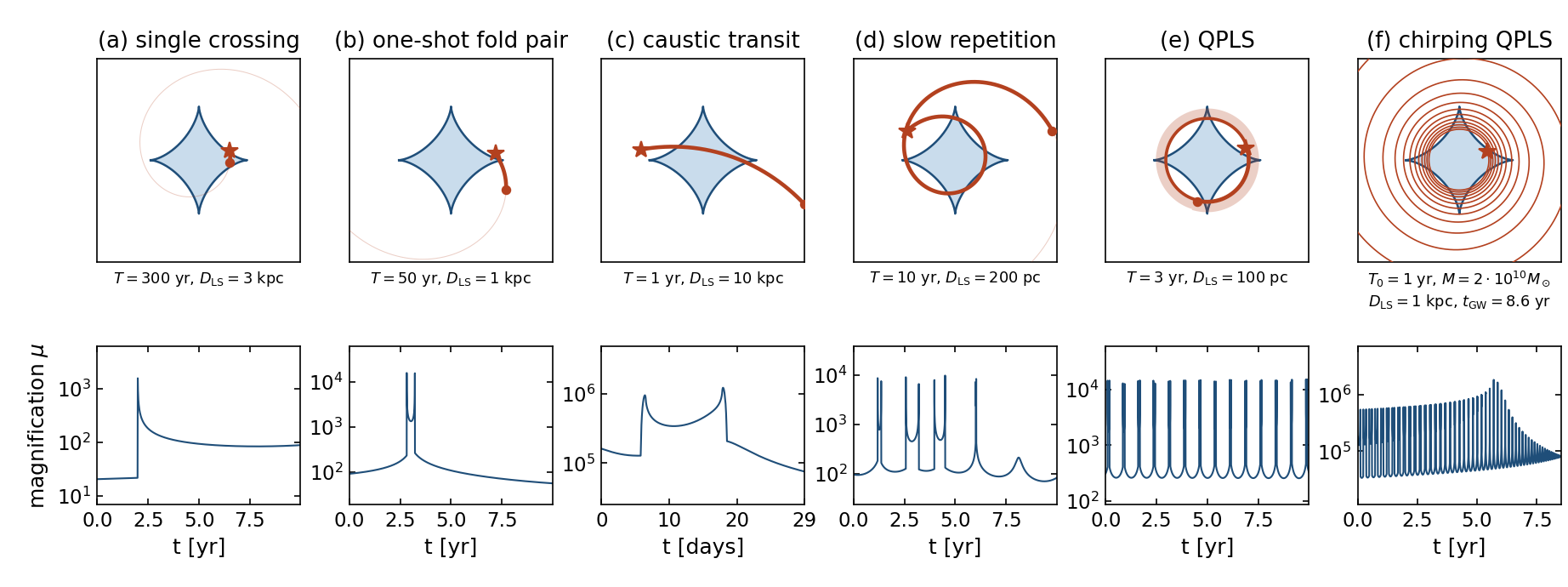}
    \caption{Six regimes of \textit{binary-lens amplification of starlight} (BLAST), from the simplest signal (a) to the richest (f), see Sec.~\ref{sec:blast}.
\textit{Top:} track of the star in the frame co-rotating with the caustic, in units of the instantaneous cusp radius; thick red line corresponds to the interval shown in the magnification curve underneath (star: start, dot: end).
\textit{Bottom:} the corresponding magnification curve for a $10\,R_\odot$ source.
Unless stated in the panel, $M=10^9\,M_\odot$, $q=1$, $T_{\rm obs}=10$ yr and $v_\star$ follows Eq.~\eqref{eq:vstar_general} (except for (e) and (f), where $v_\star=2$ km/s). Binaries are circular, oriented face-on. 
(a) \textit{Single crossing}: 
the star crosses a single fold in the observing period.
(b) \textit{One-shot fold pair}: a single arm sweeps over the star within the survey, causing two peaks.
(c) \textit{Caustic transit}: the star crosses the caustic in $t_{\rm cross}<T/4$, giving one fold pair separated by $\simeq t_{\rm cross}$. 
(d) \textit{Slow repetition}: the star drifts across the caustic during one period, so successive passages sample different parts of the arms.
(e) \textit{QPLS}: the slow star traces a near circle, with a fold pair at each of the four arm passages per orbit.
(f) \textit{Chirping QPLS}: a $10^{10}+10^{10}\,M_\odot$ binary merging within the survey (cf. Fig.~2 of \PaperI); the projected track spirals outward relative to the caustic as the caustic physically shrinks: fold pairs give way to a cusp maximum and then to decaying peaks.}
    \label{fig:frames}
\end{figure*}

\begin{figure*}
    \centering
    \includegraphics[width=0.85\linewidth]{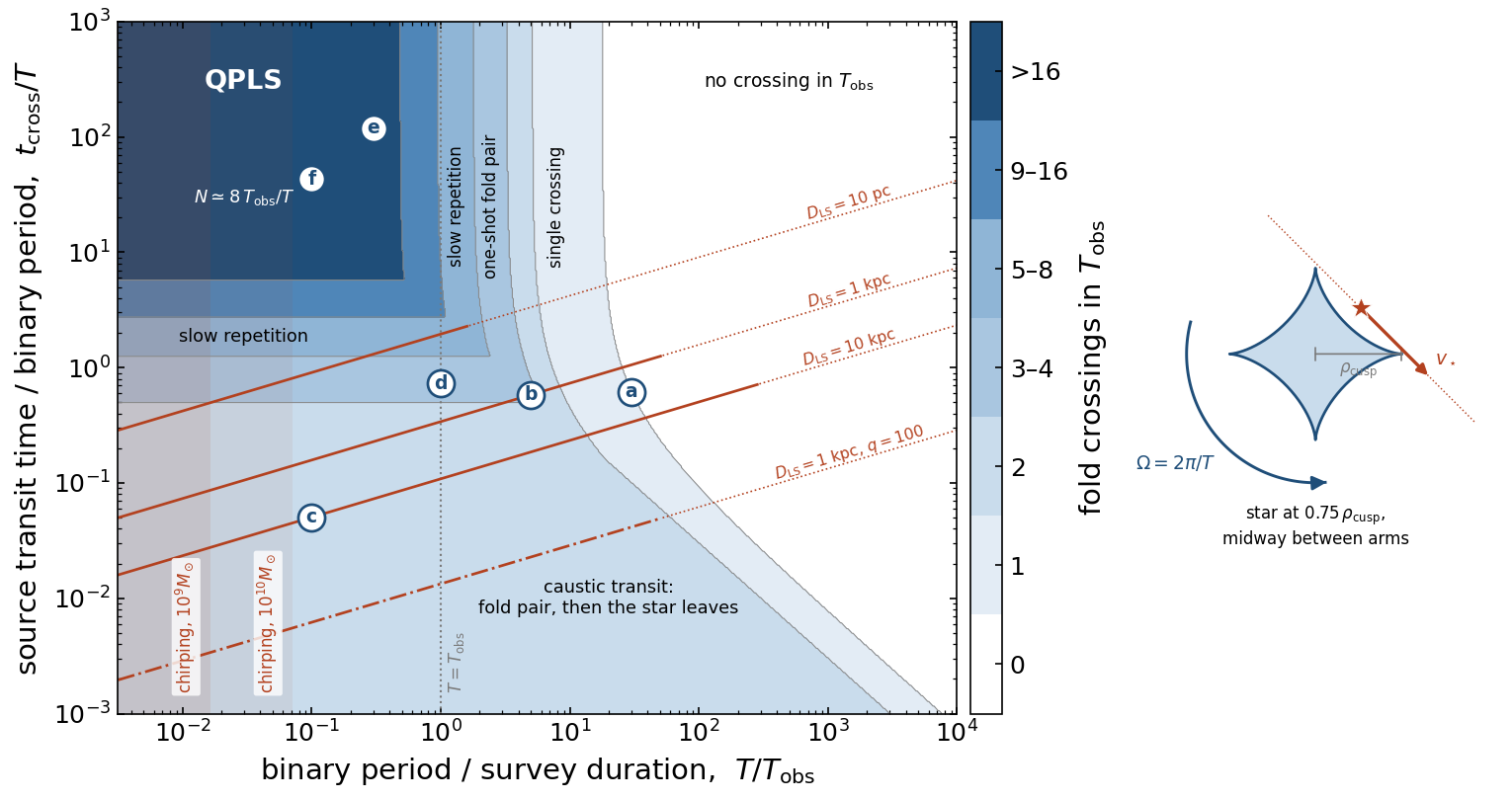}
    \caption{BLAST regimes in the plane of binary period over survey duration, $T/T_{\rm obs}$, and source transit time, Eq.~\eqref{eq:tcross_def}, over binary period, $t_{\rm cross}/T$.
Shading gives the number of fold crossings within $T_{\rm obs}=10$ yr for the reference configuration sketched on the right (see Sec.~\ref{sec:blast}): a star at $0.75\,\rho_{\rm cusp}$, midway between two arms, moving on a straight line tangentially and against the rotation.
For $t_{\rm cross}\gtrsim T$ the star is effectively at rest and the count is set by the rotation, Eq.~\eqref{eq:ncross}; for lower $t_{\rm cross}$ only the transit pair survives.
Red lines are the loci of $M=10^9\,M_\odot$ binaries for several $D_{\rm LS}$ and mass ratios, with typical $v_\star$ from Eq.~\eqref{eq:vstar_general} (solid: tight binary $d<1/\sqrt{2}$, dotted: wider binaries).
Shaded vertical strips mark binaries merging within the survey, $t_{\rm GW}<T_{\rm obs}$.
Letters mark the parameters of the examples of Fig.~\ref{fig:frames}; (e) and (f) assume a slow star ($v_\star=2$ km/s).
Note that the initial configurations in Fig.~\ref{fig:frames} differ, so the number of peaks does not necessarily coincide in both figures.
}
    \label{fig:regime_map}
\end{figure*}

\begin{figure*}
    \centering
    \includegraphics[width=0.62\linewidth]{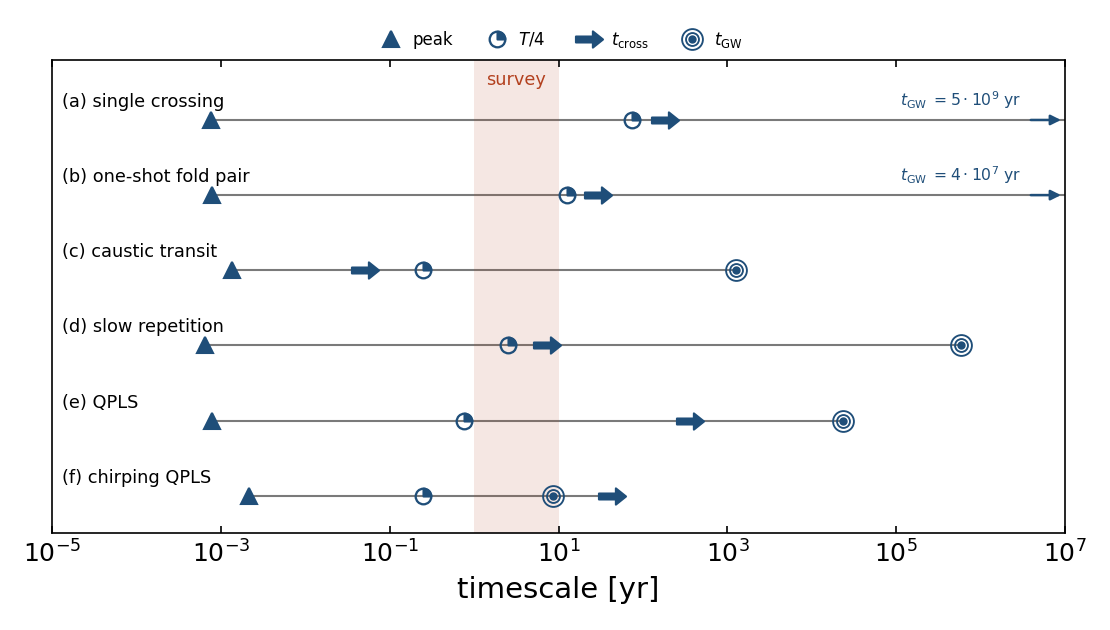}
    \caption{Characteristic timescales of the six examples of Fig.~\ref{fig:frames}: duration of a caustic-crossing
peak, $2R_{\rm src}/v_{\rm rel}$, with $v_{\rm rel}$ the speed of the caustic relative to the star at
$0.75\,\rho_{\rm cusp}$; the interval between successive arm passages, $T/4$; the source transit time across the
central caustic, $t_{\rm cross}$; and the time to merger, $t_{\rm GW}$. The shaded band spans survey durations of
1--10 yr. The ordering of $T/4$, $t_{\rm cross}$ and $T_{\rm obs}$ sets the regime: a single fold crossing or one
arm passage when $T/4 \gtrsim T_{\rm obs}$ (a, b), a transit pair when $t_{\rm cross}<T/4$ (c), a few repetitions
when $t_{\rm cross}\lesssim T_{\rm obs}$ (d), and many when $T/4 \ll T_{\rm obs} \ll t_{\rm cross}$ (e, f);
the signal chirps when $t_{\rm GW}$ falls inside the survey (f).}
    \label{fig:timescales}
\end{figure*}

\section{Binary-Lens Amplification of Starlight} \label{sec:overview}

We will start by introducing the basics of binary lenses (\ref{sec:binary-lens}) before discussing the different regimes for binary-lens amplification of starlight (BLAST), depending on the stellar velocity and orbital period (\ref{sec:blast}).

\subsection{Lensing by binary systems}\label{sec:binary-lens}

We model the SMBHB as binary lens \citep{schneider:1986} with mass $M_1$ and $M_2$ projected onto a common lens plane perpendicular to the line of sight at angular diameter distance $D_L$ to the observer. The source star is located in the same host galaxy, at a distance of $D_S$ to the observer and $D_{\rm LS}$ to the lens plane. We assume that $D_{\rm LS}\ll D_{L}\sim D_{S}$. We consider only the lensing effect by the SMBHB on the source star and ignore the effect from the other components such as stars and dark matter in the same galaxy. The Einstein radius of this lensing system is 
\begin{equation}
    \label{eq:ein_rad}
    \xi_{0} =  2  \sqrt{R_{\rm g} \frac{D_LD_{\rm LS}}{D_S}} \approx
    2\sqrt{R_{\rm g}
    D_{\rm LS}} =1.38\text{pc}\, M_{10}^{1/2}{D_{\rm kpc}^{1/2}}\,,
\end{equation}
where $R_{\rm g} = GM/c^2$, $M=M_1+M_2$ is the total mass of the SMBHB, $M_{10}=M/(10^{10}\Msun)$, and $D_{\rm kpc}=D_{\rm LS}/\rm kpc$.

Assuming weak gravitational field, the lens equation in units of the Einstein radius is  
\begin{align}
\label{eq:leqn}
\bm{y} &=\bm{x}- m_1 \frac{\bm{x}-\bm{x}_{m_1}}{|\bm{x}-\bm{x}_{m_1}|^2}-m_2 \frac{\bm{x}-\bm{x}_{m_2}}{|\bm{x}-\bm{x}_{m_2}|^2}\,,
\end{align}
where $\xi_0\bm{y}$ is the source position in the source plane, $\xi_0\bm{x}$ is the image position in the lens plane,  $\xi_0\bm{x_{m_1}}$ and $\xi_0\bm{x_{m_2}}$ are the lens positions in the lens plane, and $m_1$ and $m_2$ are dimensionless component masses 
\begin{equation}
m_{1,2} = \frac{M_{1,2}}{M_1+M_2}\,.    
\end{equation}
We compute the magnification of a point-source using the lens equation using
\begin{equation}
    \mu = \left|{\rm det }\frac{\partial \bm{y}}{\partial\bm{x}}\right|^{-1}
\end{equation}
The caustic curves cause the point-source magnification to diverge, i.e.
\begin{equation}
    \det\left(\frac{\partial \bm{y}}{\partial \bm{x}}\right) \to 0 \quad \text{(caustic)}\,.
\end{equation}
Sec.~\ref{sec:caus} discusses the properties of the caustic regions below.

To obtain the magnification of an extend source we either (1) compute the average point-source magnification over the area or (2) employ the extended-source magnification using contour integration with VBMicrolensing~\citep{Bozza2010,Bozza:2018,Bozza2021}. See Appendix~\ref{sec:method} for more details of the methods and assumptions of the binary lensing system adopted in this work. The caustics are determined by solving for source positions $\bm{y}$ for which $\mu$ is divergent.

\subsection{BLAST regimes}\label{sec:blast}

We now outline the different types of expected magnification curves%
\footnote{Observed lightcurves will contain the magnification curve multiplied by the stellar luminosity, along with the background contribution from many other sources.} 
accounting for the relative proper motion of the source star and the SMBH binary lens. The limiting cases are situations when the signal consists of at most two pairs of magnification flares as the source rapidly drifts across the caustic region (a caustic transit), and conversely when the source remains in the caustic region for several orbits producing the QPLS signals examined previously in \PaperI.

BLAST can be divided into different regimes by comparing the time scale of the SMBHB orbit $T$, source motion and survey duration $T_{\rm obs}$.\footnote{
Throughout this paper, $M$ and $T$ denote the redshifted total binary mass and observer-frame orbital period, respectively. All distances entering the lensing calculations, including $D_{\rm LS}$, $a$, $\xi_0$, and $R_{\rm src}$, are defined as $1+z$ times their corresponding proper values in the source frame. With this convention, the lensing and Keplerian equations retain their usual form. Stellar velocities are invariant under this rescaling. Empirical relations involving intrinsic masses, such as the $M$–$\sigma$ relation, are evaluated using $M/(1+z)$.
} 
Consider the caustics of a diamond-shape with 4 symmetrical caustic arms representing the primary caustic of a tight binary (see Sec.~\ref{sec:caus} for more details). The caustic has a radius of $\rho_{\rm cusp} \xi_0$, where $\rho_{\rm cusp}$ is the radius of the cusp in units of the Einstein radius (see Sec.~\ref{sec:caus_size} for more details). The caustic rotates around the barycenter of the binary lens with period $T$. Consider a star moving along a straight line with speed $v_{\star}$. The transit time for the star across the whole caustic is approximated by
\begin{equation}
    \label{eq:tcross_def}
    t_{\rm cross} = \frac{2\rho_{\rm cusp}\xi_0}{v_\star}\,.
\end{equation}
A star at rest with respect to the barycenter is swept by each of the four arms once per orbit, and each passage produces two fold crossings. The number of crossings during a survey of duration $T_{\rm obs}$ is
\begin{equation}
    \label{eq:ncross}
    N \simeq 8\,\frac{T_{\rm obs}}{T}\,, \qquad t_{\rm cross}\gg T\,,
\end{equation}
for a slowly moving star. If the source is outside of the caustic but close to it, the number of close encounters is the relevant number, $N = 4\,T_{\rm obs}/T$.

We categorize BLAST into six regimes, from the simplest to the richest signal:
\begin{enumerate}[label=(\alph*),leftmargin=0pt,itemindent=2em,labelwidth=1.5em,align=left]
    \item \textit{single crossing}: the star enters/exits the caustic a single time during the survey or it has a single close encounter outside a cusp. One fold crossing or close encounter produces a single magnification peak. 
    \item \textit{one-shot fold pair}: the star travels across 1 caustic arm during the survey, producing 2 fold crossings.
    \item \textit{caustic transit}: the star travels across the entire caustic during the survey, producing 2 fold crossings. The time separation between the fold crossings is around $t_{\rm cross}$.
    \item \textit{slow repetition}: the star drifts through the caustic within one period. The binary exhibits noticeable rotation during the survey. The star crosses 2-4 different caustic arms at different parts of the arm, producing 4-8 fold crossings. The peaks are unequally spaced and shaped.
    \item \textit{QPLS}: a star with small transverse proper motion with respect to the caustic traces a near circle around the caustic. The star crosses each of 4 caustic arms repetitively, producing a pair of fold crossings each time. Fold-crossing pairs will be approximately evenly spaced in time and have approximately the same shape. 
    \item  \textit{chirping QPLS}: in addition to small transverse velocity, the binary inspirals with its time to merger $t_{\rm GW}$ comparable to the survey duration. The caustic shrinks as the binary components are driven closer by GW emission. This produces modulation in the peak amplitude, duration, separation and multiplicity. 
\end{enumerate}
In cases (a-c) the stellar velocity sets the dominant timescale, while the orbital period rules (e,f). Regime (d) represents an intermediate case, where both timescales are comparable. Note that cases (a-f) are shown for caustic crossings, which produce the brightest signal. BLAST can also be produced from close encounters of stars with the caustics without crossings. Examples of close encounters will be presented in Sec.~\ref{sec:static_caustic}. 

Fig.~\ref{fig:frames} presents examples of the 6 regimes of BLAST. The top panels show the track of the source star in the frame co-rotating with the caustic, in units of the instantaneous cusp radius. All cases consider 
face-on
binaries for simplicity. The bottom panels show the corresponding magnification curves for a 10$R_{\odot}$ source. Examples (a)--(d) adopted values of $v_*$ based on typical values based on velocity dispersion, Eq.~\eqref{eq:vstar_general}, while examples (e) and (f) assumed the slow tail of the velocity distribution ($v_*=2$ km/s). It can be seen that examples (a)--(c) produce non-periodic fold-pairs in the magnification curves in contrast to examples (d)--(f). 

\begin{figure*}
    \includegraphics[width=1\linewidth]{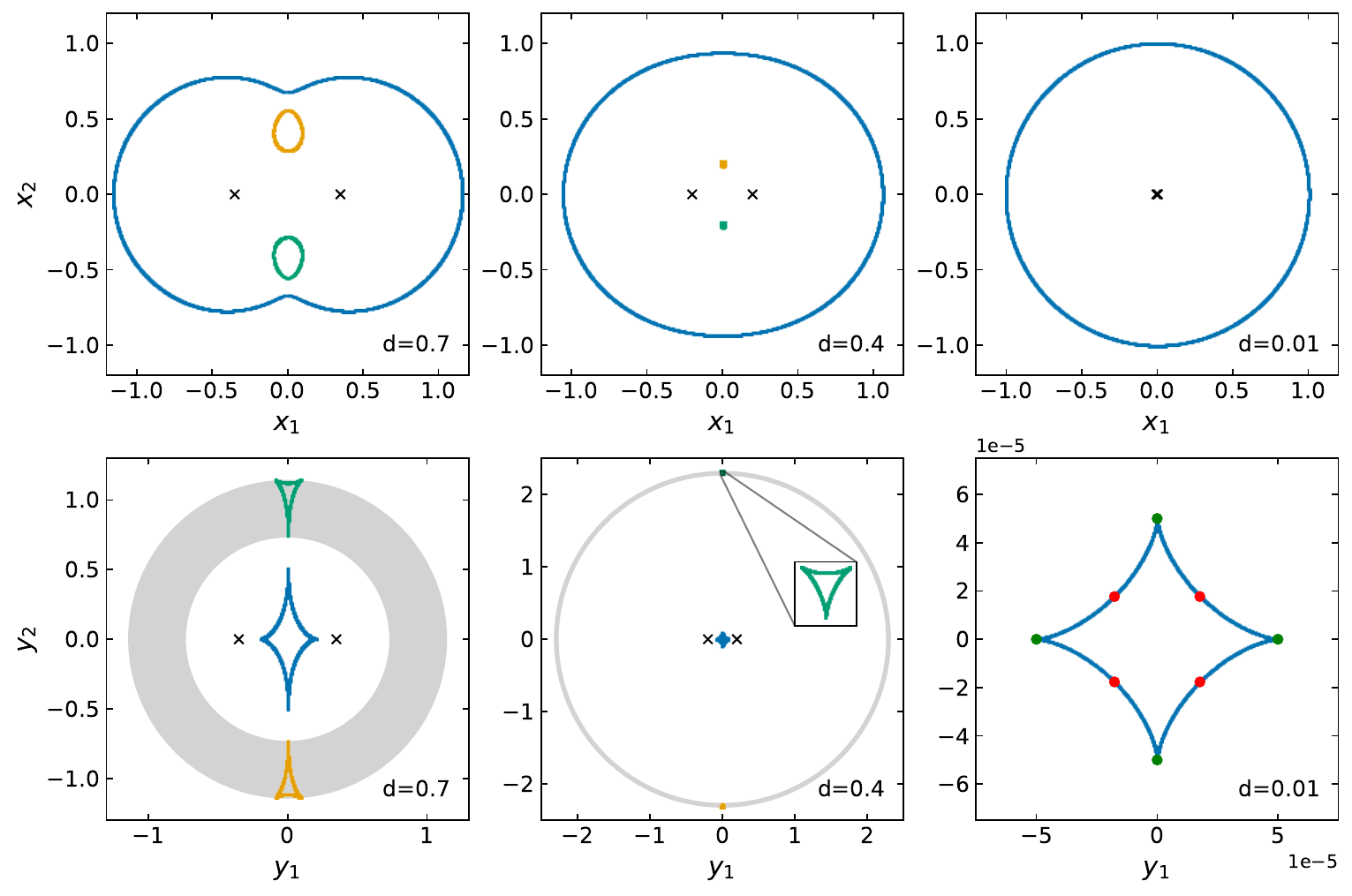}
    \caption{The critical curves and caustic curves for equal mass binary lenses and three values of $d<1$. $y_{1,2}$ and $x_{1,2}$ are the source and image position's coordinates in their respective planes, all in units of the Einstein radius (Eq.~\eqref{eq:ein_rad}). $q=1$ in this figure. The black crosses marked with $L_1$ and $L_2$ are the two SMBH lenses. The three branches of the critical curves are plotted in different colours, with the corresponding branches in the caustic curves following the same colouring scheme. In these cases, the critical and caustic curves have the same topologies independently of $d$: the critical curves form symmetrically two smaller closed curves enclosed by a larger curve, and the caustic curves are comprised of a central diamond-shaped curve and two outer triangular curves. {Note that the upper (lower) secondary critical curve maps to the lower (upper) secondary caustic.} The shaded region in the first column marks the region covered by the secondary (triangular) caustic during an orbital period (not shown in the right column). The inset in the caustic plot of the middle column shows a zoom-in of the upper secondary caustic curve. In the caustic plot of the right column, we mark the 4 cusp points with green dots and the 4 innermost fold points with red dots.\label{fig:caustic}}
\end{figure*}

Fig.~\ref{fig:regime_map} shows the number of crossings as a function $t_{\rm cross}/T$ and $T/T_{\rm obs}$ for a fixed system configuration.
We place a reference star at $0.75\rho_{\rm cusp}$ from the caustic center, midway between two caustic arms, moving along a straight line tangentially and against the orbital rotation, such that we have mostly fold crossings (sketched on the right of Fig.~\ref{fig:regime_map}). The caustic crossings stop when the star leaves the annulus bounded by the innermost fold (the red dots in the bottom right panel of Fig.~\ref{fig:caustic}) and the cusp (the green dots in the bottom right panel of Fig.~\ref{fig:caustic}), after $\simeq 0.33 t_{\rm cross}$. The shading gives the number of fold crossings, which vary for the different regimes. The red lines are the loci of the $M=10^9\Msun$ binaries with different source-lens distances $D_{\rm LS}$ and mass ratio $q\equiv m_1/m_2$. These loci are computed assuming $v_{\rm *}$ is set by the mean velocity dispersion in Eq.~\eqref{eq:vstar_general}. It can be seen that most parts of the loci intersect with regimes with zero or few peaks. The shaded vertical strips on the left mark the region where the binaries merge within the survey duration, i.e. $t_{\rm GW} < T_{\rm obs}$ for binaries with total mass $10^{9} \Msun$ and $10^{10} \Msun$. The letters (a)-(f) correspond to examples of the 6 regimes presented in Fig.~\ref{fig:frames}. 
Note that the shading refers to the reference star defined above, while the examples of Fig.~\ref{fig:frames} have other initial conditions, often changing the number of magnification peaks.

Fig.~\ref{fig:timescales} shows the characteristic timescales of the six examples of Fig.~\ref{fig:frames}: duration of a peak, the interval between successive arm passages ($\sim T/4$), the transit time of the source across the caustic, $t_{\rm cross}$, and the time to merger $t_{\rm GW}$. The shaded band shows survey durations of 1-10 yr. It can be seen that the ordering of the these time scales is different for the six regimes. 
For cases (a) and (b), $t_{\rm cross},T/4\gtrsim T_{\rm obs}$ such that there is small amount of motion of both the source and the binary during the survey duration. 
For regime (c), $t_{\rm cross}\lesssim T/4$ such that the source does not cross multiple caustic arms and produce repeated fold-pairs and $t_{\rm cross}<T_{\rm obs}$ such that the source is able to travel across the entire caustic within the survey duration. 
For case (d), the three time scales are of the same order such that the source crosses multiple caustic arms, producing a few repetitions in the magnification curve. For regimes (e) and (f), $T/4\ll T_{\rm obs} \ll t_{\rm cross}$ such that the star moves slowly across the caustic and the caustic rotate for many periods during the survey duration. This produces a series of quasi-periodic peaks in the magnification curve. 
For regime (f), we have additionally $t_{\rm GW} \sim T_{\rm obs}$ such that the binary is chirping, producing the modulation in the amplitude, duration and separation of peaks in the magnification curve.

\section{The critical and caustic curves}\label{sec:caus}
The critical (caustic) curves consist of image (source) positions where the magnification is divergent of a point-source. We find the critical curve by evaluating image positions $\bm{x}$ for which ${\rm det} (A) = 0$, where the matrix elements of $A$ are defined by 
\begin{equation}
    \label{eq:T_deriv}
   A_{ij} = \frac{\partial y_j}{\partial x_i},
\end{equation}
The corresponding source positions $\bm{y}$ of the caustic curve can be obtained by substituting the critical curve into the lens equation Eq.~\eqref{eq:leqn}.

It is convenient to express the positions of the image, source and lenses in terms of complex numbers: $z \equiv x_1+x_2i$, $z_{s} \equiv y_1+y_2i$ and $z_{1,2} \equiv (\bm{x}_{m_{1,2}})_1+(\bm{x}_{m_{1,2}})_2i$, following~\citet{Witt1990}. The lens equation Eq.~\eqref{eq:leqn} is rewritten as 
\begin{equation}
    \label{eq:binary_leqn_complex}
    z_s = z- \frac{m_1}{z^*-z_1^*}- \frac{m_2}{z^*-z_2^*}.
\end{equation}
Solving the critical curve using $\det (A)=0$ is equivalent to solving the equation \citep{maneghetti:2022book}
\begin{equation}
    \label{eq:detA0_condition}
    \lambda(z)\equiv\frac{m_1}{(z-z_1)^2}+\frac{m_2}{(z-z_2)^2} = e^{-i\phi},
\end{equation}
for $z$ where $\phi \in[0,2\pi]$ is an arbitrary phase. We set the lens positions $z_1 = (d/2,0)$ and $z_2=-z_1$, such that $d$ is the dimensionless binary separation in units of the Einstein radius, see Eq.~\eqref{eq:d_dimensionless_scale} below. Eq.~\eqref{eq:detA0_condition} can be converted to a polynomial of $z$ of degree 4
\begin{equation}
    \label{eq:crit_poly}
  z^4-\left(\frac{d^2}{2}+e^{i \phi}\right)z^2-d(m_1-m_2)e^{i\phi}z+\frac{d^2}{16}(d^2-4e^{i\phi})=0,
\end{equation}
For each value of $\phi\in[0,2\pi]$, solving for all solutions of $z$ in Eq.~\eqref{eq:crit_poly} gives the critical curves. Substituting the image positions of critical curves in the lens equation Eq.~\eqref{eq:binary_leqn_complex} gives the caustic curves. 

By the fundamental theorem of algebra, Eq.~\eqref{eq:crit_poly} has four complex roots for each value of $\phi$. The critical curves thus consist of 4 branches of continuous curves, which form closed loops because of the periodicity of the $e^{i\phi}$ term in $\lambda(z)$ \citep{Danvek2015}. Each loop might contain multiple branches of solutions of Eq.~\eqref{eq:crit_poly}. The number of branches of solutions contained by a certain loop $C$ can be determined by the magnitude of the winding number $d_C$ of the function $\lambda(z)$:
\begin{equation}
    \label{eq:winding_no}
    d_C = \frac{1}{2\pi i}\oint_C \frac{\lambda'(z)}{\lambda(z)}dz.
\end{equation}
Using Cauchy's argument principle,
\begin{equation}
    \label{eq:argument_principle}
    d_C = N_{\rm zero}-P,
\end{equation}
where $N_{\rm zero}$ is the number of zeros counted with their multiplicity and $P$ is the number of poles counted with their order. For $N$ number of loops in the critical curves labeled as $C_1,{C_2},...,{C_{N}}$, the sum of the magnitude of the winding number of all loops will be 4, the total number of roots of Eq.~\eqref{eq:crit_poly}:
\begin{equation}
    \label{eq:loop_sum}
    \sum_i \left|d_{C_{i}}\right| = 4.
\end{equation}
$\lambda(z)$ has two second-order poles at $z=z_1$ and $z=z_2$, i.e. the positions of the lenses. Solving for $\lambda(z)=0$, it is found that there are always two non-repeating solutions of $z$ at
\begin{equation}
    \label{eq:z_zeros}
    z=\frac{d}{1+q}\left(\frac{1-q}{2} \pm i\sqrt{q}\right),
\end{equation}
Hence $\lambda(z)$ has two simple zeros given by Eq.~\eqref{eq:z_zeros}. 

With the existence of two double-poles and two simple zeros, it can be seen that the maximum number of loops for Eq.~\eqref{eq:loop_sum} to hold is 3, because it is not possible to create 4 loops with winding number 1. Similar arguments have been used in~\citet{Rhie2001}.

\begin{figure}
    \centering
    \includegraphics[width=1\linewidth]{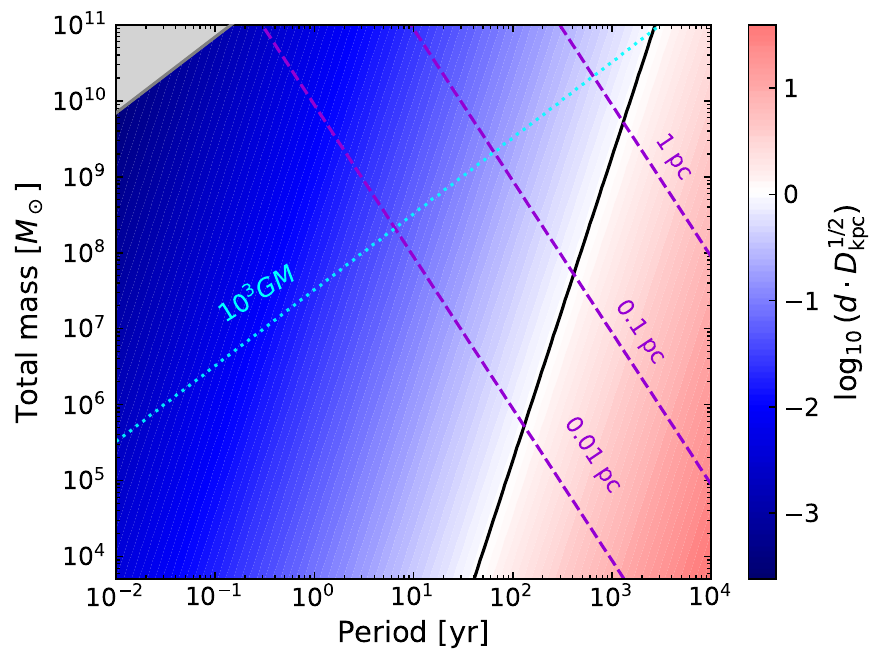}
    \caption{Dimensionless orbital separation $d=a/\xi_0$ (Eq.~\ref{eq:d_dimensionless_scale}) of a binary lens if the source is at 1\,kpc. Dashed lines represent constant physical orbital separation. The dotted line indicates the transition to GW-dominated inspiral at $a=10^3 GM/c^2$~\citep{Haiman_Kocsis_Menou2009}. }
    \label{fig:caustic_size}
\end{figure}

The topology of the critical and caustic curve are determined by two dimensionless quantities: 
\begin{enumerate}[leftmargin=0pt,itemindent=2em,labelwidth=1.5em,align=left]
    \item the 
{symmetric mass ratio 
\begin{equation}
    \eta = m_1 m_2 = \frac{q}{(1+q)^2}
\end{equation}
where $q\equiv m_1/m_2$ is the mass ratio,} and
\item the dimensionless binary separation in units of Einstein-radii
\begin{align}\label{eq:d_dimensionless_scale}
    d \equiv \frac{R_{\rm sep}}{\xi_0} \approx
    \frac{a(1\pm e)}{2\sqrt{R_{\rm g} D_{\rm LS}}} = 0.0075 (1\pm e)\, \frac{T_{\rm yr}^{2/3}}{M_{10}^{1/6}D_{\rm kpc}^{1/2}}\,,
\end{align}
where $R_{\rm sep}$ is the separation between the binary components. Here $R_{\rm g}=GM/c^2$ and Kepler's law have been assumed for $a = 0.01
\, \text{pc} \; M_{10}
^{1/3}T_{\rm yr}^{{2}/{3}}$. The plus/minus sign corresponds to the separation of the binary at apoapsis/periapsis passages. Figure \ref{fig:caustic_size} shows how $d$ varies for the different masses of the SMBHB lens and orbital period. 

\end{enumerate}
 
\begin{figure*}
    \includegraphics[width=1\linewidth]{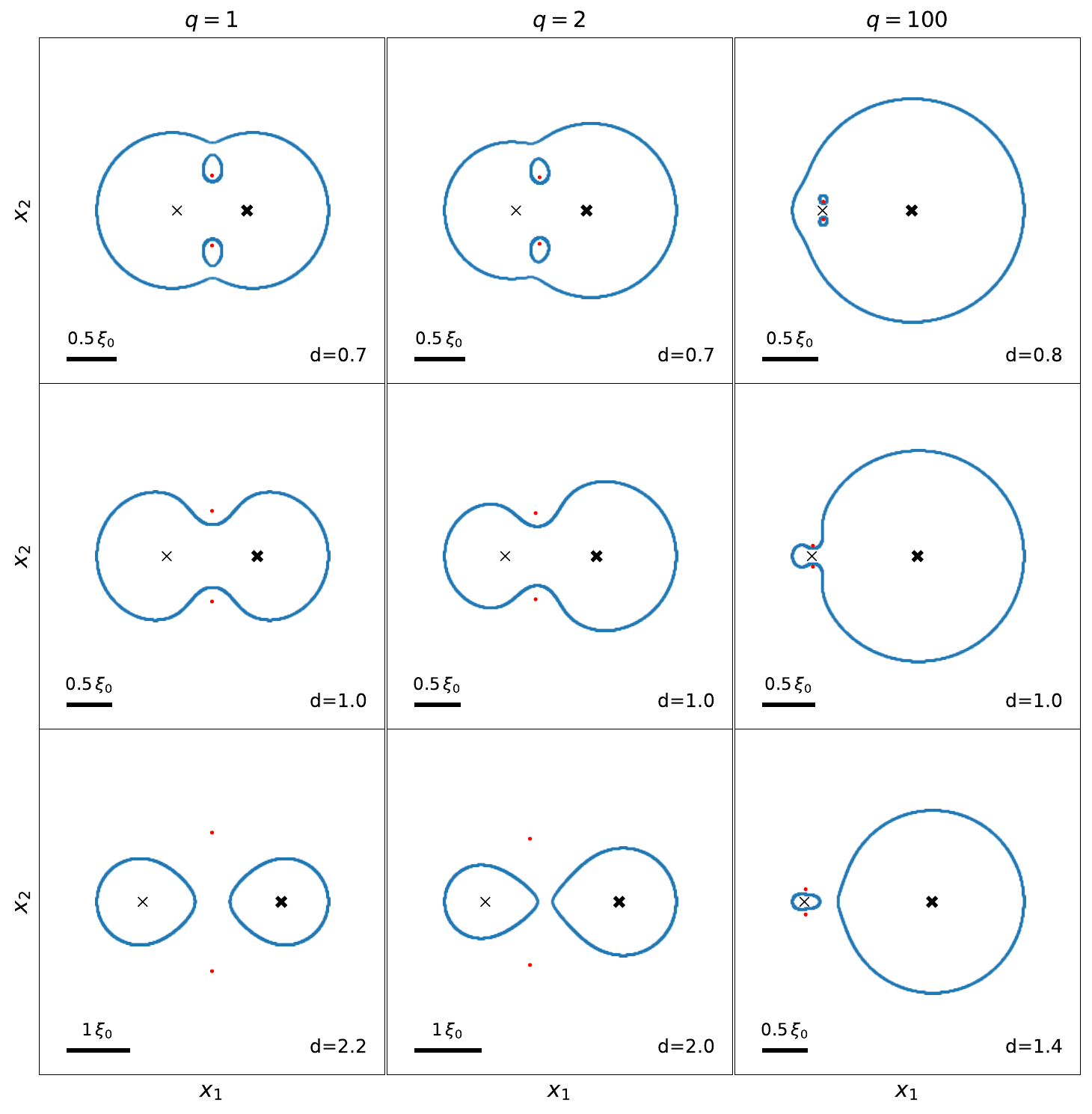}
    \caption{Example critical curves from all topologies for different mass ratios $q=m_1/m_2$. The black crosses mark the positions of the two lenses, with the thicker ones representing the primary component (i.e. the heavier mass when $q\neq1$). The red dots mark the positions of the zeros of $\lambda(z)$ found in Eq.~\eqref{eq:z_zeros}.  The black bar in each panel indicates the length scale in units of the Einstein radius. \label{fig:crit_all}} 
\end{figure*}

\begin{figure*}
    \includegraphics[width=1\linewidth]{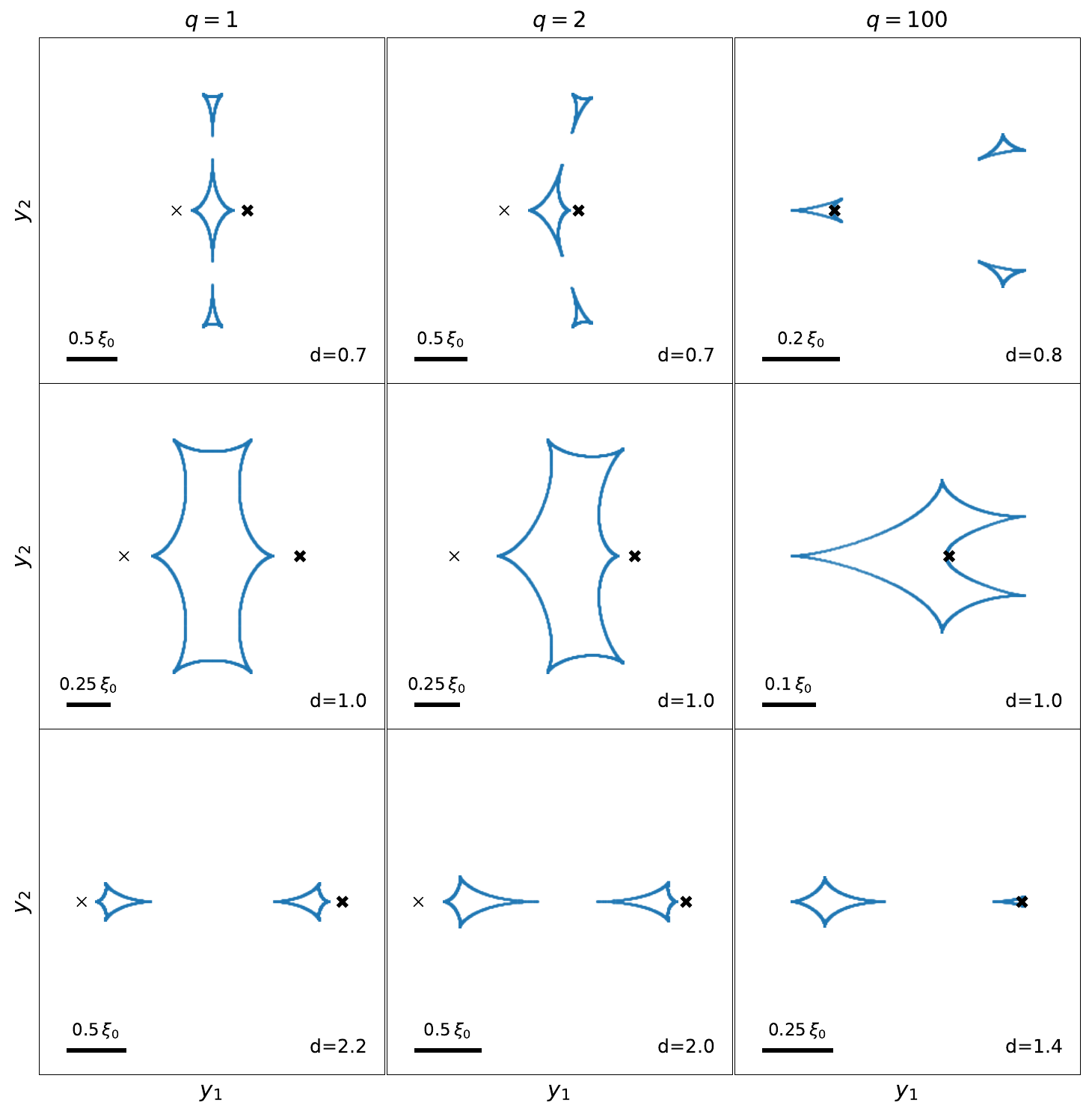}
    \caption{Example caustic curves from all topologies for different mass ratios  $q=m_1/m_2$ corresponding to the critical curves in Fig.~\ref{fig:crit_all}. The black crosses mark the position of the two lenses with the thicker ones representing the primary component (i.e. the heavier mass when $q\neq1$). Note that for the third column of $q=100$, we only mark the heavier lens position for better demonstration. Due to the large mass ratio, the caustic curve is much nearer the more massive lens.  The black bar in each panel indicates the length scale in units of the Einstein radius, Eq.~\eqref{eq:ein_rad}.\label{fig:caustic_all}}
\end{figure*}

Figures~\ref{fig:crit_all} and~\ref{fig:caustic_all} show how changing $\eta$ and $d$ affects the critical curves and caustics. There are always three different topologies of the critical and caustic curves for a fixed $q$ and varying $d$ and vice versa \citep{schneider:1986,Schneider:1993}. We denote the three topologies \textit{``close''}, \textit{``intermediate''}, and \textit{``wide''} binaries following~\citet{Dominik:1999}. For an equal-mass binary where $\eta=1/4$ , the topology transition occurs at $d=1/\sqrt{2}$ and $d=2$ \citep{schneider:1986}. The panels are arranged by $d$ and $q$: each row corresponds to a given topology (close, intermediate, wide) and each column to a mass ratio (equal, $\eta=2/9$, $\eta=0.01$).

The black crosses in Figures~\ref{fig:crit_all} and~\ref{fig:caustic_all} mark the positions of the lenses ($z_1$ and $z_2$) and the red dots in Fig.~\ref{fig:crit_all} mark the positions of the two simple zeros from Eq.~\eqref{eq:z_zeros}. Clearly for the ``close'' topology (first row of Fig.~\ref{fig:crit_all}), the two inner curves (secondary critical curves) each loop around a simple zero. From Eq.\eqref{eq:argument_principle}, the winding number is 1. Hence each inner curve represents a single branch of solution of Eq.~\eqref{eq:crit_poly}. The outer curve (primary critical curve) loops around two simple zeros and two double pole. The magnitude of the winding number is thus 2. Hence the outer curve represents two branches of the solution of Eq.~\eqref{eq:crit_poly}. The sum of the winding numbers is 4, consistent with Eq.~\eqref{eq:loop_sum}. Similar arguments can be made with the other two topologies.


\section{Analytical solutions and sizes of the critical and caustic curves} \label{sec:caus_size}
Eq.~\eqref{eq:crit_poly} can be solved analytically under certain limits for the critical curves and the corresponding caustic curves. In this section, we present {the different types of solutions} and estimate the size of the caustic curves.

\subsection{Tight Binaries}\label{sec:size_tight_binary}

In this work, we will denote binaries with $d\ll1$ as ``tight'' binaries, which belongs to the `close' binary topology in Figs.~\ref{fig:crit_all} and~\ref{fig:caustic_all}.  
Fig.~\ref{fig:caustic} shows how the critical curves and caustic curves change as $d$ decreases toward small values in the close binary topology ($d< 1/\sqrt{2}$) for $q=1$ ($\eta=1/4$). The critical curves consist of a  primary larger loop around both lenses and two secondary smaller loops along the $x_2$ axis. The caustic consist of a diamond-shaped primary loop around the center of mass and two triangular secondary loops along the $y_2$ axis.

As $d$ decreases across the columns, the outer primary critical curve moves closer to the Einstein ring, the inner secondary critical curves shrink in size and become closer to the center of mass. Both the central primary caustic curves and the secondary (triangular) caustic curves shrink, while the secondary caustic curves divert further from the center of mass. A source near the central primary caustic curve has highly magnified images near the corresponding primary 
critical curve, i.e. the Einstein ring. A source near the secondary caustic curve has magnified images near the corresponding secondary critical curves,near the center of mass. 
As the binary lens rotates, the secondary caustic curves span an annular area as seen in the shaded region of the first two columns of the caustic curve plots.

The third column has $d\ll1$. In this case, we only plot the central primary 
caustic curve that has a diamond shape. The corresponding critical curve is similar to the Einstein ring. The four end points of the caustic curve as marked by green dots are the \textit{cusps}, and the rest of the points on the caustic curve are the \textit{folds} which are continuous curves without end points. For example, the four red dots mark the positions of the four innermost fold points.

\subsubsection{The primary critical and caustic curve}

A seen in Fig.~\ref{fig:caustic}, the binary lens approximates to a point-mass lens and the primary critical curve approximates to the Einstein ring to the zeroth order. We solve for the critical and caustic curves perturbatively, similarly to~\citet{Bozza2000,Bozza:2000za}. We use an ansatz for $z$ by expanding around the unit circle (the Einstein ring)
\begin{equation}
    \label{eq:z_crit_ansatz}
    z_{\rm crit,tight}\approx e^{i\theta} + a_1 d + a_2d^2 + a_3d^3 + \mathcal{O}(d^4), 
\end{equation}
where $\theta$ is approximately the polar angle of position $\bm{x}$ on the Einstein ring. Substituting Eq.~\eqref{eq:z_crit_ansatz} into Eq.~\eqref{eq:crit_poly} and solve the equation up to $\mathcal{O}(d^3)$. The $\mathcal{O}(d^0)$ equation gives the relation between $\phi$ and $\theta$:
\begin{equation}
    \label{eq:phi_theta_relation}
    \phi=2\theta+2k\pi, \,\,\, k\in \mathbb{Z}.
\end{equation}
Substituting Eq.~\eqref{eq:phi_theta_relation}, the $\mathcal{O}(d^1)$ equation gives the value of $a_1$:
\begin{equation}
    \label{eq:z_crit_a1}
   a_1={\frac12 \delta}
\end{equation}
Substituting Eqs.~\eqref{eq:phi_theta_relation} and \eqref{eq:z_crit_a1}, the $\mathcal{O}(d^2)$ equation gives the value of $a_2$:
\begin{equation}
    \label{eq:z_crit_a2}
     a_2={\frac32 \eta e^{-i\theta}}
\end{equation}
Substituting Eqs.~\eqref{eq:phi_theta_relation}--\eqref{eq:z_crit_a2}, the $\mathcal{O}(d^3)$ equation gives the value of $a_3$:
\begin{equation}
    \label{eq:z_crit_a3}
    a_3= {-}2 e^{-2 i \theta} {\eta\delta}.
\end{equation}

Substituting Eqs.~\eqref{eq:phi_theta_relation}--\eqref{eq:z_crit_a3} into Eq.~\eqref{eq:z_crit_ansatz} gives the positions of the primary critical curves. The corresponding caustic curves can be found using the lens equation Eq.~\eqref{eq:binary_leqn_complex},
\begin{align}
    \label{eq:caustic_zs_approx_final}
    z_{\rm caus,tight}
    &\approx 
    \frac{\delta d}{2}
+
{2\eta d^2(\cos^3 \theta - i \sin^3 \theta)}
\notag \\
&-
{\eta \delta d^3 (2 e^{-2i \theta} + e^{4i\theta})}
+\mathcal{O}(d^4),
\end{align}
where we introduced another dimensionless mass ratio parameter
\begin{equation}
    \delta = \frac{q-1}{q+1}\,.
\end{equation}
Substituting $\theta=[0,2\pi]$ in Eq.~\eqref{eq:caustic_zs_approx_final} gives the primary diamond-shaped caustic curve (bottom right panel of Fig.~\ref{fig:caustic}). The $\mathcal{O}(d)$ term gives the position of the center of mass of the binary, which corresponds to the center of the caustic curve. The $\mathcal{O}(d^2)$ term gives approximately the radius of the caustic from its center 
\begin{equation}
    \label{eq:radius_caustic}
    \rho_{\rm caus,tight} \approx \frac{\eta d^2}{\sqrt{2}}\left(5+3\cos(4\theta)\right)^{1/2} + \mathcal{O}(d^3).
\end{equation}
Computing the stationary points of $\rho_{\rm caus}$ with respect to $\theta$ gives $\theta=k\pi/4$ where $k\in \mathbb{Z}$. The 4 cusp points are maxima of the caustic radius, located at $\theta=(0,\pi/2,\pi,3\pi/2)$ (green dots in the bottom right panel of Fig.~\ref{fig:caustic}) and the 4 innermost fold points are the minima of the caustic radius, located at $\theta=(\pi/4,3\pi/4,5\pi/4,7\pi/4)$ (red dots in the bottom right panel of Fig.~\ref{fig:caustic}). The radius of the cusp is\footnote{Note that this radius is in units of the Einstein radius, we will keep this convention in the analytical calculations in Sec.~\ref{sec:caus_size}--\ref{sec:mag}.}
\begin{equation}
    \label{eq:radius_cusp}
    \rho_{\rm cusp,tight} \approx 2\eta d^2,
\end{equation}
and the radius of the innermost fold is
\begin{equation}
    \label{eq:radius_fold}
    \rho_{\rm fold,tight} \approx \eta d^2.
\end{equation}

\subsubsection{The secondary critical and caustic curves}

We consider the solutions of the secondary critical and caustic curves in the tight binary regime. It can be seen from Fig.~\ref{fig:caustic_all} that the two secondary critical curve (the two inner circles) are centered around the two zeros of $\lambda(z)$ as worked out in Eq.~\eqref{eq:z_zeros}, which are the dominant $\mathcal{O}(d)$ terms. We thus write an ansatz of the secondary critical curves to solve Eq.~\eqref{eq:crit_poly} up to $\mathcal{O}(d^4)$
\begin{equation}
    \label{eq:z_ansatz_sec}
    z_{\rm crit,sec} = d\left(-\frac{\delta}{2} \pm i\sqrt{\eta}\right)+b_2 d^2+b_3d^3+\mathcal{O}(d^4),
\end{equation}
where the $\mathcal{O}(d)$ term is adopted from Eq.~\eqref{eq:z_zeros}. The $\mathcal{O}(1)$, $\mathcal{O}(d)$ and $\mathcal{O}(d^2)$equations are automatically satisfied. The $\mathcal{O}(d^3)$ equation gives 
\begin{equation}
    \label{eq:b2_sol}
    b_2 =0.
\end{equation}
Substituting Eq.~\eqref{eq:b2_sol}, the $\mathcal{O}(d^4)$ equation gives 
\begin{equation}
    \label{eq:b3_sol}
    b_3= \frac{e^{-i\phi}}{2}
\left[4\eta\delta\pm i\sqrt{\eta}(1-8\eta)\right].
\end{equation}
Substituting Eqs.~\eqref{eq:z_ansatz_sec}--\eqref{eq:b3_sol} into Eq.~\eqref{eq:binary_leqn_complex} gives the solutions of the secondary caustic curves (the two outer triangular shapes)
\begin{align}
z_{\rm caus,sec} =&(\delta\mp2i\sqrt{\eta})
\left(\frac{1}{d}-\frac{d}{2}\right)
\notag\\
&+d^3e^{-i\phi}
\Bigl[
\delta\eta(e^{3i\phi}+2)
\mp\frac{i}{4}\sqrt{\eta}(1-8\eta)(e^{3i\phi}-2)
\Bigr]
\notag\\
&+\mathcal{O}(d^4).
\label{eq:sec_caus_sol}
\end{align}
The first term of Eq.~\eqref{eq:sec_caus_sol} give the centers of the secondary caustics. The distance of the secondary caustics from the barycenter can be estimated as
\begin{equation}
    \label{eq:sec_distance}
    R_{\rm sec} = \frac{1}{d} + \mathcal{O}(d).
\end{equation}
The radius of the secondary caustics can be computed from the $\mathcal{O}(d^3)$ of Eq.~\eqref{eq:sec_caus_sol}
\begin{align}
    \label{eq:rad_sec_caus}
    \rho_{\rm caus,sec} = \frac{\sqrt{\eta}\,d^3}{4} \sqrt{5-4\cos\left(3\phi \pm 8 \alpha\right)},
\end{align}
where $\alpha$ is given by 
\begin{equation}
    \label{eq:alpha_sec}
    \tan(\alpha) \equiv1/\sqrt{q}.
\end{equation}
The minima of the caustic radius are at 
\begin{equation}
    \label{eq:phi_sec_minima}
    \phi 
    =\frac{\mp8\alpha +2k\pi}{3}, \quad k\in \mathbb{Z},
\end{equation}
which give the radius of the secondary caustics at the innermost folds
\begin{equation}
    \label{eq:rad_sec_fold}
    \rho_{\rm sec,fold} {=} \frac{\sqrt{\eta}\,d^3}{4} .
\end{equation}
The maxima of the caustic radius are at 
\begin{equation}
    \label{eq:phi_sec_maxima}
    \phi 
    =\frac{\mp8\alpha +(2k+1)\pi}{3}, \quad k\in \mathbb{Z},
\end{equation}
which give the radius of the secondary caustics at the cusps
\begin{equation}
 \label{eq:rad_sec_cusp}
    \rho_{\rm sec,cusp} =  3\,\rho_{\rm sec,fold} 
\end{equation}

\subsection{Caustic curve of a ultrawide binary}

In this work, we refer to a binary with $d\gg1$ as an ``ultrawide'' binary, which belongs to the wide binary topology. It is expected that the primary critical curves are approximately two Einstein rings around each lens respectively, at $\bm{x}=(\pm d/2,0)$. The radius of the critical curve around lens 1 (lens 2) is approximately the Einstein radius of a point-mass lens of mass $M_1$ ($M_2$). In units of $\xi_0$ of the binary lens, the radius is $\sqrt{m_1}$ ($\sqrt{m_2}$) for lens 1 (lens 2). Similarly to Eq.~\eqref{eq:z_crit_ansatz}, we pick an ansatz for the critical curve around lens 1
\begin{align}
    \label{eq:z_ansatz_crit_wide}
    z_{\rm crit,wide} &= \frac{d}{2} + \sqrt{m_1} e^{i\theta}+ \frac{c_1}{d} + \frac{c_2}{d^2} \notag \\
    &+ \frac{c_3}{d^3}+\frac{c_4}{d^4} + \mathcal{O}(d^{-5}),
\end{align}
where $\theta$ is the polar angle of position $\bm{x}$ on the Einstein ring around lens 1, with the origin being the position of lens 1. Substituting Eq.~\eqref{eq:z_ansatz_crit_wide} into Eq.~\eqref{eq:crit_poly}, the $\mathcal{O}(d^3)$ and $\mathcal{O}(d^4)$ equations are automatically satisfied. The $\mathcal{O}(d^2)$ equation gives
\begin{equation}
    \label{eq:phi_theta_relation_wide}
    \phi=2\theta+2k\pi, \,\,\, k\in \mathbb{Z},
\end{equation}
which is the same as Eq.~\eqref{eq:phi_theta_relation} in the tight binary case. 
Substituting Eq.~\eqref{eq:phi_theta_relation_wide}, the $\mathcal{O}(d)$ equation gives 
\begin{equation}
    \label{eq:b1_sol_wide}
    c_1=0.
\end{equation}
Substituting Eqs.~\eqref{eq:phi_theta_relation_wide}--\eqref{eq:b1_sol_wide}, the $\mathcal{O}(1)$ equation gives 
\begin{equation}
    \label{eq:b2_sol_wide}
    c_2=\frac{e^{3 i \theta}}{4}\sqrt{2\eta(1-\delta)}
\end{equation}
Substituting Eqs.~\eqref{eq:phi_theta_relation_wide}--\eqref{eq:b2_sol_wide}, the $\mathcal{O}(d^{-1})$ equation gives 
\begin{equation}
    \label{eq:b3_sol_wide}
    c_3=-\eta e^{4 i \theta}.
\end{equation}
Substituting Eqs.~\eqref{eq:phi_theta_relation_wide}--\eqref{eq:b3_sol_wide}, the $\mathcal{O}(d^{-2})$ equation gives 
\begin{equation}
    \label{eq:b4_sol_wide}
    c_4=\frac{3e^{5 i \theta}(5+3\delta)\sqrt{2\eta(1-\delta)}}{32}.
\end{equation}
Substituting Eqs.~\eqref{eq:phi_theta_relation_wide}--\eqref{eq:b4_sol_wide} into Eq.~\eqref{eq:z_ansatz_crit_wide} gives the primary critical curve around lens 1. The corresponding caustics is found using the lens equation Eq.~\eqref{eq:binary_leqn_complex}
\begin{align}
    \label{eq:caus_zs_wide}
    z_{\rm caus,wide} &= \frac{d}{2}
-\frac{1-\delta}{2d} \notag \\
&+\frac{\sqrt{2\eta(1-\delta)}\left(\cos^3\theta-i\sin^3\theta\right)}{d^2}
+\mathcal{O}(d^{-3}).
\end{align}
Substituting $\theta=[0,2\pi]$ in, Eq.~\eqref{eq:caus_zs_wide} gives the primary diamond-shaped caustic curve around lens 1. The first two terms give the center of symmetry of the caustic curve. 

The radius of the caustic around lens 1 is given by the $\mathcal{O}(d^{-2})$ term in Eq.~\eqref{eq:caus_zs_wide}
\begin{equation}
    \label{eq:rad_caus_wide}
    \rho_{\rm wide} = \frac{\sqrt{2\eta(1-\delta)}}{d^2}\left(1-\frac{3}{4}\sin^2(2\theta)\right)^{1/2}.
\end{equation}
The maxima of Eq.~\eqref{eq:rad_caus_wide} correspond to the 4 cusp points at $\theta=(0,\pi/2,\pi,3\pi/2)$ and the minima of Eq.~\eqref{eq:rad_caus_wide} correspond to the 4 innermost fold points at $\theta=(\pi/4,3\pi/4,5\pi/4,7\pi/4)$. The radius of the cusp is 
\begin{equation}
    \label{eq:radius_cusp_wide}
    \rho_{\rm cusp,wide} = \frac{\sqrt{2\eta(1-\delta)}}{d^2},
\end{equation}
the radius of the innermost fold is 
\begin{equation}
    \label{eq:radius_fold_wide}
    \rho_{\rm fold,wide} =  \frac{\sqrt{2\eta(1-\delta)}}{2d^2}.
\end{equation}
The radii of the caustic curve around lens 2 can be obtained by substituting $\delta$ with $-\delta$ in Eqs.~\eqref{eq:radius_cusp_wide} and \eqref{eq:radius_fold_wide}.

\section{Magnification of sources near the caustic curve} \label{sec:mag}
In this section, we give analytical estimates of source
magnification near caustic curves. We expand the lens system near the critical point where the magnification is infinite, or $\det A = 0$, where the matrix elements of $A$ are defined by 
\begin{equation}
    \label{eq:T_deriv2}
   A_{ij} = \frac{\partial y_j}{\partial x_i} = \delta_{ij}-\psi_{ij} = \partial_{i}\partial_{j}t_d\equiv T_{ij},
\end{equation}
where $\psi_{ij} \equiv\partial_i\partial_j\psi$ and $\partial_i \equiv \partial/\partial x_{i}$. $\psi$ is the gravitational lensing potential and $t_d$ is the dimensionless time delay, see App.~\ref{app:leqn} for more details. For higher orders of derivatives of the potential, this gives tensors of higher rank, e.g. $T_{ijk} = -\partial_i \partial_j\partial_k \psi$ and $T_{ijkl} = -\partial_i \partial_j\partial_k \partial_l\psi$.

Consider a critical point $\bm{x}^{(0)}$ at which rank of $A$ is 1 and $\det A=0$. The corresponding caustic point is at $\bm{y}^{(0)}$ on the source plane. We work in the coordinate system in which the origin of the image plane (source plane) is at $\bm{x}^{(0)}$ ($\bm{y}^{(0)}$), the $\bm{x}$ and $\bm{y}$ axes are parallel, and $A$ is diagonal at $\bm{x}^{(0)}$. $A = {\rm diag} (T_{11},0)$ where $T_{11}$ and other tensors or $T$ are evaluated in the new coordinate system. 
\subsection{Magnification near a fold}
For a source near a fold, the total magnification is dominated by two bright images near the critical curve, which only exist if the source is inside the caustic curve. Both images have approximately the same magnification of \citep{Schneider:1992}
\begin{equation}
    \label{eq:mu_fold_general}
    \mu_{\rm fold} = \frac{1}{T_{11} \sqrt{2T_{222} y_2}}.
\end{equation}
Note that $T_{11}$ and $T_{\rm 222}$ here refer to their absolute values. We will keep this convention for all the tensors of $T$ in this work. For $A = {\rm diag}(T_{11},0)$, $T_{11} = {\rm tr} A$. From Eq.~\eqref{eq:T_deriv2}, ${\rm tr} A =2-\nabla^2\psi$. For a binary lens, $\nabla^2\psi = 0$ following Poisson's equation and thus $T_{11}=2$. 

For an extended source near the fold, the magnification is computed using Eq.~\eqref{eq:mu_e} by integrating the point-source magnification in Eq.~\eqref{eq:mu_fold_general}. Note that we are taking the magnitude of the magnification and summing up the contribution from both images. Assuming that the source is circular with radius $r$, with uniform brightness and at distance $l$ ($l>0$ indicates the interior of the caustic) from the fold, the total magnification is given by \citep{Miralda-Escude1991} 
\begin{equation}
    \mu_{\rm ext,fold} = \frac{\mu_0}{\sqrt{r}}\,F\left(\frac{l}{r}-1\right),
\end{equation}
where 
\begin{align}
F(y_0) &= \frac{2}{\pi}\int_{\max(-y_0,0)}^{2} \mathrm{d}y \left[  \frac{y(2-y)}{y+y_0}\right]^{1/2}\label{eq:F_def_mag2}
\end{align}

and 
\begin{equation}
    \label{eq:mu_0_def}
    \mu_0 = \frac{2}{T_{11} \sqrt{2T_{222} }}=\frac{1}{\sqrt{2T_{222}}}.
\end{equation}
The maximum of $\mu_{\rm ext,fold}$ occurs at $l=0.65r$ with
\begin{equation}
    \label{eq:mu_fold_xtend_max}
    \mu_{\rm ext,fold} = 1.4\frac{\mu_0}{r^{1/2}}.
\end{equation}

We will now apply these results to the primary and secondary caustics of tight bianries, before addressing ultrawide binaries.

\subsubsection{Tight binary (primary caustics)}

For the primary caustic of a tight binary with $d\ll1$, substituting $\theta=\pi/4$ into Eqs.~\eqref{eq:z_crit_ansatz}--\eqref{eq:z_crit_a3} gives the position of an innermost fold in the lens plane 
\begin{equation}
\bm{x}_{\rm tight,fold} \approx
\begin{pmatrix}
\displaystyle
\frac{1}{\sqrt{2}}
+\frac{d\delta}{2}
+\frac{3\sqrt{2}\,d^2\eta}{4}
\\
\displaystyle
\frac{1}{\sqrt{2}}
-\frac{3d^2\eta}{2\sqrt{2}}
\end{pmatrix}.
\end{equation}
Diagonalizing $A$ at this position gives 
\begin{equation}
    \label{eq:T_222_tight}
    T_{\rm222} \approx 3\eta d^2.
\end{equation}

The maximum magnification of a star during a fold crossing is
\begin{align}
    \label{eq:fold_mag_ext_theory}
    \mu_{\rm tight,fold} &\approx 
    0.57 d^{-1}r^{-1/2} 
    \eta^{-1/2} \notag\\
    &=0.57 \frac{\xi_0}{a(1\pm e)} \sqrt{\frac{\xi_0}{R_{\rm src}}} \eta^{-1/2}
    \notag \\
    &=2\cdot 10^{5}(1\pm e)^{-1} \eta^{-1/2}M_{10}^{5/12}
 T_{\rm yr}^{-2/3} 
 R_{10}^{-1/2}D_{\rm kpc}^{3/4},
\end{align}
where $a$ and $e$ are the semimajor axis and eccentricity of the SMBHB, $R_{10}=(1+z)R_{\rm src} /(10 R_{\odot})$ is the redshifted radius of the source in units of 10 solar radii\footnote{Note that while including this redshift dependence simplifies the equations, it should be noted that the stellar radius in the source frame is approximately independent of redshift.}, $D_{\rm kpc} = D_{\rm LS}/{\rm 1\, kpc}$ is the lens-source distance in units of kpc, the $(1\pm e)$ dependence corresponds to caustic crossing near apastron and periastron passage respectively.

Eq.~\eqref{eq:fold_mag_ext_theory} assumes that the radius of the source is smaller than that of the caustic curve. Using Eq.~\eqref{eq:radius_fold}, it is required that 
\begin{align}
    \label{eq:tight_size_condition}
 R_{\rm src} \leq \eta d^2\xi_0\approx& 3,500R_{\odot} \, (1\pm e)^2\eta 
 M_{10}^{1/6}     T_{\rm  yr}^{4/3} D_{\rm kpc}^{-1/2}.    
\end{align}

\subsubsection{Tight binary (secondary caustics)}
For the secondary caustics of a tight binary with $d\ll1$, we consider the case for $q=1$ for simplicity because the orientation of secondary caustics depend on $q$ (see Eq.~\eqref{eq:alpha_sec}). 

For $q=1$, we take the minus sign in Eq.~\eqref{eq:sec_caus_sol}, and $\alpha=\pi/4$ from Eq.~\eqref{eq:alpha_sec}. Substituting $\phi=0$ into Eqs.~\eqref{eq:z_ansatz_sec}--\eqref{eq:b3_sol} gives the position of an innermost fold in the lens plane:
\begin{equation}
    \label{eq:x_pos_sec_fold}
    \bm{x}_{\rm sec,fold} \approx 
    \begin{pmatrix}
        \displaystyle
    0\\ \displaystyle
    -\frac{1}{4}d(2-d^2)
    \end{pmatrix},
\end{equation}
Diagonalizing $A$ at this position gives
\begin{equation}
\label{eq:T222_sec}
    T_{222} \approx \frac{4}{d^3}.
\end{equation}
The maximum magnification of a star during a fold crossing is 
\begin{align}
    \label{eq:mag_fold_sec}
    \mu_{\rm sec,fold} &\approx 0.5 d^{3/2}r^{-1/2} 
    = 0.5 \left(\frac{a(1\pm e)}{\xi_0}\right)^{3/2}\sqrt{\frac{\xi_0}{R_{\rm src}}} \notag \\
    & = 25 (1\pm e)^{3/2}T_{\rm  yr}
 R_{10}^{-1/2}D_{\rm pc}^{-1/2}, 
\end{align}
where $D_{\rm pc} = D_{\rm LS}/$pc. Note that the magnification is approximately independent of the mass of the binaries. 

Similarly, Eq.~\eqref{eq:mag_fold_sec} assumes that 
\begin{equation}
    \label{eq:size_condition_sec}
    R_{\rm src} \leq d^3 \xi_0/8 \approx3300 R_{\odot} (1\pm e)^3 T_{\rm yr}^2 D_{\rm pc}^{-1}.
\end{equation}

\subsubsection{Ultrawide binary}

For an ultrawide binary with $d\gg1$, consider the caustic around the lens with mass $M_1$. Substituting $\theta = \pi/4$ into Eqs.~\eqref{eq:z_ansatz_crit_wide}--\eqref{eq:b4_sol_wide} gives the position of an innermost fold in the lens plane
\begin{equation}
    \label{eq:x_fold_wide}
    \bm{x}_{\rm wide,fold} \approx 
    \begin{pmatrix}
        \displaystyle
        \frac{d}{2}
        +
        \frac{\sqrt{1+\delta}}{2}
        -
        \frac{\sqrt{\eta(1-\delta)}}{4d^2}
        \\[6pt]
        \displaystyle
        -
        \frac{\sqrt{1+\delta}}{2}
        -
        \frac{\sqrt{\eta(1-\delta)}}{4d^2}
    \end{pmatrix}.
\end{equation}

Diagonalizing $A$ at this position gives 
\begin{equation}
    \label{eq:T_222_wide}
   T_{222}
    \approx
    \frac{3}{d^2\sqrt{q(1+q)}}.
\end{equation}

The maximum magnification of a star during a fold crossing is 
\begin{align}
    \label{eq:mu_fold}
    \mu_{\rm wide,fold}
    &\approx
    0.57\,d\,r^{-1/2}
    \left[\frac{8\eta}{(1-\delta)^3}\right]^{1/4}
    \notag \\
    &=
    0.57
    \frac{a(1\pm e)}{\xi_0}
    \sqrt{\frac{\xi_0}{R_{\rm src}}}
    \left[\frac{8\eta}{(1-\delta)^3}\right]^{1/4}
    \notag \\
    &=
    6\cdot10^3(1\pm e)
    \left[\frac{8\eta}{(1-\delta)^3}\right]^{1/4}
    \notag \\
    &\quad\times
    M_{10}^{1/12}
    T_{\rm kyr}^{2/3}
    R_{10}^{-1/2}
    D_{\rm pc}^{-1/4},
\end{align}
where $T_{\rm kyr}=T/{\rm kyr}$.

Similarly, Eq.~\eqref{eq:mu_fold} assumes that 
\begin{align}
    \label{eq:size_condition_wide}
    R_{\rm src} \leq \frac{\sqrt{2\eta(1-\delta)}}{2d^2}\xi_0 \approx& 3400 R_{\odot} \frac{\sqrt{2\eta(1-\delta)}}{2} \notag
    \\ &\times M_{10}^{5/6}T_{\rm kyr}^{-4/3} D_{pc}^{3/2}
\end{align}
\subsection{Magnification near a cusp}
The caustic curve around a cusp can be approximated by a semi-cubic curve given by \citep{Schneider:1992}
\begin{align}
    \label{eq:cusp_equations}
    y_1^3 = Ky_2^2,
\end{align}
where $y_1$ is along the axis of symmetry of the cusp and $K$ is given by
\begin{equation}
    \label{eq:K_cusp_def}
    K = \frac{27T_{11}^2(T_{122}^2-\frac{1}{3} T_{2222}T_{11})}{8 T_{122}^3}.
\end{equation}

For a point source near a cusp, three bright images near the critical curve exist if the source is inside the caustic curve. The total magnification of the three images is given by \citep{Pejcha:2009}
\begin{align}
    \mu_{\rm cusp,inside} =
\frac{2}{ T_{122}}
\sqrt{\frac{y_1}{y_1^3-Ky_2^2}}
\cos\left[
\frac{1}{3}\arcsin
\sqrt{\frac{Ky_2^2}{y_1^3}}
\right].
   \label{eq:cusp_mag_inside}
\end{align}
One bright images near the critical curve exists if the source is outside the caustic curve with a magnification of \citep{Pejcha:2009}
\begin{align}
    &\mu_{\rm cusp, out} = \frac{K^{-1/3}}{2T_{122}} \notag \\
&\times\frac{
\left[\sqrt{y_2^2-K^{-1}y_1^3}+y_2\right]^{1/3}
+
\left[\sqrt{y_2^2-K^{-1}y_1^3}-y_2\right]^{1/3}
}{
\sqrt{y_2^2-K^{-1}y_1^3}
}.
    \label{eq:cusp_mag_outside}
\end{align}
For an extended source, it is found numerically that the maximum magnification near a cusp is obtained when the source is approximately tangent to the caustic curves near the cusp from inside \citep{Pejcha:2009}. Assuming $K>0$ without loss of generality, the cusp opens in the positive $y_1-$direction. Consider a circular source of radius $r$ centered at $(y_1,y_2)=(l,0)$. We assume that the size of the source is much smaller than that of the cusp. If the source is approximately tangent to the cusp, we have 
\begin{equation}
    \label{eq:cusp_tangent}
    l =\lambda K^{1/3}r^{2/3},
\end{equation}
where $\lambda\sim1$ is some unknown constant. As will be shown later, $K\sim\rho_{\rm cusp}$ for all three limiting cases considered in Sec.~\ref{sec:caus_size}. We assume that the source is much smaller than the caustic, i.e. $K\gg r$. From Eq.~\eqref{eq:cusp_tangent}, we find that 
\begin{equation}
\label{eq:cusp_assump}
    \frac{r}{l} = \frac{1}{\lambda} \left(\frac{r}{K}\right)^{1/3} \ll1.
\end{equation}
We estimate the extended-source magnification near a cusp by integrating the point-source magnification from Eqs.~\eqref{eq:cusp_mag_inside} and \eqref{eq:cusp_mag_outside} 
\begin{align}
    \label{eq:cusp_mag_ext_int}
    \mu_{\rm cusp,ext} &= \frac{1}{\pi r^2}\iint_{(r_1^2+r_2^2) <r^2}\mu_{\rm cusp}(r_1+l,r_2)\,dr_1dr_2 \notag \\
    &\approx \frac{1}{\pi r^2} \iint_{(r_1^2+r_2^2) <r^2}\mu_{\rm cusp}(l,r_2) \,dr_1dr_2 \notag \\ 
    &= \frac{4}{\pi r^2} \int_{0}^{r} \sqrt{r^2-r_2^2} \,\mu_{\rm cusp}(\lambda K^{1/3}r^{2/3},r_2)\,dr_2,
\end{align}
where we have applied Eq.~\eqref{eq:cusp_assump} in the second line, applied Eq.~\eqref{eq:cusp_tangent} and used the symmetry of the cusp in the last line. The cusp magnification, $\mu_{\rm cusp}$, is given by Eq.~\eqref{eq:cusp_mag_inside} if the source at $(l,r_2)$ satisfies $r_2<K^{-1/2}l^{3/2}$, and by Eq.~\eqref{eq:cusp_mag_outside} otherwise. 

Applying the change of variable of $r_2' =r_2/r$, Eq.~\eqref{eq:cusp_mag_ext_int} can be rewritten as 
\begin{equation}
    \label{eq:cusp_ext_simplied_int}
    \mu_{\rm cusp,ext} = \frac{1}{\pi} r^{-2/3}T_{122}^{-1}K^{-1/3} G(\lambda),
\end{equation}
where $G(\lambda)$ is given by
\begin{align}
\label{eq:G_def_cusp_int}
G(\lambda)
=&
\int_0^{\lambda^{3/2}}
8\sqrt{
\frac{\lambda\bigl(1-(r_2')^2\bigr)}
{\lambda^3-(r_2')^2}
}
\cos\!\left[
\frac{1}{3}
\arcsin\!\left(
\frac{r_2'}{\lambda^{3/2}}
\right)
\right]
\,\mathrm{d}r_2'
\notag\\
&+
\int_{\lambda^{3/2}}^{1}
\frac{2\sqrt{1-(r_2')^2}}
{\sqrt{(r_2')^2-\lambda^3}}
\Biggl[
\Bigl(
-r_2'+\sqrt{(r_2')^2-\lambda^3}
\Bigr)^{1/3}
\notag\\
&+
\Bigl(
r_2'+\sqrt{(r_2')^2-\lambda^3}
\Bigr)^{1/3}
\Biggr]
\,\mathrm{d}r_2' .
\end{align}
It is found numerically that $G(\lambda)$ has a maximum value of 9.556 at $\lambda=0.706$. Hence, the maximum magnification for an extended source near a cusp is given by 
\begin{equation}
    \label{eq:cups_max_general}
    \mu_{\rm cusp,max} \approx3 r^{-2/3}T_{122}^{-1}K^{-1/3}.
\end{equation}

We will now apply these results to the primary and secondary caustics of tight binaries, before addressing ultrawide binaries.

\subsubsection{Tight binary (primary caustics)}
For the primary caustic of a tight binary with $d\ll1$, substituting $\theta=0$ into Eqs.~\eqref{eq:z_crit_ansatz}--\eqref{eq:z_crit_a3} gives the position of a cusp in the lens plane. 
\begin{equation}
    \label{eq:cusp_tight_pos}
    \bm{x}_{\rm tight,cusp} \approx 
    \begin{pmatrix}
        \displaystyle 
        1+\frac{d\delta}{2}+\frac{3d^2\eta}{2} \\
        \displaystyle 0
    \end{pmatrix}
\end{equation}
Diagonalizing $A$ at this position gives 
\begin{equation}
    \label{eq:T122_tight}
    T_{122} \approx2,
\end{equation}
and 
\begin{equation}
    \label{eq:K_tight}
    K\approx\frac{27\,d^2\eta}{4}.
\end{equation}
The maximum magnification near the cusp is 
\begin{align}
    \label{eq:cusp_mag_tight}
    \mu_{\rm cusp,tight} &\approx  0.8\,
{d^{-2/3}\eta^{-1/3}r^{-2/3}} \notag \\
    & = 7\cdot10^5 \eta^{-1/3}
M_{10}^{4/9} R_{10}^{-2/3}D_{\rm kpc}^{2/3}T_{\rm yr}^{-4/9}.
\end{align}
Similarly, Eq.~\eqref{eq:tight_size_condition} is required for Eq.~\eqref{eq:cusp_mag_tight} to hold. 
\subsubsection{Tight binaries (secondary caustics)}
For the secondary caustics of a tight binary with $d\ll1$ and $q=1$, we take the minus sign in Eq.~\eqref{eq:sec_caus_sol}. Substituting $\phi=\pi$ into Eqs.~\eqref{eq:z_ansatz_sec}--\eqref{eq:b3_sol} gives the position of a cusp in the lens plane:
\begin{equation}
    \label{eq:cusp_pos_sec}
    \bm{x}_{\rm cups,sec} = 
    \begin{pmatrix}
        \displaystyle 
        0 \\ 
        \displaystyle 
        -\frac{1}{4}d(2+d^2)
    \end{pmatrix}
\end{equation}
Diagonalizing $A$ at this position gives
\begin{equation}
    \label{eq:T122_sec}
    T_{122} \approx \frac{4}{d^3},
\end{equation}
and 
\begin{equation}
    \label{eq:K_sec}
    K \approx\frac{27d^3}{8}.
\end{equation}
The maximum magnficiation near the cusp is 
\begin{align}
    \label{eq:cusp_mag_sec}
    \mu_{\rm cusp,sec} &\approx 0.5 d^2 r^{-2/3} \notag \\
    & =100 T_{\rm yr}^{4/3} D_{\rm pc}^{-2/3}R_{10}^{-2/3}.
\end{align}
Similarly, Eq.~\eqref{eq:size_condition_sec} is required for Eq.~\eqref{eq:cusp_mag_sec} to hold.

\subsubsection{Ultrawide binaries}

For an ultrawide binary with $d\gg1$, consider the caustic around the lens with mass $M_1$. Substituting $\theta = 0$ into Eqs.~\eqref{eq:z_ansatz_crit_wide}--\eqref{eq:b4_sol_wide} gives the position of a cusp in the lens plane 
\begin{equation}
    \label{eq:cusp_pos_wide}
    \bm{x}_{\rm cusp,wide}=
    \begin{pmatrix}
        \displaystyle
        \frac{d}{2}
+\sqrt{\frac{1+\delta}{2}}
+\frac{\sqrt{2\eta(1-\delta)}}{4d^2} \\
\displaystyle
0
    \end{pmatrix}
\end{equation}
Diagonalizing $A$ at this position gives
\begin{equation}
    \label{eq:T122_wide}
    T_{122} \approx \sqrt{\frac{2}{1+\delta}}
\left[2-\frac{3(1-\delta)}{2d^2}\right],
\end{equation}
and 
\begin{equation}
    \label{eq:K_wide}
    K \approx\frac{27\sqrt{2\eta(1-\delta)}}{8d^2}.
\end{equation}
The maximum magnification near the cusp is 
\begin{align}
    \label{eq:mu_cusp_wide}
    \mu_{\rm cusp,wide }&\approx 0.8 \left(\frac{1+\delta}{1-\delta}\right)^{1/3}d^{2/3}r^{-2/3} \notag \\
    &=2\cdot10^4 \left(\frac{1+\delta}{1-\delta}\right)^{1/3} M_{10}^{2/9}
    T_{\rm kyr}^{4/9}
    R_{10}^{-2/3}
\end{align}

\section{Caustic transit magnification curves}\label{sec:static_caustic}

We now consider the caustic transits in which the source's velocity dominates the signal. As explained briefly in Sec.~\ref{sec:blast}, BLAST phenomenology depends very strongly on three timescales: the duration of the survey, the binary's orbital period, and source's caustic crossing time. For the purposes of this section we will consider the binary to be static, that is $t_{\rm cross}\ll T$. The caustic crossing time is largely determined by the distribution of transverse source velocities. 

To estimate the typical transverse speed of the star, we assume that the stars in the galaxy follow a singular isothermal sphere distribution (see e.g. \citealt{binney2008galactic}). The stellar population is spherical and non-rotating with isotropic distribution. The stellar number density is distributed as 
\begin{equation}
    \label{eq:SIS_density}
    n_*(D) \propto D^{-2},
\end{equation}
where $D$ is the distance from the center of the galaxy. For a singular isothermal sphere distribution, the gravitational acceleration is given by $2\sigma_v^2/D$ \citep{binney2008galactic}, where $\sigma_v$ is the one-dimensional velocity dispersion derived from a $M-\sigma$ relation as in Eq.~\eqref{eq:M_sigma}. Adding the contribution from the massive black holes as well, the total gravitational acceleration of a star is given by
\begin{equation}
    \label{eq:grav_acc}
    g(D) = \frac{2\sigma_v^2}{D}+\frac{GM}{D^2}.
\end{equation}
Applying the isotropic Jeans equation \citep{Cappellari:2026},
\begin{equation}
    \label{eq:jeans}
    \frac{d}{dD} \left({n_*(D)\sigma_{\rm 1D}^2(D)}\right) = -n_*(D)g(D),
\end{equation}
where $\sigma_{\rm 1D}$ is the local one-dimensional velocity dispersion. Requiring $\sigma_{\rm 1D}\to\sigma_v$ at large radius gives
\begin{equation}
    \label{eq:one_d_v}
    \sigma_{\rm 1D} = \sqrt{\sigma_v^2+\frac{GM}{3D}}.
\end{equation}
For an isotropic velocity distribution, assuming each component of the velocity follows an independent Gaussian distribution, the mean transverse speed of the star in the source plane is given by
\begin{equation}
\label{eq:vstar_general}
v_* = \sqrt{\frac{\pi}{2}} \sigma_{\rm 1D}\approx
\begin{cases}
\begin{aligned}
&\sqrt{\frac{\pi}{2}}\sigma_v\\
&\quad=550\,M_{10}^{0.23}Z^{-1}\,\mathrm{\frac{km}{s}},
\end{aligned}
& D_{\mathrm{LS}}\gg \frac13 D_{\mathrm{infl}},
\\[4ex]
\begin{aligned}
&\sqrt{\frac{\pi}{6}}\sqrt{\frac{GM}{D_{\mathrm{LS}}}}\\
&\quad=150\,M_{10}^{1/2}D_{\mathrm{kpc}}^{-1/2} 
  \,\mathrm{\frac{km}{s}},
\end{aligned}
& D_{\mathrm{LS}}\ll \frac13 D_{\mathrm{infl}}.
\end{cases}
\end{equation}
Here we defined 
\begin{equation}
    Z=(1+z)^{0.23}
\end{equation}
such that $\sigma_v \propto M^{0.23}/Z$ in the $M$-$\sigma$ relation (see Eq.~\eqref{eq:M_sigma}) and
substituted Eq.~\eqref{eq:one_d_v}. The asymptotic expression uses the radius of the sphere of influence of the massive black holes as 
\begin{equation}
    \label{eq:sphere_of_infl}
    D_{\rm infl} = \frac{GM}{\sigma_v^2} \approx0.23 \, {\rm kpc} \, M_{10}^{0.54}Z^2.
\end{equation}

Substituting  $v_*$ into Eq.~\eqref{eq:tcross_def}, for the primary caustic of a tight binary we have 
\begin{equation}
\label{eq:tcross}
t_{\rm cross}=
\begin{cases}
        0.56\,{\rm yr}\, \eta M_{10}^{-0.06}T_{\rm yr}^{4/3} D_{\rm kpc}^{-1/2}Z,&D_{\rm LS} \gg \frac13 D_{\rm infl} \\[2ex]
       2.05\,{\rm yr} \, \eta T_{\rm yr}^{4/3}M_{10}^{-1/3},& D_{\rm LS}\ll\frac13 D_{\rm infl}
\end{cases}
\end{equation}

The ratio $t_{\rm cross}/T$ is given by
\begin{align}
    \label{eq:cross_period_ratio}
    \frac{t_{\rm cross}}{T}  = \begin{cases}
       0.56 \eta\,M_{10}^{-0.06}T_{\rm yr}^{1/3} D_{\rm kpc}^{-1/2}Z,&D_{\rm LS} \gg \frac13 D_{\rm infl} \\[2ex]
        2.05 \eta \, T_{\rm yr}^{1/3}M_{10}^{-1/3},& D_{\rm LS}\ll \frac13 D_{\rm infl}
    \end{cases}
\end{align}
where $t_{\rm cross}/T$ is fixed inside the sphere of influence and decreases with larger $D_{\rm LS}$ outside the sphere of influence. 

The peak duration of each caustic crossing in this case is estimated using the total time during which part of the star overlaps with single caustic curve~\citep{Venumadhav:2017pps}
\begin{equation}
    \label{eq:peak_T}
    T_{\rm peak} = \frac{2R_{\rm src}}{v_*} = 
    \begin{cases}
       0.3\,{\rm day}\,M_{10}^{-0.23}R_{10}Z,&D_{\rm LS} \gg \frac13 D_{\rm infl} \\[2ex]
       1.07\,{\rm day} \, M_{10}^{-1/2}D_{\rm kpc}^{1/2} R_{10},& D_{\rm LS}\ll \frac13 D_{\rm infl}
\end{cases}
\end{equation}

Consider a static binary lens system of $T= 10$ yr, $q=1$, $M =10^{10} \Msun$, $D_{\rm LS} = 10 $kpc. $t_{\rm cross}/T =0.1$ using Eq.~\eqref{eq:cross_period_ratio}, and we approximate the caustic curve as static during the crossing of the star for simplicity. This system forms a tight binary with $d=0.01$. We will explore the role of the initial position and velocity on the morphology of the magnification curve.

\begin{figure*}
    \centering
    \includegraphics[width=1\linewidth]{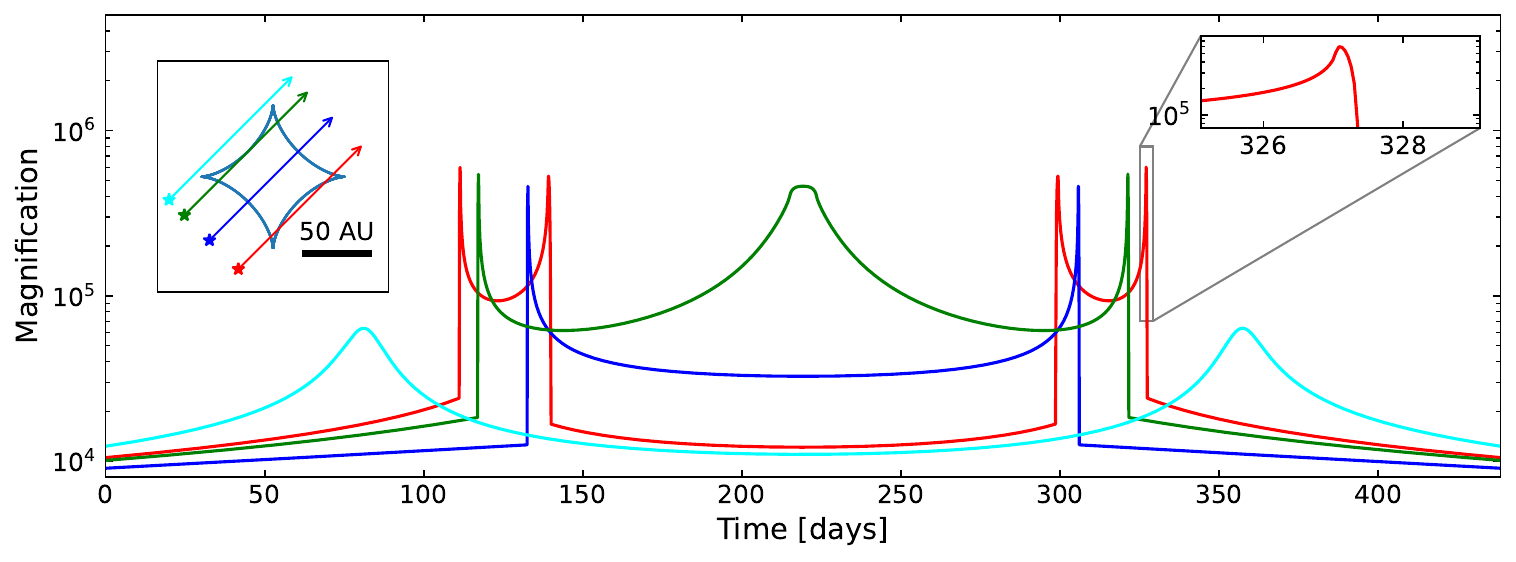}
    \caption{Magnification curves of stars moving across a static caustic curve at $45\degree$ with respect to the $y_1$ axis. Here $M=10^{10} \Msun, T=10 \,{\rm yr}, D_{\rm LS}=10 \, {\rm kpc}, R_{\rm src}=10R_{\odot}$. The left inset shows the four trajectories of the star with the caustic curve (the black bar shows the length scale of the inset). The colours of the trajectories match accordingly with that of the magnification curves. The right inset shows a zoom-in of the magnification curve near one of the fold crossings.}\label{fig:caus_static_lightcurve_fold}
\end{figure*}

Fig.~\ref{fig:caus_static_lightcurve_fold} shows four example magnification curves of a bright star with radius $10R_{\odot}$ traveling past a static caustic curve and the corresponding trajectories. Along the ``red'' trajectory, the star crosses the caustic curve near a fold four times, which corresponds the maximum number of peaks in the magnification curve in this scenario. The peak duration is estimated to be 0.3 day using Eq.~\eqref{eq:peak_T}, see the right inset in Fig.~\ref{fig:caus_static_lightcurve_fold}. Each peak is asymmetric which is characteristic of fold crossing because of the absence of highly magnified image for a source outside a fold \citep{Miralda-Escude1991}. Consecutive fold crossings at the same caustic arm have an intermediate trough between them with relatively high magnification. The 'blue' trajectory crosses two innermost fold points, creating two similar peaks. The ``green'' trajectory produces three peaks with the middle one having a much longer duration than the prediction of Eq.~\eqref{eq:peak_T}. Eq.~\eqref{eq:peak_T} predicts the peak duration by assuming that the star moves orthogonal to the caustic curve (which can be approximated as a straight line locally, see e.g.~\citealt{Schneider:1992}). The ``green'' trajectory moves along the fold line instead, creating the wider and more symmetric peak. The ``cyan'' trajectory is a close encounter with the caustic instead of a caustic crossing. The magnification curve produces two lower peaks as a result of the two cusp flybies.

\begin{figure*}
    \centering
    \includegraphics[width=1\linewidth]{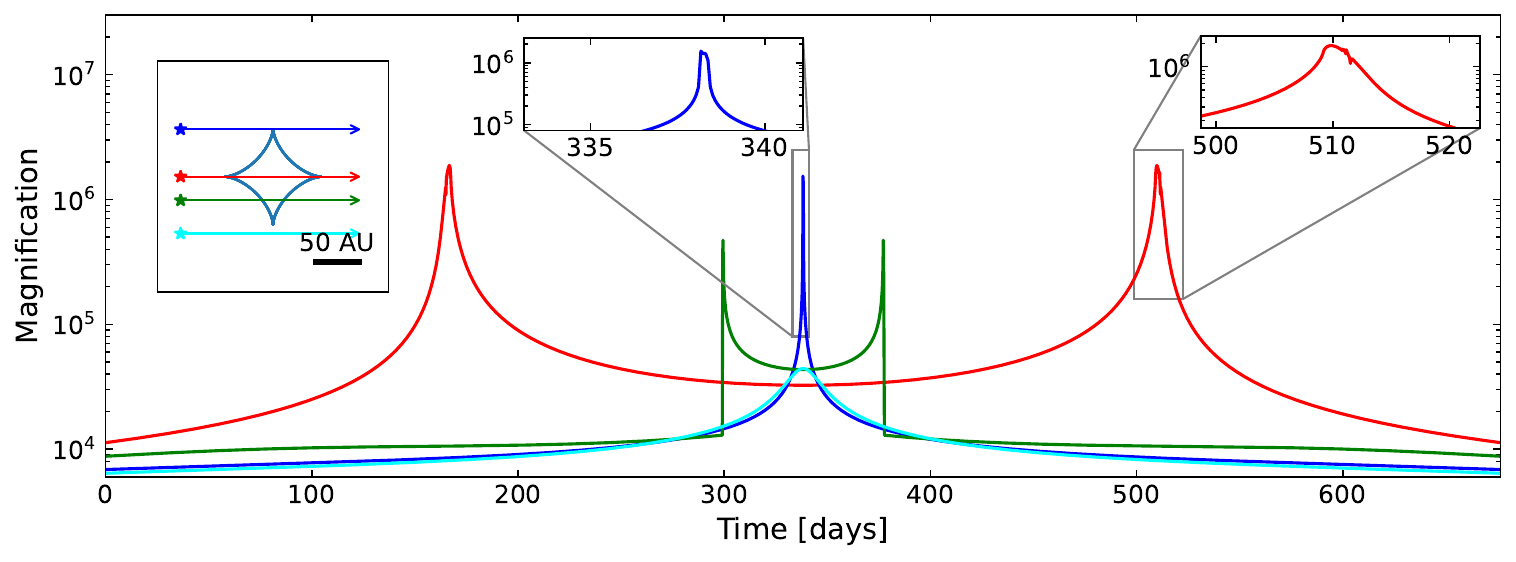}
    \caption{Magnification curves of stars moving across a static caustic curve along the $y_1$ axis direction. Here $M=10^{10} \Msun, T=10 \,{\rm yr}, D_{\rm LS}=10 \, {\rm kpc}, R_{\rm src}=10R_{\odot}$. The left inset shows the four trajectories of the star with the caustic curve (the black bar shows the length scale of the inset). The colours of the trajectories match accordingly with that of the magnification curves. The middle and right inset show zoom-ins of the magnification curve near two of the cusp crossings.}\label{fig:caus_static_lightcurve_cusp}
\end{figure*}

Fig.~\ref{fig:caus_static_lightcurve_cusp} shows another four example magnification curves for stars traveling past the same caustic curve and the corresponding trajectories. The ``green'' trajectory crosses the caustic curve at two fold points, producing two peaks. The ``red'' trajectory crosses two cusp points along the cusp axis, producing two peaks. The peaks are more symmetric than the fold crossing peaks because of the existence of one highly magnified image outside the cusp. It is also noticed that the peak duration of the magnification curve is underestimated by the Eq.~\eqref{eq:peak_T} because of the different geometry of images near the cusp. The 'blue' trajectory crosses near the cusp point orthogonal to the cusp axis, producing a symmetric peak with smaller width than the ``red'' trajectory. The ``cyan'' trajectory represents a close encounter with the caustic. The magnification curve has one lower peak corresponding to the cusp flyby.

\section{\qpls{} magnification curves: exploring the parameter space of the lens-source system}
\label{sec:different_lightcurves}

In contrast to caustic transits, \qpls{} involves many repeated crossings due to the binary system's motion. The resulting magnification curves display a very rich structure, which depends strongly on the parameters controlling the properties of the binary components and their orbit, as well as the properties of the source. We will now explore how variations of lens/source properties affect the shape of the magnification curves, generalizing the results presented in \PaperI.

We will focus mainly on the central caustic of tight binaries, which have shorter period and higher magnification except for Sec.~\ref{sec:intermediate-sep}.
For simplicity, we assume that the SMBHB is oriented face-on except in Sec.~\ref{sec:inclination}. We also assume that the star is static except for Secs.~\ref{sec:moving} and \ref{sec:multi-source}. The magnification during a caustic crossing is computed numerically, but matches the analytical expressions given in Sec.~\ref{sec:mag}.

For a source with diameter $2R_{\rm src}$ located near the cusp, the crossing time at the primary caustic is given by
\begin{align}\label{eq:caus_cross_time_detailed}
t_{\rm mag, tight} &= \frac{2 R_{\rm src}}{\omega \rho_{\rm cusp,tight}\xi_0}{ =
\frac{r}{\omega \eta d^2} }
= 
\frac{R_{\rm src} \xi_0 }{\omega R_{\rm sep}^2\eta} 
\notag \\ &=  \frac{R_{\rm src} \xi_0 }{\sqrt{GMa(1-e^2)}\eta}
= \frac{T}{\eta}\frac{R_{\rm src} \xi_0 }{2\pi a^2 \sqrt{1-e^2}}
\notag\\
&= \frac{T}{2\pi \eta}\frac{r}{d_a^2 \sqrt{1-e^2}}
\notag\\
&=4\,{\rm hr}\,\frac{R_{10} {M_{10}}^{-1/6} T_{\rm yr}^{-1/3} D_{\rm kpc}^{1/2}}{\eta\sqrt{1-e^2}},
\end{align}
where in the third equality, we identified the specific angular momentum with $\omega R_{\rm sep}^2 = \sqrt{GMa(1-e^2)}$ and noted that it is a constant of motion, {and defined the constant $d_a = a/\xi_0$}. In this calculation we neglect the change of the SMBHB separation, valid at apoapsis and periapsis. 
The result is equal to $T/\eta$ times a geometrical factor: the ratio of the source radius times Einstein radius over the area of the SMBH orbit.\footnote{
The crossing time at the secondary caustic is given by 
\begin{align}
 \label{eq:caus_cross_sec}
    t_{\rm mag,sec} &= \frac{2R_{\rm src}}{\omega R_{\rm sec}\xi_0}  =\frac{2R_{\rm src}d}{\omega \xi_0} \notag \\
    &= 12  {\rm s} \, M_{10}^{-2/3}T_{\rm yr}^{5/3} 
 R_{10}D_{\rm pc}^{-1} \frac{(1\pm e)^3}{\sqrt{1-e^2}} \, .
\end{align}
}

In this section, we extended the modeling of \qpls{} magnification curves of \PaperI by exploring the parameter space of the lens-source system. We investigated the effect of the mass ratio, separation, inclination, eccentricity, precession of the binary lenses in Sec.~\ref{sec:unequal_mass}--\ref{sec:eccentric}. In Sec.~\ref{sec:moving}, we investigate the interplay of lens orbit and source motion. In Sec.~\ref{sec:multi-source}, we explore the imprint of multiple sources, first at different distances from the lens and then forming a binary system.

\subsection{Unequal masses}\label{sec:unequal_mass}

In this section, we consider the SMBHB with unequal masses. Eq.~\eqref{eq:radius_fold} and Eq.~\eqref{eq:fold_mag_ext_theory} show that binaries with unequal masses produces smaller caustics but higher magnification. As seen in Figure~\ref{fig:caustic_all}, a larger mass ratio makes the caustic curve more asymmetric. For a tight binary with $q\gg1$ ($\eta \to 0$), the primary caustic might still display strong asymmetry.

\begin{figure*}
    \includegraphics[width = \linewidth]{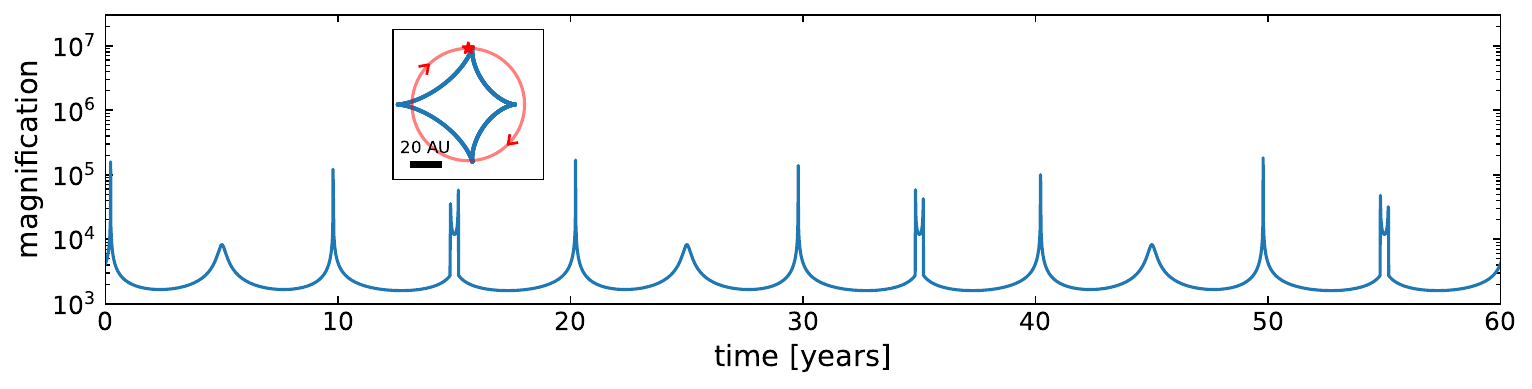}
    \caption{The magnification curve of a \qpls{} system with unequal mass ratio. Here $M=5\cdot10^9 \Msun$, $q=100$, $T =20$ yr, $D_{\rm LS}=200 $ pc, $R_{\rm src} =10 R_{\odot}$. The inset shows the trajectory of the star in the co-rotating frame of the caustic curve. The black bar labels the length scale of the caustic curve.}\label{fig:unequal_lightcurve}
\end{figure*}

Consider a \qpls{} system with $M=5\cdot10^9 \Msun$, $q=100$ ($\eta=0.01$), $T =20$ yr, $D_{\rm LS}=200 $ pc, $R_{\rm src} =10 R_{\odot}$. This system forms a tight binary with $d=0.14$. The SMBHB is $6\cdot10^6$ years before merger hence the caustic curve is approximately fixed in size . Fig.~\ref{fig:unequal_lightcurve} shows the magnification curve of this system. The inset shows the trajectory of the star in the co-rotating frame of the caustic. With the large mass ratio, the caustic curve displays a strong asymmetry as compared to that with $q=1$ (see the inset of Fig.~\ref{fig:unequal_lightcurve}). For the magnification curve during each period, there are two singly-peaked cusp crossings with the highest magnification (when the star crosses the upper and lower caustic arm), one doubly-peaked fold crossing (when the star crosses the left caustic arm), and one lower peak (when the star flies by the right caustic arm).

\subsection{Intermediate-separation binary} \label{sec:intermediate-sep}

Deviations from the tight-binary regime can modify the shape of magnification curves due to the asymmetry of the caustics. For an equal-mass binary with $2> d> 1/\sqrt{2}$, it is in the 'intermediate' binary topology. The caustic curve have different length along $y_1$ and $y_2$ axis, as seen in Fig.~\ref{fig:caustic_all}. This will also result in a varying peak amplitude of caustic crossings in a \qpls{} system, similarly to the case with an unequal-mass binary in Sec.~\ref{sec:unequal_mass}.

\begin{figure*}
    \includegraphics[width=\linewidth]{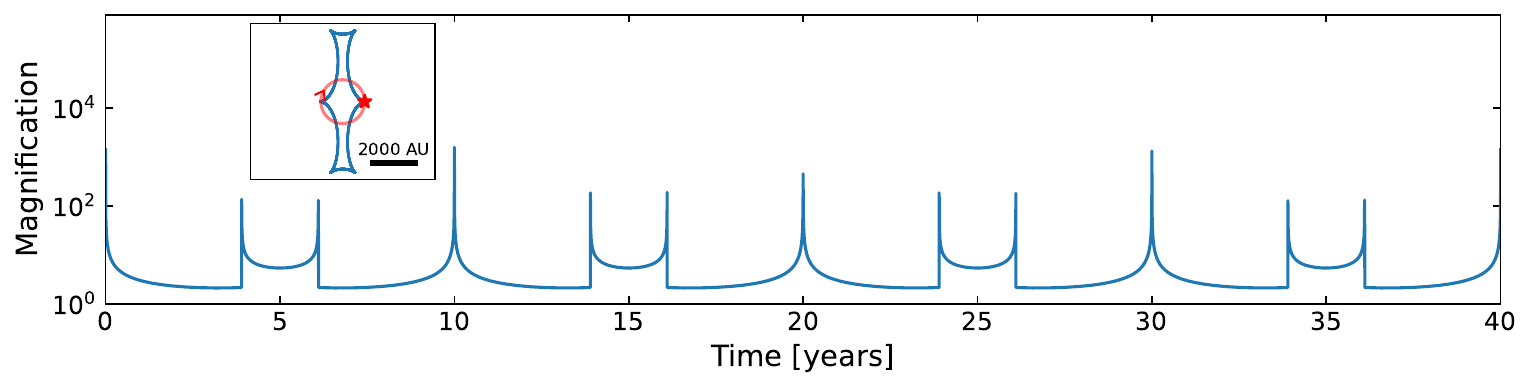}  \caption{The magnification curve of a \qpls{} system with an binary at ``intermediate'' separation $d=0.85$. Here $T= 20$ yr, $M =10^{8} \Msun$, $D_{\rm LS} = 20 $pc, $R_{\rm src} = 10R_{\odot}$. The inset shows the trajectory of the star in the co-rotating frame of the caustic. The black bar shows the length scale of the caustic plot. }\label{fig:intermediate_lightcurve}
\end{figure*}

Consider a \qpls{} system with $T= 20$ yr, $M =10^{8} \Msun$, $D_{\rm LS} = 20 $pc, $R_{\rm src} = 10R_{\odot}$. This forms a intermediate-binary system with $d=0.85$. The SMBHB is $2\cdot10^8$ years before merger hence the caustic curve is approximately fixed in size . 
Fig.~\ref{fig:intermediate_lightcurve} shows the trajectory of the star around the caustic and the magnification curve. The caustic curve has a larger size along the $y_2-$axis, resulting in the varying peak amplitude and number of peaks in the magnification curve.

\subsection{Inclined binary}\label{sec:inclination}

In general, the binary does not lie in the plane perpendicular to the line of sight. Let us consider a binary with standard orbital elements: longitude of ascending node $\Omega$, inclination $\iota$, argument of periapsis $\omega$ and true anomaly $\nu$. The orbital elements are defined with respect to the reference direction $x-$axis and the reference plane $xy-$ plane. The image plane is the $xy-$ plane and the line of sight is along the $z-$axis. In the $xyz-$coordinate system, the relative position vector of the binary is given by
\begin{equation}\label{eq:r_general}
\mathbf r
=
r
\begin{bmatrix}
\cos\Omega\,\cos(\omega+\nu)
-
\sin\Omega\,\sin(\omega+\nu)\cos\iota
\\[6pt]
\sin\Omega\,\cos(\omega+\nu)
+
\cos\Omega\,\sin(\omega+\nu)\cos\iota
\\[6pt]
\sin(\omega+\nu)\sin\iota
\end{bmatrix},
\end{equation}
where $r$ is the orbital radius given by
\begin{equation}
    \label{eq:orbital_radius}
    r=\frac{a(1-e^2)}{1+e\cos\nu}.
\end{equation}
The caustic curve is determined by $\bm{r}_{\perp}$, which the position of $\bm{r}$ projected onto the image plane ($xy-$plane):
\begin{equation}\label{eq:r_perp}
\mathbf r_{\perp}
=
r
\begin{bmatrix}
\cos\Omega\,\cos(\omega+\nu)
-
\sin\Omega\,\sin(\omega+\nu)\cos\iota
\\[6pt]
\sin\Omega\,\cos(\omega+\nu)
+
\cos\Omega\,\sin(\omega+\nu)\cos\iota
\end{bmatrix},
\end{equation}
from which $d=|\bm{r}_{\perp}|/\xi_0$ is computed and sets the topology and size of the caustics. Consider the simple case of a quasi-circular binary with $\Omega=0$, $d$ varies between $a/\xi_0$ and $a\cos{\iota}/\xi_0$:
\begin{equation}
    \label{eq:d_inclinded}
    d= \frac{a}{\xi_0}\sqrt{\cos^2{\nu}+\sin^2\nu\cos^2\iota},
\end{equation}
The consequence is that $d$ changes during each orbital period of the binary, and hence the size of the caustic curve. This will be manifested in the features of the magnification curve.

\begin{figure*}
  \includegraphics[width=\linewidth]{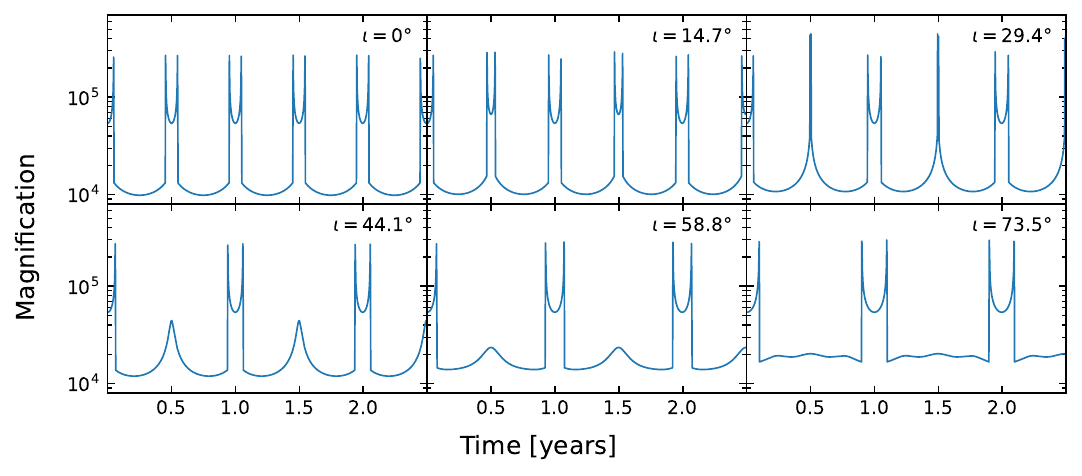}
    \includegraphics[scale = 0.8]{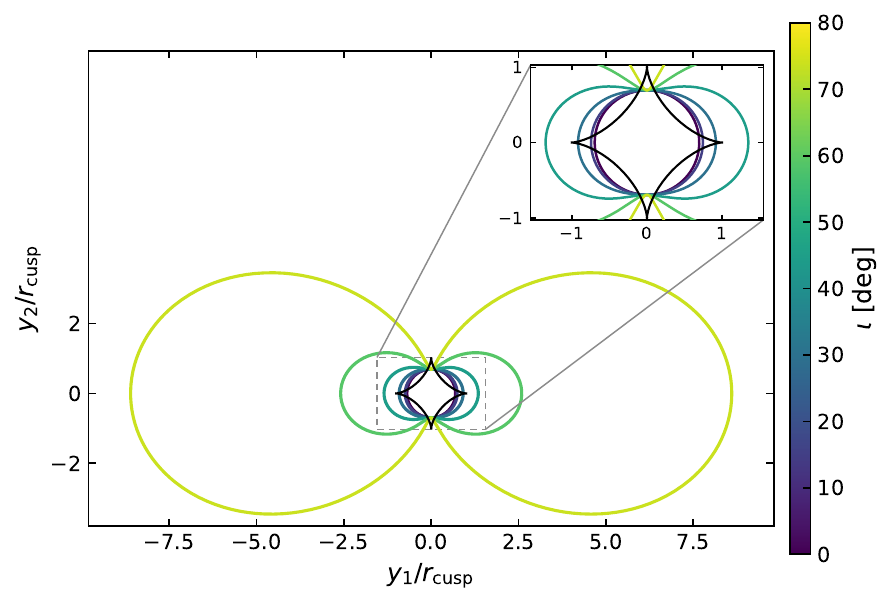}
    \caption{\textit{Top}: the magnification curves of \qpls{} systems with 6 different inclinations of the binary. We have $T= 2$ yr, $M =10^{10} \Msun$, $D_{\rm LS} = 1 $kpc, $R_{\rm src} = 10R_{\odot}$. \textit{Bottom}: The trajectories of star in the co-rotating frame of the caustic curve, in units of the instantaneous cusp radius, corresponding to the 6 cases in the top panel. The colour of the trajectories indicate the inclination angle. The inset shows a zoom-in at the central region where trajectories with smaller binary inclination angles lie.}\label{fig:caus_corotate_inclinations}
\end{figure*}

Consider an example \qpls{} system with $T= 2$ yr, $M =10^{10} \Msun$, $D_{\rm LS} = 1 $kpc, $R_{\rm src} = 10R_{\odot}$. The SMBHB is 176 years before merger and is approximately not inspiralling during the observational time. We compute the magnification curves for 6 different inclination angles of the binary, as in the top panel of Fig.~\ref{fig:caus_corotate_inclinations}. The bottom panel shows the trajectories of the star with different binary inclination angles in the co-rotating frame of the caustic curve, in units of the instantaneous cusp radius. This change of frame is possible because the caustic curve only changes its size but not its shape during the evolution. Under this frame, the caustic curve is fixed and we can observe where the caustic crossing events occur along the caustic curve.

For the zero-inclination case, each caustic crossing event has the same amplitude, width and number of peaks because the caustic crossing events occur at the same location along each caustic arm. As the inclination angle increases, two of the caustic crossings of each orbital period (when the binary is aligned with the lens plane) have the roughly the same peak amplitude but increasing peak width. These two crossing peaks correspond to caustic crossings at the upper and lower caustic arms, where the instantaneous size of the caustic is the largest ($d$ is maximum at $\nu=0$ in Eq.~\eqref{eq:d_inclinded}). The other two of the caustic crossings initially gain narrower width and higher amplitude, becoming singly-peaked cusp crossings at a certain stage. These two crossings correspond to the those at the left and right caustic arms, where the instantaneous size of the caustic is the smallest ($d$ is minimum at $\nu=\pi/2$ in Eq.~\eqref{eq:d_inclinded}). As the inclination angle further increases, the star only flies by the cusp point outside the caustic curve and these two sub-leading peaks eventually disappear. In summary, the number of caustic crossings per period of the binary decreases from 4 to 2 as the inclination angle of the binary increases.

\subsection{Eccentricity}\label{sec:eccentric}

In an eccentric binary the separation between components changes over the orbital period. This causes the projected size of the caustic to change, even if the binary is face-on. $d$ varies between $(1\pm e)a/\xi_0$. The changing size of the caustic curve will produce a varying peak amplitude in the magnification curve. \PaperI demonstrated two examples with eccentricity, specifically in reducing the timescale over which the binary evolves towards merger. 
In this work, we will study the case for which the orbit is quasi-periodic (i.e.~no inspiral evolution). We will also discuss the effect of precession at first post-Newtonian (1PN) order.

\subsubsection{Without precession}\label{sec:no_prec}
We first ignore the effect of apsidal precession and simulate the Newtonian orbit using REBOUND \citep{rebound,reboundias15}. Consider a \qpls{} system with $T= 0.5$ yr, $M =6\cdot10^{8} \Msun$, $D_{\rm LS} = 1
$kpc, $R_{\rm src} = 10R_{\odot}$. We evaluate the magnification curves for four eccentricity values: $e=0,0.17,0.33,0.5$.
{Even for the most eccentric case ($e=0.5$),
the SMBHB is approximately 170 years before merger,
so we neglect orbital decay over the observational
period.} 
For each case, the source is positioned to be close to the cusp radius of the binary during apoapsis passage. 

\begin{figure*}
    \includegraphics[scale=0.57]{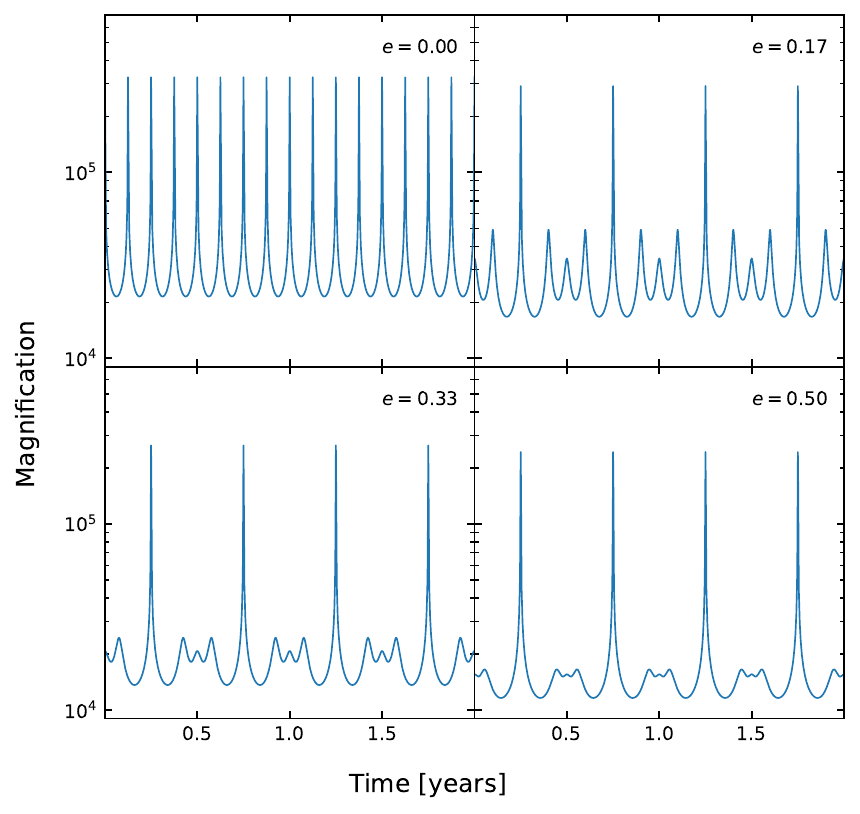}
    \includegraphics[scale=0.63]{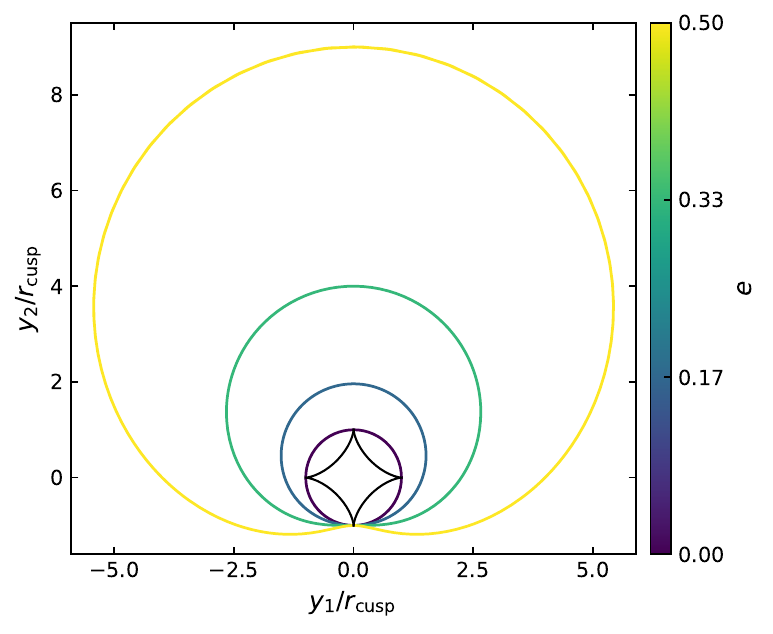}
    \caption{\textit{Left}: magnification curves of \qpls{} systems with 4 different eccentricities, ignoring precession. Here $T= 0.5$ yr, $M =6\cdot10^{8} \Msun$, $D_{\rm LS} = 1
$kpc, $R_{\rm src} = 10R_{\odot}$. \textit{Right}: the corresponding trajectories of the source star in the co-rotating frame of the caustic curve in units of the instantaneous cusp radius during a single period of the massive binary. The colour indicates the eccentricity of each trajectory. \label{fig:lightcurve_eccentric}}
\end{figure*}

The left panel of Fig.~\ref{fig:lightcurve_eccentric} shows the magnification curves and the right panel shows the trajectories of the source star in the co-rotating frame of the caustic curve. The distance of the source to the center of mass is re-scaled with the size of the caustic curve at each time in the right panel. For the circular case, there are four single-peaked cusp crossings for each period. For eccentric cases, the star flies by the cusp outside the caustic except during apoapsis passages. As a result, there is one primary peak (caustic crossing) and three sub-leading peaks (cusp flybies) in the magnification curve during each period of the binary. As the eccentricity increases, the sub-leading peaks become weaker and negligible eventually. In summary, as eccentricity increases, number of caustic crossings per period decreases from 4 to 1. Note the difference from the case with increasing inclinations studied in Sec.~\ref{sec:inclination}, where the number of caustic crossings period decreases to 2. This comparison shows the an example of potential degeneracies between different \qpls{} systems which require further studies.

\subsubsection{Periaxis precession} \label{sec:precession}

We now consider the same \qpls{} eccentric system, including general-relativistic periastron precession at the first post-Newtonian (1PN) order. An eccentric SMBHB experiences periaxis precession and the argument of periapsis $\omega$ changes at a rate of (see e.g.~\citealt{Naoz2013})
\begin{equation}
    \label{eq:precession_rate}
    \frac{d\omega}{dt} = \frac{3 G^{3/2} (M_1 + M_2)^{3/2}}{a^{5/2} c^2 (1 - e^2)},
\end{equation}
which causes the orientation of the caustic curve with the maximum size during each period to change over time. The time taken for the argument of periapsis to precess by $2\pi$ is given by 
\begin{equation}
    \label{eq:t_prec}
    T_{\rm prec} = \frac{2\pi}{d\omega/dt}= 7.3 \, {\rm yr}(1-e^2) M_{10}^{-2/3}T_{\rm yr}^{5/3}.
\end{equation}

We simulate the motion of the binary using REBOUND with REBOUNDx \citep{rebound,reboundias15,reboundx}. The binary has $e=0.5$ and $M =6\cdot10^{8} \Msun$ as in the previous case without precession. The semi-major axis of the binary is $531$ AU such that a Keplerian binary of the $M$, $a$ and $e$ will have a period of 0.5 years as in the previous case without precession. The binary is initialized at periapsis, with a separation of $a(1-e)$. The initial velocity of the binary is tuned slightly around the Keplerian velocity such that the separation at apoapsis is $a(1+e)$. $T_{\rm prec}$ is found numerically to be 12 years while Eq.~\eqref{eq:t_prec} predicts a precession time scale of 11 years.

\begin{figure*}
    \includegraphics[scale=0.63]{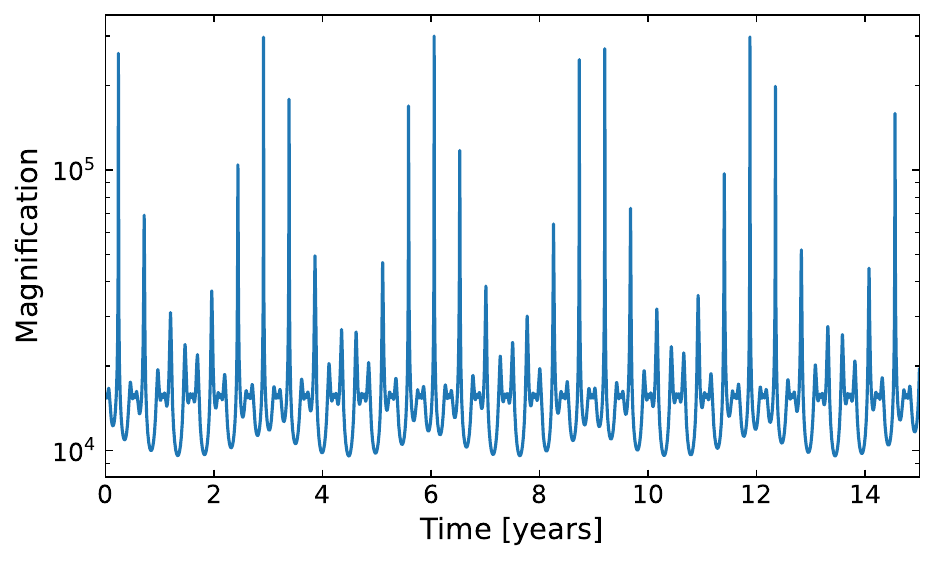}
    \includegraphics[scale=0.63]{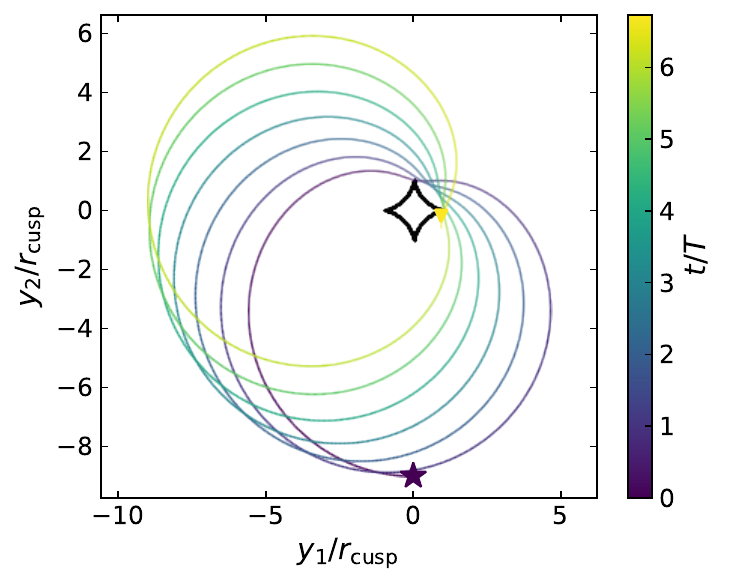}
    \caption{\textit{Left}: the 
    magnification curve of a \qpls{} system with an eccentric binary including the effect of 1PN precession. Here $T= 0.5$ yr, $e=0.5$, $M =6\cdot10^{8} \Msun$, $D_{\rm LS} = 1
$kpc, $R_{\rm src} = 10R_{\odot}$. $T_{\rm prec}$ is found numerically to be 12 years while Eq.~\eqref{eq:t_prec} predicts a precession time scale of 11 years. \textit{Right}: the trajectory of the source star in the co-rotating frame of the caustic curve in units of the instantaneous cusp radius for 3.5 years. The colour indicates the time normalized with the period of the SMBHB. }
 \label{fig:lightcurve_precession}
\end{figure*}

The left panel of Fig.~\ref{fig:lightcurve_precession} shows the magnification curve and the right panel shows a part of the trajectory of the source star in the co-rotating frame of the caustic curve in units of the instantaneous cusp radius. The separation between adjacent peaks is around 0.52 years, which is the period of the SMBHB, similarly to the $e=0.50$ case in Fig.~\ref{fig:lightcurve_eccentric} ignoring precession. Due to the presence of precession, there is additional modulation in the peak amplitude. The period of this modulation is around $3$ years, which corresponds to $T_{\rm prec}/4$. The factor of 4 is due to the presence of 4 arms/cusps in the caustic; the star gets close to one of the cusp points around every 3 years, as evident in the right panel of Fig.~\ref{fig:lightcurve_precession}. Neglecting precession can thus result in highly inaccurate modeling of the magnification curves because the $T_{\rm prec}/4\lesssim T_{\rm obs}$ for a survey of a duration of $\sim 10$ years. For the other two examples of eccentric binaries presented in Fig.~\ref{fig:lightcurve_eccentric} ($e=0.17$ and $e=0.33$), $T_{\rm prec}$ is estimated to be 13 and 14 years respectively, by Eq.~\eqref{eq:t_prec}. It is expected that the magnification curves of these two cases will also demonstrate strong deviation from what was presented in Fig.~\ref{fig:lightcurve_eccentric} which neglected precession.

\subsection{Motion of a single star in \qpls{}} \label{sec:moving}

As discussed in \PaperI, the duration of a \qpls{} magnification curve is limited by the motion of the source, because the star remains within the area bounded by the caustic curve for a limited time. In this section, instead of assuming that the star is stationary, we consider the more realistic case of the star moving with a velocity with respect to the caustic curve and observe the changes made to the \qpls{} magnification curve. This corresponds to the ``slow repetition'' regime described in Fig.~\ref{fig:regime_map} and Sec.~\ref{sec:blast}.

\begin{figure*}
    \centering
    \includegraphics[scale=0.6]{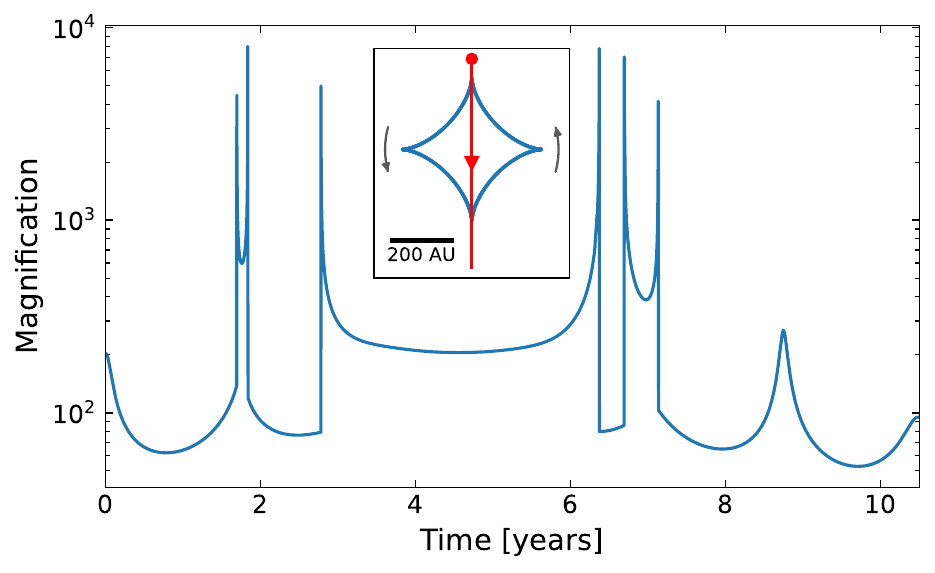}\includegraphics[scale=0.5]{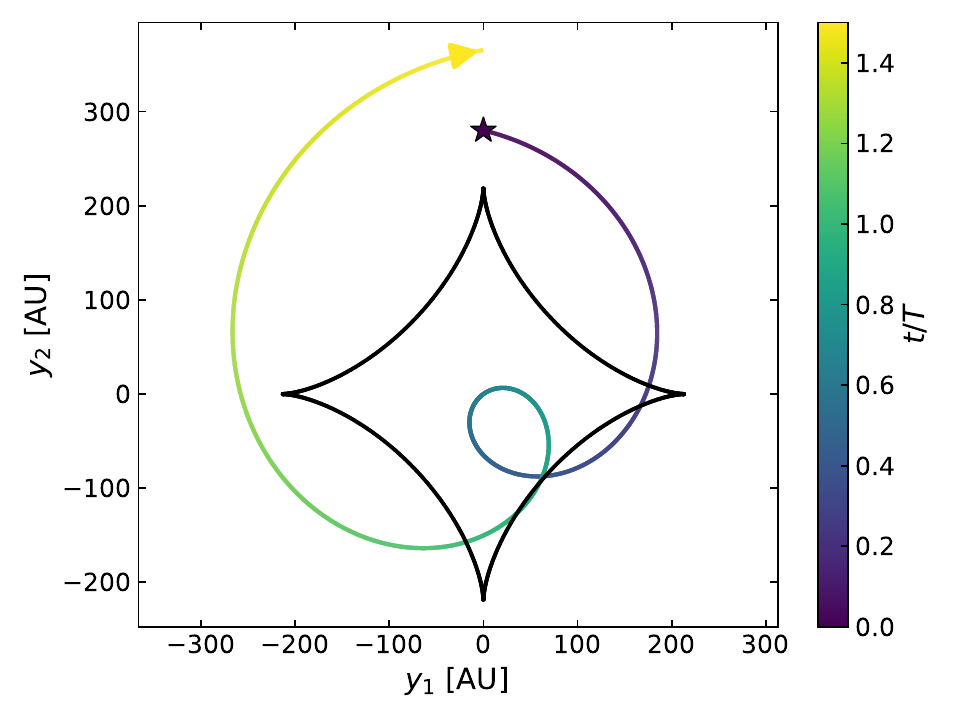}
    \caption{\textit{Left}: the magnification curve of a \qpls{} system where the star is moving with a certain velocity. Here $T= 7$ yr, $M =6\cdot10^{8} \Msun$, $D_{\rm LS} = 100 $pc, $R_{\rm src} = 10R_{\odot}$. The source star is moving at 300 km/s along the red line while the caustic is rotating, as seen in the inset. \textit{Right}: the trajectory of the source star in the co-rotating frame of the caustic curve. The colour indicates time normalized with the period of the SMBHB.}
    \label{fig:movingstar_lightcurve}
\end{figure*}

Consider a \qpls{} system with $T= 7$ yr, $M =6\cdot10^{8} \Msun$, $D_{\rm LS} = 100 $pc, $R_{\rm src} = 10R_{\odot}$. This forms a tight binary system with $d=0.14$. The SMBHB is $10^5$ years to merger hence the caustic curve is approximately fixed in size and only rotating during the time span of the \qpls{} signal. The star is at $v_{*} = 300$ km/s inside the source plane, as predicted by Eq.~\eqref{eq:vstar_general}. The left panel of Fig.~\ref{fig:movingstar_lightcurve} shows the magnification curve with the inset demonstrating the trajectory of the star. The right panel shows the trajectory star in the co-rotating frame of the caustic. It can be seen that the caustic crossing peaks are not evenly separated by $T/4$ in time as in the case of a stationary source. Because the star crosses the caustic curve at different parts, caustic crossings have different amplitude and number of peaks even though the caustic curve is fixed in size.

\subsection{Multiple sources}\label{sec:multi-source}
In this section, we consider the scenario where two stars behind the SMBHB contribute to the \qpls{} magnification curve. We will consider first the case of two unrelated stars on separate source planes, and then the case where two stars form themselves a binary system.

\subsubsection{Two sources on different source planes}\label{sec:mult_source_diff_plane}
\begin{figure*}
    \includegraphics[width=\linewidth]{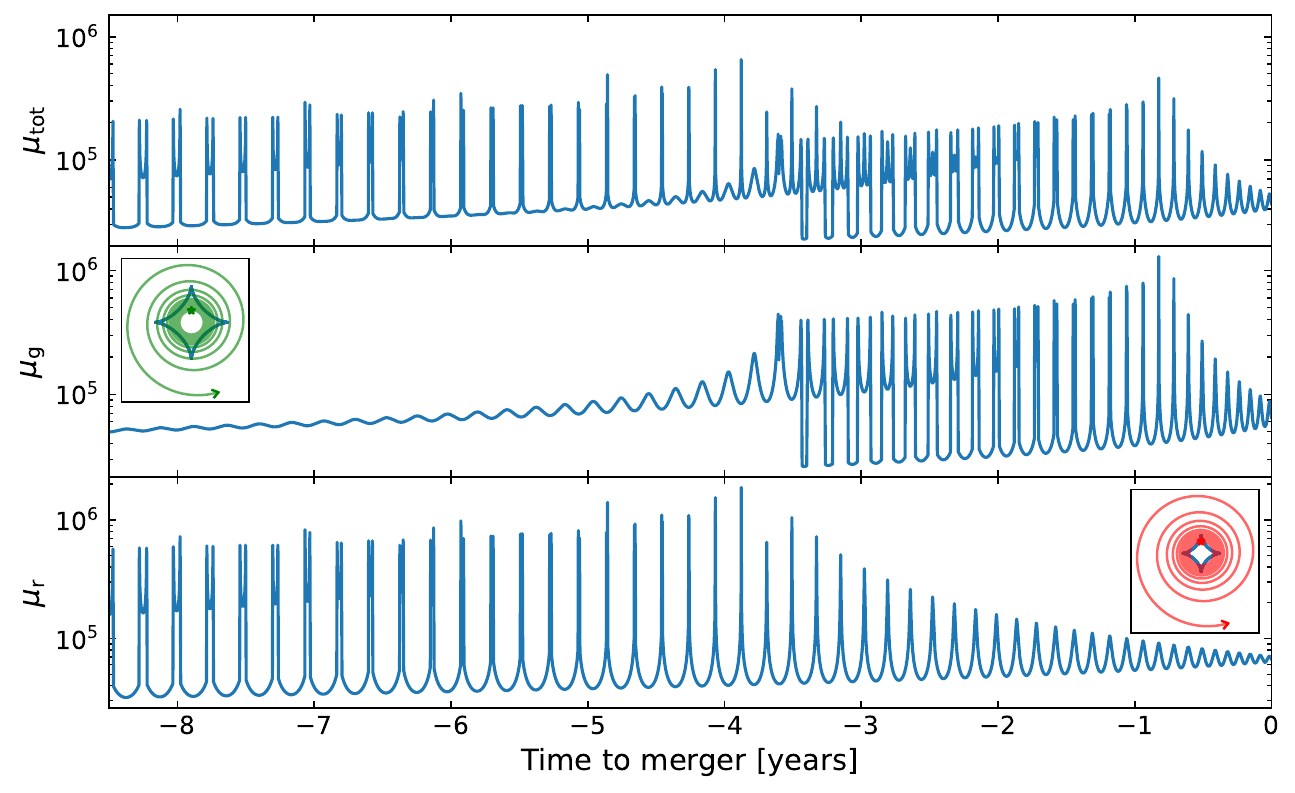}\caption{The magnification curves of a \qpls{} system with sources on different source planes. The two sources marked as ``green'' and ``red'', positioned at different source planes, with $D_{\rm LS} = 0.5$ kpc and $1$ kpc, respectively. Here $T= 1$ yr, $M =2\cdot10^{10} \Msun$, $R_{\rm src} = 10R_{\odot}$. \textit{Top}: the overall magnification $\mu_{\rm tot}$. \textit{Middle}: the individual magnification curve $\mu_{\rm g}$ of the ``green'' star. \textit{Bottom}: the individual magnification curve of the ``red'' star $\mu_{\rm r}$. We assume that the intrinsic luminosity of the ``green'' star is the same as the ``red'' star. The overall magnification is given by $\mu_{\rm tot} = {\frac12(\mu_{\rm g}+\mu_{\rm r})}$.    
    The insets of the middle and bottom panel shows the trajectory of the star in the co-rotating frame of the caustic, in units of instantaneous cusp radius.}\label{fig:mult_diffplane_lightcurve}
\end{figure*}

Assume that we have two source stars on different source planes in the same host galaxy, at different $D_{\rm LS}$ from the SMBHB lenses. Consider a \qpls{} system with $T= 1$ yr, $M =2\cdot10^{10} \Msun$, $R_{\rm src} = 10R_{\odot}$. One of the stars is at $D_{\rm LS} = 0.5$ kpc (marked as the ``green'' star in the inset of the middle panel of Fig.~\ref{fig:mult_diffplane_lightcurve}) and the other star is at $D_{\rm LS} = 1$ kpc (marked as the ``red'' star in the inset of the bottom panel of Fig.~\ref{fig:mult_diffplane_lightcurve}). The stars are assumed to be stationary, not forming a binary star system because of the large distance between them. 

Fig.~\ref{fig:mult_diffplane_lightcurve} shows the magnification curves of the overall system (top panel), the ``green`` star (middle panel) and the ``red'' star (bottom panel). We assume that the intrinsic luminosity of the ``green'' star and the ``red'' star is the same. The overall magnification is computed by 
\begin{equation}
    \label{eq:mu_overall}
    \mu_{\rm tot} = \frac{\mu_{\rm g} L_g +\mu_{\rm r} L_r}{L_{\rm g}+L_{\rm r}},
\end{equation}
where $\mu_{\rm g}$ and $\mu_{\rm r}$ are the magnification and $L_{\rm g}$ and $L_{\rm r}$ are the intrinsic luminosity of the ``green'' and ``red'' star respectively. 

The insets of Fig.~\ref{fig:mult_diffplane_lightcurve} show the trajectories of the stars in the corotating frame of the caustic, in units of the instantaneous cusp radius. The ``red'' star is near the cusp of the caustic curve initially hence its magnification curve displays strong amplitude modulation in earlier times. The amplitude modulation grows weaker as the binary inspirals and the caustic curve shrinks, getting further away from the ``red'' star. The ``green'' star is much nearer the center of mass than the ``red'' star. It only marginally crosses the innermost fold points initially, as indicated by the weak amplitude modulation of its individual magnification curve in earlier times. In later times as the binary inspirals, the ``green'' star starts to cross the fold and eventually the cusp of the caustic curve, displaying stronger amplitude modulation. We thus see the superposition of two periodic magnification patterns in the overall magnification curve at different stages of the inspiralling of the SMBHB. This clearly indicates the presence of multiple \qpls{} sources behind the SMBHB.

\subsubsection{Binary stars as sources}\label{sec:mult_star_binary}
\begin{figure*}
    \centering\includegraphics[width = \linewidth]{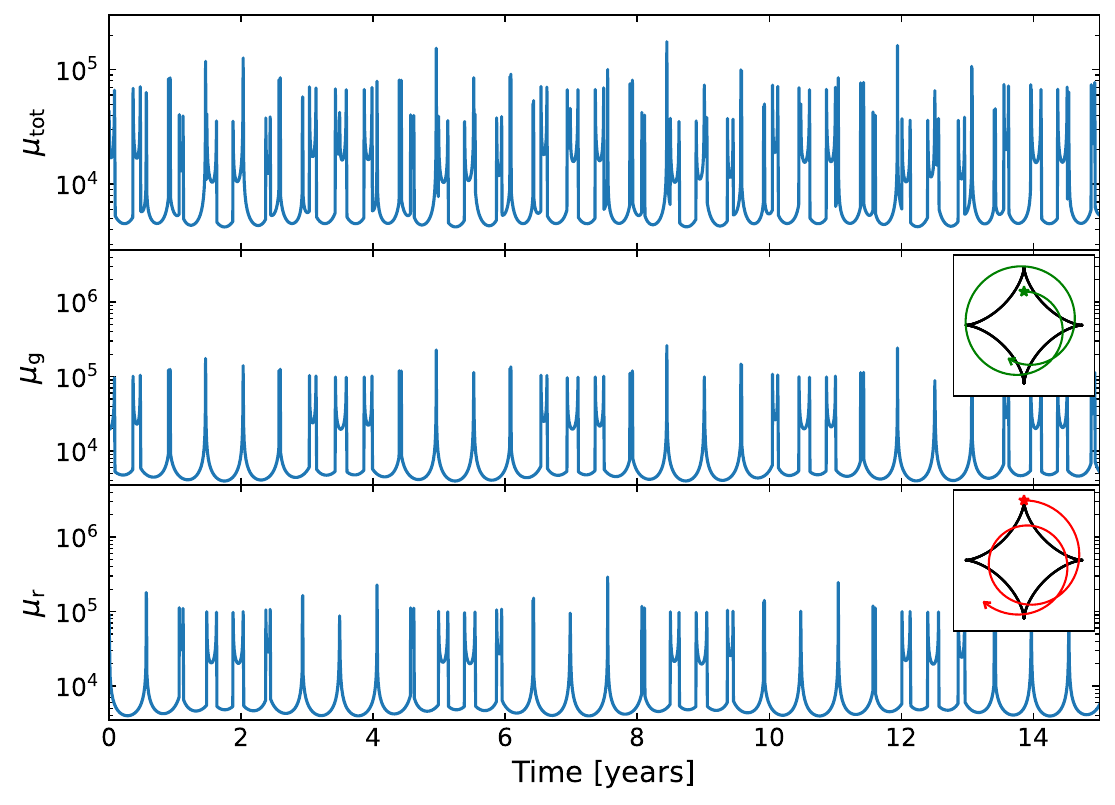}
    \caption{ This figure shows the magnification curves and the trajectory of the sources of a \qpls{} system where a binary star is lensed by the same SMBHB. Here $T= 2$ yr, $M =10^{9} \Msun$, $D_{\rm LS} = 1 $kpc, $R_{\rm src} = 10R_{\odot}$. The two sources , the ``red'' and ``green'' stars are on the same source plane and assumed form a binary system with a period of 3.6 years (each star has a mass of $10\Msun$). The rotation of the binary star is retrograde with respect to the caustic. \textit{Top}: the overall magnification curve $\mu_{\rm tot}$. \textit{Middle}: the individual magnification curve $\mu_{\rm g}$ of the ``green'' star. The inset shows the trajectory of the ``green'' star in the co-rotating frame of the caustic in units of the instantaneous cusp radius. \textit{Bottom }: the individual magnification curve $\mu_{\rm r}$ of the ``red'' star. The inset shows parts of the trajectories of the ``red'' star in the co-rotating frame of the caustic in units of the instantaneous cusp radius. We assume that the intrinsic luminosity of the ``green'' star is twice of that of the ``red'' star. The overall magnification is given by $\mu_{\rm tot} = {\frac23\mu_{\rm g} + \frac13\mu_{\rm r}}$.}\label{fig:binary_shorter_lightcurve}
\end{figure*}

Assuming that the two stars are close to each other, they might form a binary system and orbit around each other. The motion of the stars with respect to the caustic curves will produce a more complicated magnification curve. Consider a \qpls{} system with $T= 2$ yr, $M =10^{9} \Msun$, $D_{\rm LS} = 1 $kpc, $R_{\rm src} = 10R_{\odot}$, where the two source stars are at the same $D_{\rm LS}$. The SMBHB is 8000 years to merger hence the caustic curve is approximately fixed in size and only rotating. The binary star has a period of 3.6 years and its rotation is retrograde with respect to the SMBHB. We label the binary star as ``green'' and ``red'' in the insets of the middle and bottom panel of Fig.~\ref{fig:binary_shorter_lightcurve}.

Fig.~\ref{fig:binary_shorter_lightcurve} shows the magnification curves of the overall system (top panel), the``green'' star (middle panel), and the ``red'' star (bottom panel). We assume that the intrinsic luminosity of the ``green'' star is twice of that of the ``red'' star. The overall magnification is computed by Eq.~\eqref{eq:mu_overall}. The insets show parts of the trajectories of the stars in the corotating frame of the caustic, in units of the instantaneous cusp radius. 

In the magnification curves of the individual stars, it can be seen that the time separation between neighboring cusp crossing single-peaks is approximately $T/4$, as a result of the rotation of the caustic curve, similar to the case when the star is stationary. The magnification curve switches between singly-peaked cusp crossings and doubly-peaked fold crossings. Each group of fold crossings occurs approximately every 3.6 years, which corresponds to the period of the binary star. This is evident in the plots of the trajectories of the stars, where the star switches between cusp crossings and fold crossings depending on its distance from the center of mass. Because of the phase difference between the individual magnification curves, the overall magnification curve $\mu_{\rm tot}$ shows a more complicated structure, with more number of peaks within the same time duration.

\section{Spectral representation of QPLS magnification curves}
\label{sec:spectrum}

In this section, we consider the Fourier transform on the \qpls{} magnification curves and analyze their frequency spectra. We analyze the time series of the magnification curve $\mu_n = \mu(n\Delta t)$, comprising of $N$ uniformly spaced samples. We subtract the mean and apply a Hann window,
\begin{equation}
\label{eq:Hann}
w_n=\frac{1}{2}\left[1-\cos\left(\frac{2\pi n}{N-1}\right)\right],
\end{equation}
to reduce spectral leakage. We compute the discrete Fourier transform
using NumPy's function for real-input Fast Fourier Transform (FFT) \texttt{numpy.fft.rfft}\footnote{\url{https://numpy.org/devdocs/reference/generated/numpy.fft.rfft.html}}:
\begin{equation}
\widetilde{\mu}_k
=\sum_{n=0}^{N-1}w_n(\mu_n-\bar{\mu})
\exp\left(-2\pi i\frac{kn}{N}\right),
\end{equation}
where $f_k=k/N\Delta t$. No zero-padding is applied. We plot the frequency spectrum as 
\begin{equation}
    \label{eq:frequency_amplitude}
    A(f_k) = \frac{2}{\sum_n w_n}\left|\widetilde{\mu}_k\right| \approx \frac{4}{N} \left|\widetilde{\mu}_k\right| .
\end{equation}

\subsection{Spectra of periodic \qpls{} signals by quasi-circular binaries}
We first investigate the simplest case of a \qpls{} signal produced by a non-inspiralling quasi-circular binary. This is a simple case, as the signal will be approximately periodic, with constant peak amplitude and peak frequency. 

\begin{figure*}
  \includegraphics[width = \linewidth]{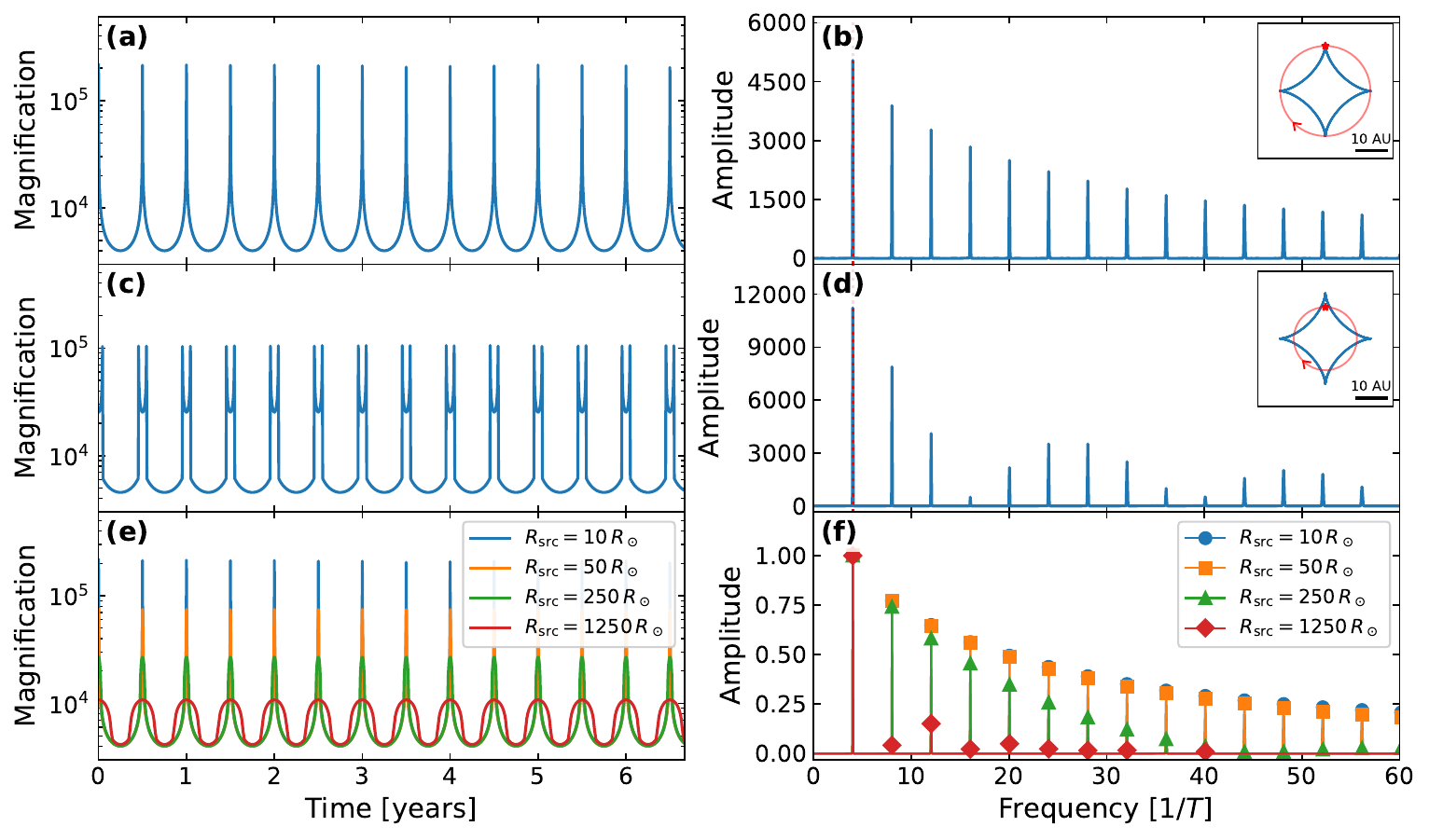}
    \caption{The magnification curves and Fourier spectra of QPLS systems by quasi-circular binaries. (a): the magnification curve of repeating cusp crossings for a \qpls{} system with $T=2$ yr, $M=10^9M{\odot}$, $D_{\rm LS} = 1$ kpc, $R_{\rm src} = 10 R_{\odot}$. (b): the Fourier spectrum of the magnification curve of repeating cusp crossings from panel. The frequency is in units of $1/T$. (a). The inset shows the trajectory of the star in the corotating frame of the caustic. The black line indicates length scale. (c): the magnification curve of repeating fold crossings for a \qpls{} system with the same parameters as panel (a). (d): the Fourier spectrum of the magnification curve of repeating cusp crossings from panel (c). The frequency is in units of $1/T$. The inset shows the trajectory of the star in the corotating frame of the caustic. The black line indicates length scale. (e): the magnification curves of \qpls{} systems with cusp crossings for different source radii. The other parameters and the source position are the same as panel (a). (f): the Fourier spectra of the magnification curves from panel (E). The frequency is in units of $1/T$. Note that the amplitude of spectra of different $R_{\rm src}$ are re-normalized such that all spectra have the same amplitude at the fundamental mode $4/T$.  }\label{fig:FTT_circular_all}
\end{figure*}

Consider a \qpls{} system with $T=2$ yr, $M=10^9\Msun$, $D_{\rm LS} = 1$ kpc, $R_{\rm src} = 10 R_{\odot}$. The binary is 8000 years before merger and can be approximated as not inspiralling. Panel (a) and (c) of Fig.~\ref{fig:FTT_circular_all} show the \qpls{} signals for cusp crossings and fold crossings respectively. Panel (b) and (d) show the respective spectra. Note that the Fourier spectrum is computed from the magnification curve of an extended duration of 100 years. The insets of the spectra show the trajectories of the stars in the corotating frame of the caustic. The red vertical dotted line marks $4/T$, which coincides with the fundamental frequency mode. The rest of the peaks correspond to the higher harmonics, which encode the shape of the magnification curve around each peak. This is as expected because the star crosses the caustic curve 4 times during each period. 
The fold-crossing spectrum has more modulation in the higher harmonics because the double-peaks during fold crossings produce an extra periodic structure.

To understand the structure of the spectrum, especially that of the higher harmonics, we plotted the magnification curve of the same \qpls{} system as in panels (a)-(d) of Fig.~\ref{fig:FTT_circular_all} but varying the size of the source. Panel (e) of Fig.~\ref{fig:FTT_circular_all} shows the magnification curves of different source radii. The peak amplitude decreases and peak width increases as the source radius increases, as expected from Eq.~\eqref{eq:cusp_mag_tight} and Eq.~\eqref{eq:caus_cross_time_detailed}. Panel (f) show the corresponding Fourier spectra. It can be seen that the modes of higher harmonics become weaker as the source radius increases. This suggests that the higher harmonics are caused by the narrow and cusp-shaped peak in the magnification curves.

\subsection{Spectra of QPLS signals by eccentric binaries}
We investigate the spectra of the QPLS signals by eccentric binaries presented in Sec.~\ref{sec:eccentric}. 

\subsubsection{Without precession}
\begin{figure*}
    \includegraphics[width = \linewidth]{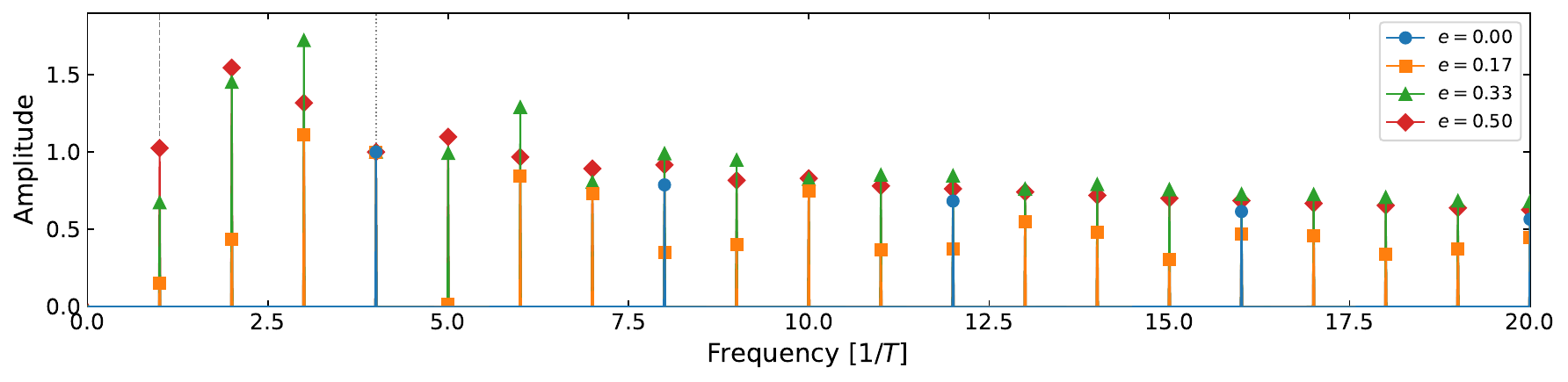}
    \caption{The Fourier spectra of the magnification curves produced by \qpls{} systems with different eccentricities (ignoring precession), corresponding to Fig.~\ref{fig:lightcurve_eccentric}. The frequency is in units of $1/T$. The amplitude is normalized such that the amplitude at $4/T$ is the same for all cases. The grey dashed vertical line corresponds to $1/T$ and the grey dotted vertical line corresponds to $4/T$.}\label{fig:FFT_eccentric}
\end{figure*}

{Fig.~\ref{fig:FFT_eccentric} shows the Fourier spectra of the magnification curves produced by \qpls{} systems with different eccentricities ignoring precession, corresponding to Fig.~\ref{fig:lightcurve_eccentric} in Sec.~\ref{sec:no_prec}. The amplitude is normalized such that the amplitude at $4/T$ is the same for all cases. The grey dashed vertical line corresponds to $1/T$ and the grey dotted vertical line corresponds to $4/T$. It can be seen that the fundamental mode for the $e=0$ case is at $4/T$ because there are 4 repeating peaks per period in the magnification curve (see Fig.~\ref{fig:lightcurve_eccentric}). For the three eccentric cases, there is one primary peak per period in the magnification curves, hence the fundamental mode is at $1/T$. Increasing eccentricity results in a stronger $1/T$ mode because of the weaker sub-leading peaks in the magnification curves (see Fig.~\ref{fig:lightcurve_eccentric}). }

\subsubsection{Periaxis precession}
\begin{figure*}
    \centering
    \includegraphics[width=1\linewidth]{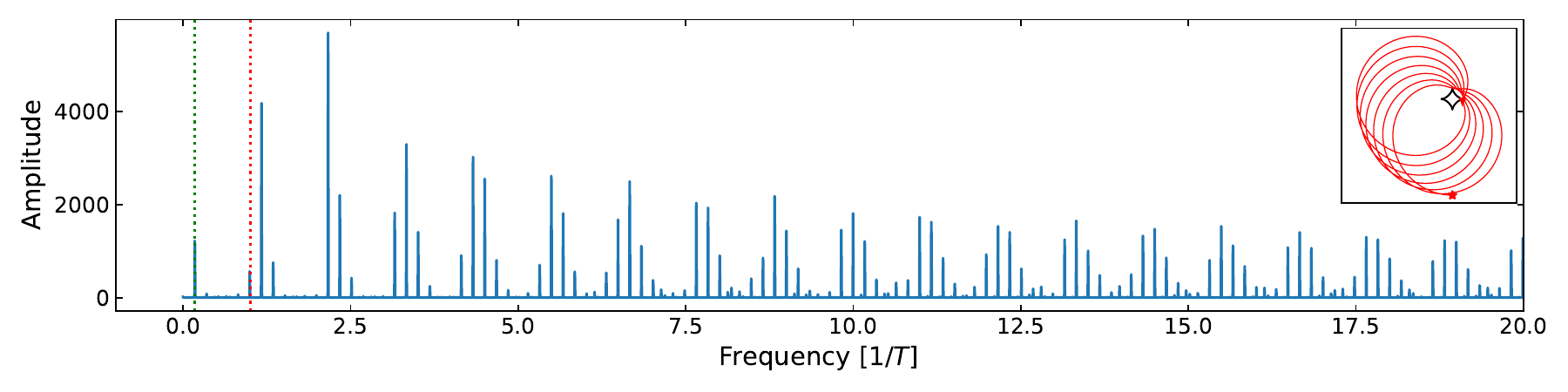}
    \caption{The Fourier spectrum of \qpls{} system where the eccentric SMBHB is undergoing periaxis precession precession, corresponding to Fig.~\ref{fig:lightcurve_precession}. The green vertical dotted line corresponds to $4/T_{\rm prec}$ and the red vertical dotted line corresponds to $1/T$. The inset shows the trajectory of the star in the corotating frame of the caustic, in units of the instantaneous cusp radius. }
    \label{fig:FFT_precessions}
\end{figure*}

We investigate the spectrum of the \qpls{} signal produced by a precessing SMBHB displayed in Sec.~\ref{sec:precession} and Fig.~\ref{fig:lightcurve_precession}. Fig.~\ref{fig:FFT_precessions} show the Fourier spectrum of the corresponding signal. Note that the Fourier spectrum is computed from the magnification curve of an extended duration of 100 years. The green vertical dotted line corresponds to $4/T_{\rm prec}$ and the red vertical dotted line corresponds to $1/T$. \{The existence of these modes are expected based on the discussion of the magnification curve in Sec.~\ref{sec:precession}. The $1/T$ mode is present because of the one primary peak per binary period in the magnification curve for an eccentric binary. The $4/T_{\rm prec}$ mode is related to the precession time scale of the binary. The factor of 4 is due to the symmetry of caustic about the 4 arms.

\subsection{Spectrum of the \qpls{} signal by a binary stellar source}

\begin{figure*}
    \centering
    \includegraphics[width=1\linewidth]{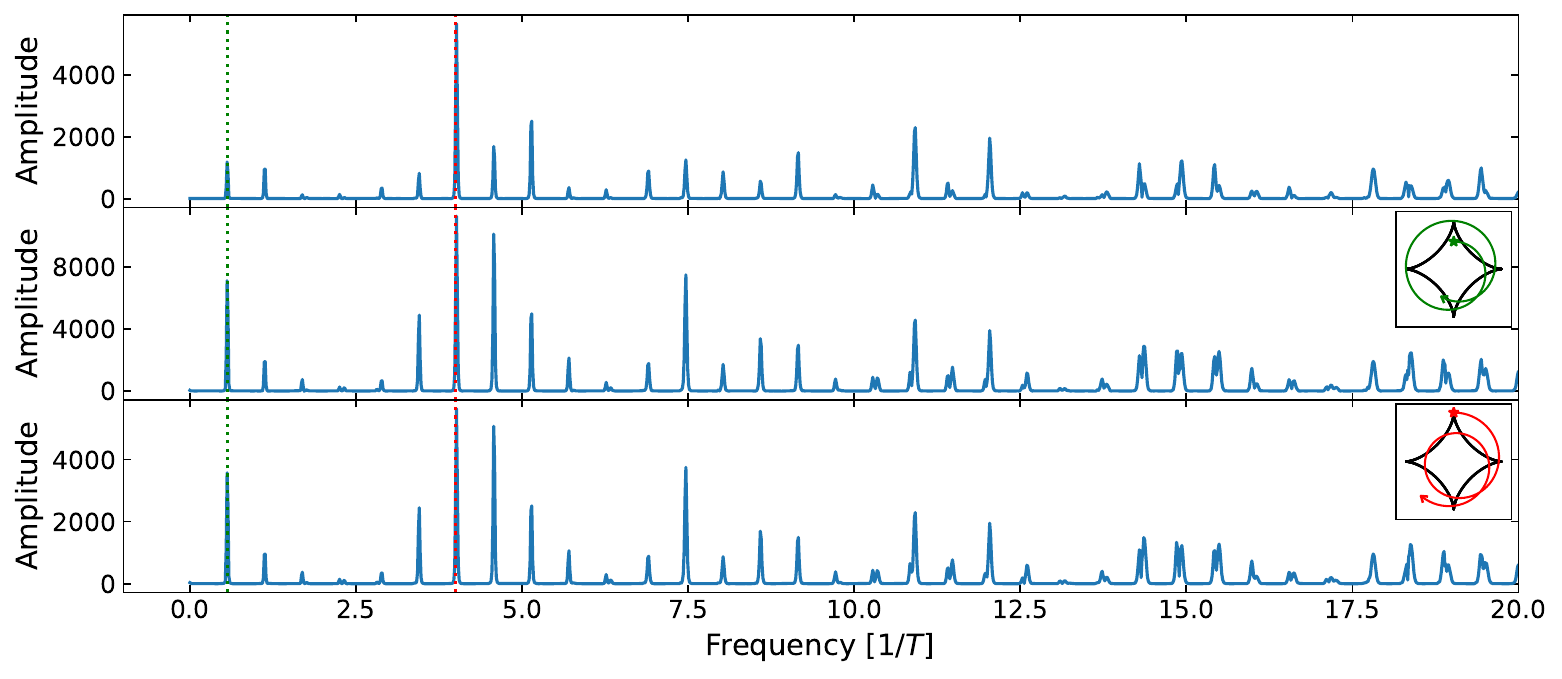}
    \caption{Fourier spectra of a \qpls{} system where a binary star is lensed by the same SMBHB in the same galaxy, corresponding to the magnification curves in Fig.~\ref{fig:binary_shorter_lightcurve}. The top, middle and bottom panel show the Fourier spectrum of the overall magnification curve $\mu_{\rm tot}$, the individual magnification curve $\mu_{\rm g}$ of the ``green'' star and the individual magnification curve $\mu_{\rm r}$ of the ``red'' star. The insets show the trajectories of the stars in the corotating frame of the caustic in units of the instantaneous cusp radius. The green vertical dotted line corresponds to the orbital frequency of the binary star and the red vertical dotted line corresponds to $4/T$. }
    \label{fig:FFT_binary_star}
\end{figure*}

We investigate the spectrum of the \qpls{} signal in Fig.~\ref{fig:binary_shorter_lightcurve}, which is a binary star lensed by the same SMBHB in the same galaxy. The top two panels of Fig.~\ref{fig:FFT_binary_star} show the spectra of the individual magnification curves of the ``red'' and ``green'' star respectively. The bottom panel of Fig.~\ref{fig:FFT_binary_star} shows the spectrum of the overall magnification curve. Note that the spectra are produced by an extended signal of a duration of 150 years. The green vertical dotted line corresponds to the orbital frequency of the binary star and the red vertical dotted line corresponds to $4\times$ the orbital frequency of the SMBHB. As explained in Sec.~\ref{sec:mult_star_binary}, the presence of these two modes are expected in all three spectra in Fig.~\ref{fig:binary_shorter_lightcurve} due to the orbital motion of both the SMBHB and the binary star. The fundamental mode corresponding to the motion of the stars is weaker in the overall spectrum. The orbital phase difference between the two stars of $\sim \pi$ results in the suppression of this mode since $\mu_{\rm tot}$ is a linear superposition of $\mu_{\rm g}$ and $\mu_{\rm r}$.

\subsection{The spectrum of the \qpls{} signal from an inspiralling binary lens}\label{sec:insp}
\begin{figure*}
    \centering
    \includegraphics[width=1\linewidth]{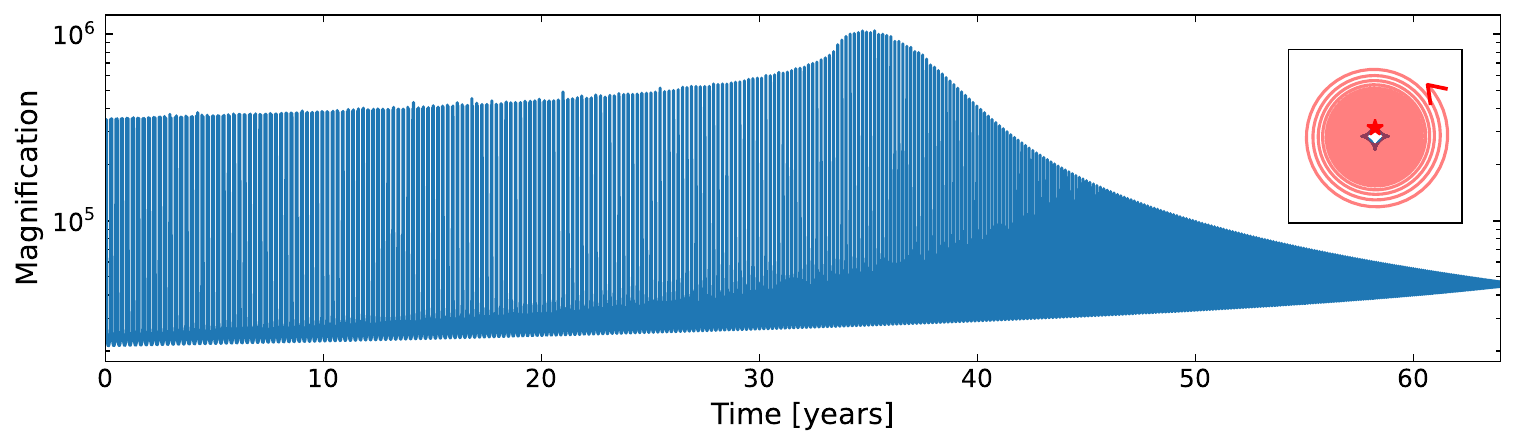}
    \includegraphics[width=1\linewidth]{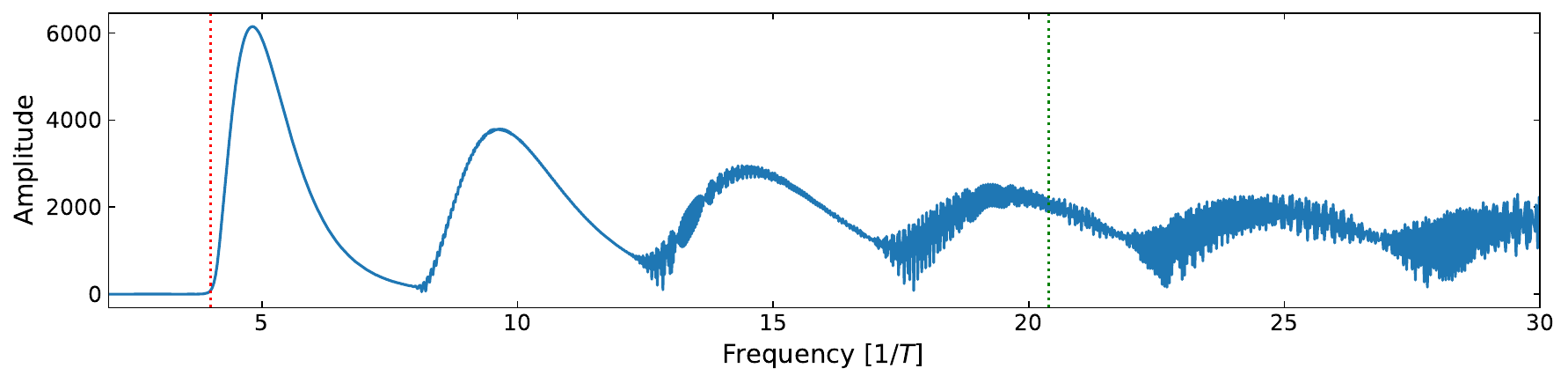}
    \caption{\textit{Top}: the magnification curve of a \qpls{} system where the SMBHB is 65 years before merger, with $T=1$ yr, $M=6\cdot10^9M{\odot}$, $D_{\rm LS} = 1$ kpc, $R_{\rm src} = 10 R_{\odot}$. The inset shows the trajectory of the star in the co-rotating frame of the caustic, with the distance of the star to the center of mass rescaled with the size of the caustic curve at each time. \textit{Bottom}: the corresponding Fourier spectrum. The red vertical dotted line corresponds to $4\times$ the initial orbital frequency of the binary lens and the green vertical dotted line corresponds to $4\times$ the final orbital frequency of the binary lens. }\label{fig:FFT_inspiral}
\end{figure*}

Consider a \qpls{} system with $T=1$ yr, $M=6\cdot10^9\Msun$, $D_{\rm LS} = 1$ kpc, $R_{\rm src} = 10 R_{\odot}$. The binary is about 65 years before merger. The top panel of Fig.~\ref{fig:FFT_inspiral} shows the magnification curve and the bottom panel shows the corresponding Fourier spectrum. For the \qpls{} signal with increasing frequency, it is expected that the fundamental mode becomes a widened peak ranging from the initial to the final frequency. The red vertical dotted line corresponds to $4\times$ the initial orbital frequency of the binary lens and the green vertical dotted line corresponds to $4\times$ the final orbital frequency of the binary lens. The green line extends beyond the first peak because the tail of the peak corresponding the higher frequencies overlap with the peaks from the higher harmonics. The \qpls{} signal also demonstrates diminishing peak amplitude towards the end, resulting in the lower amplitude in the tail of the first peak in the frequency spectrum. See App.~\ref{app:spectra_insp} for more details.

\section{Conclusion}

We have demonstrated the phenomenology of Binary-lens amplification of starlight (BLAST): how a bright star amplified by a SMBHB leads to a wide range of signatures, from several magnification flares to quasi-periodic magnification curves with different morphologies. BLAST is a more general framework built on \PaperI that incorporates \qpls{} as a special case with the richest signals. We categorized BLAST into different regimes by comparing different time scales: the orbit of the SMBHB, the motion of the source and the duration of the survey. Different regimes have different number of peaks in the magnification curves.

In complement to the estimates from \PaperI, we worked out a complete set of analytical estimates of the caustic region size, as well as the peak magnification during caustic crossings for different topologies of caustics, unequal mass ratios, and different configurations (i.e. fold and cusp crossings). These estimates improve our understanding of magnification curves, their properties and connection to BLAST systems in different regimes. Additionally, the estimates of BLAST rates from \PaperI were limited to a subset of SMBHB populations, e.g. only tight binaries with similar component masses. The updated set of analytical estimates, will allow us to extend the previous rate estimates to a broader range of binaries.
The estimates will also allow us to address observational effects: e.g.  magnification estimates can be used to apply selection cuts to caustic crossing events, given the intrinsic luminosity of sources and detection threshold of the survey.

We presented example magnification curves during caustic transits, where the star moves rapidly across the entire caustic within the survey duration. Depending on the trajectories of the star, the magnification curves can have different shapes and numbers of peaks for the same caustic. Even without periodic peak repetition from which the orbital motion of the binary can be inferred directly, these magnification curves can still be used to probe the size and geometry of the caustics and the underlying populations of lenses and sources. {\PaperI demonstrated that non-periodic caustic crossings (single crossings, one-shot fold pairs and caustic transits) produce abundant observable sources because these systems do not require small stellar velocities as in QPLS; the longer periods of these binaries also suggest that they spend more time near the current frequency band during the binary evolution.} In future work we will expand the study of these non-periodic caustic crossings including caustic transits. For example, we can build a catalog of caustic crossings and identify the morphology classes similarly to~\citet{Liebig2015} which studied static and equal-mass binaries. The morphology classes might be exhaustive as in~\citet{Liebig2015} because of the additional complexity added by the motion of the binary. By connecting with the rate estimates, we will focus on the most common morphology classes that have a good prospect of being observed.

We also investigated the different morphologies of the magnification curves of \qpls{} systems. We considered the separate effect of the relative proper motion of the source, either a single or a binary star, the relative inclination, the mass ratio of the binary, the eccentricity and the precession of the SMBHB. These effect largely transform the morphology of the magnification curves in terms of number of peaks, peak duration and peak amplitude. Some magnification curves demonstrate complex structure as a result of the superposition of multiple periodic structures, e.g. the magnifications curves produced by precession binary lenses or orbiting binary sources. A Fourier transform was performed to relate the modes of the magnification curves with the relevant motion of the sources and lenses. Potential degeneracies have also been identified. For instance, both the magnification curves of an inclined binary and an eccentric binary exhibit a reduction in the number of peaks during each period of the binary. This could cause ambiguity when inferring the orbital period of the binary from the observed magnification curves, particularly if the observational cadence is too sparse.

{Future work will further explore the morphology of the magnification curves for different lens-source configurations and address potential degeneracies.}. Moreover, to estimate the observed lightcurves, future work will incorporate effects such as limb darkening and parallax and convert the magnification to observable fluxes in different filters of the surveys using the intrinsic luminosity and spectrum of the stars. Other realistic effects such as the cadence of survey, limited exposure time, and unresolved light from the galaxy will also be considered. We expect that these detailed analyses BLAST phenomena will, in time, lead to detailed searches in time-domain data, opening a new window into sub-parsec SMBHB in quiescent galaxies.

\section*{Acknowledgements}
We are grateful to S. Ustundag and P. Vega del Castillo for helpful discussions.
MZ acknowledges support from grant RYC2024-049805-I, funded by MICIU/AEI/10.13039/501100011033 and by the European Social Fund Plus (ESF+), and from the Ayudas de Excelencia RYC-MaX 2024 programme of the Spanish National Research Council (CSIC). This publication has been funded within the framework of the R\&D\&I Project CEX2025-001574-S, funded by MICIU/AEI/10.13039/501100011033. The research presented in this publication falls within the research line Origin and Composition of the Universe: Astroparticles and Cosmology (Astro/Cosmo). This work is supported by ERC grant GLOW (101230608). Funded by the European Union. Views and opinions expressed are however those of the author(s) only and do not necessarily reflect those of the European Union or the European Research Council Executive Agency. Neither the European Union nor the granting authority can be held responsible for them.
This manuscript has been assigned the institutional preprint number IFT-UAM/CSIC-26-152.

\bibliographystyle{mnras}
\bibliography{gw_lensing}
\appendix
\section{Methods and approximations}\label{sec:method}
In this appendix, we will outline the basic methods and approximations used when computing the magnification of stars lensed by the SMBHB.
\subsection{Weak field approximation}

In this work, we use the weak-field lensing approximation. This is justified because $\Phi/c^2 \sim R_g/\xi_0 \sim 3\cdot10^{-4} \sqrt{M_{10}/D_{\rm kpc}}$ is small for our system. Here $\Phi$ is the gravitational potential, evaluated on the main critical curve (Einstein radius). 
Strong field corrections could become important for sources near the secondary (triangle) caustic, as images form much closer to the binary lens, which can be used as a test of gravity \citep{Zhong2024}. We will defer this matter to a future study.

\subsection{Point-mass lenses}

In this work, we only consider the lensing effect from the SMBHB and ignore the lensing effect of the host galaxy.
If we model the galaxy with an isothermal singular sphere, the Einstein radius is given by \citep{Schneider:1992}
\begin{equation}
    \label{eq:ein_rad_galaxy}
    \xi_{\rm galaxy} = \frac{4 \pi \sigma^2D{_L}D_{\rm LS}}{c^2 D_{S}}.
\end{equation}
Using Eq.~\eqref{eq:ein_rad}, and assuming a $M-\sigma$ relation of 
\citep{Kormendy2013} 
\begin{equation}    \label{eq:M_sigma}    \frac{M}{{1+z}} = 3.1\times 10^8 \Msun\left(\frac{\sigma_v}{200 \,{\rm km/s}}\right)^{4.38},
\end{equation}
where $\sigma_v$ is the velocity dispersion of the stars {and the $1+z$ factor in the denominator is needed because $M$ in this paper is the redshifted total mass defined as the intrinsic mass times $(1+z)$}. The ratio of Einstein radius of the SMBHB (Eq.~\eqref{eq:ein_rad}) to that of the galaxy is given by 
\begin{align}
    \label{eq:ein_rad_ratio}
    \frac{\xi_{\rm BH}}{\xi_{\rm galaxy}} =\frac{c}{2 \pi \sigma_v^2}    \sqrt{\frac{GM}{D_{\rm LS}}}
    =
    51\,M_{10}^{0.04} D_{\rm kpc}^{-1/2}Z^2,
\end{align}
which shows that the deflection of light for images near the primary caustic of the tight binary is dominated by the SMBHB. The local convergence and shear due to the singular isothermal sphere lens (Eq.~\ref{eq:SIS_density}) are
\begin{equation}
    \kappa_{\rm galaxy}(\xi_{\rm BH}) = \left|\gamma_{\rm galaxy}(\xi_{\rm BH})\right| = \frac{\xi_{\rm galaxy}}{2\xi_{\rm BH}},
\end{equation}
which indicates a weak contribution to the principal lensing scale. A departure from spherical symmetry might introduce an additional shear, which can modify the caustic structure of the tight binary \citep{Dominik:1999}.

{The short lens–source separation also suppresses the optical depth associated with stellar microlenses.} The reason is that sources in the same galaxy as the SMBHB lens lead to very low convergence for the stellar-mass objects
\begin{equation}\label{eq:kappa_*}
\kappa_* \equiv \frac{\Sigma_{*}}{\Sigma_{\rm cr}} = 6\cdot 10^{-5} \frac{\Sigma_*}{10^5 \Msun/{\rm pc}^2} D_{\rm kpc}\,,
\end{equation}
where $\Sigma_{\rm cr}=\frac{{\mathrm{c}^2} D_S}{4\pi G D_L D_{\rm LS}}=1.66\cdot 10^9 \Msun/{\rm pc}^2\,D_{\rm kpc}^{-1}$ is the critical surface density for lensing \citep{Schneider:1992} and $\Sigma_*$ is surface density of stars in the galaxy at distance $D_{\rm LS}$ from the center. $\kappa_*$ is very small even for the given reference value, which is very close to an empirical upper limit observed across many systems~\citep{maximum_surf_density}. This situation is in stark contrast to galaxy/cluster strong lenses, where $D_L, D_S, D_{\rm LS}\sim \mathcal{O}(\text{Gpc})$ reduce $\Sigma_{\rm cr}$ by a factor $\sim 10^6$, widening the caustic curves and reducing the maximum magnification by several orders of magnitude, even for modest $\Sigma_*\sim 20 \Msun/{\rm pc}^2$~\citep{Diego:2017drh,Oguri:2017ock,Venumadhav:2017pps,Muller:2024pwn}. 

Note that although $\kappa_*$ is small, the microlenses might modify the caustic structure hence the magnification curves, especially for denser stellar population or larger $D_{\rm LS}$ values. In this study, we will focus on the phenomenology produced by isolated binary lenses. We will defer further study incorporating the smooth galactic potential and stellar microlenses jointly with the binary lens in the future. 

In particular, for a centered singular isothermal sphere, the Einstein ring of the combined monopole has dimensionless radius $r_E=\frac12[s+(s^2+4)^{1/2}]$ where $s=\xi_{\rm galaxy}/\xi_{\rm BH}$ and the tight binary caustic radius changes as $\rho_{\rm cusp, gal}/\rho_{\rm cusp, BH}=r_E^{-3}$. Note that $\rho_{\rm cusp, gal}$ is the cusp radius of combined system of singular isothermal sphere with a SMBHB. For the fiducial value of $s=1/51$ in Eq.~\eqref{eq:ein_rad_ratio}, $\rho_{\rm cusp, gal}/\rho_{\rm cusp, BH}=0.971$, a reduction of only $2.9\%$. However the non-axisymmetric case requires that the external tidal shear satisfies $\gamma_{\rm ext}\ll \eta d^2$.

\subsection{Single-plane lenses}

In this work, we project both lenses to a single lens plane instead of evaluating the lens equations with two lens planes where the two point lenses sit. Following the double-plane prescription of~\citet{Schneider:1993}, we label the lens planes of the closer and farther SMBHs to the observer, respectively, as the lens-planes-1 and -2, and set the origin of lens-plane-1 to the position of the closer SMBH and define $\bm{x}^{(2)}_{m_2}$ to be the position of the farther SMBH in lens-plane-2. The two lens planes are parallel, and they are perpedicular to the line-of-sight to the observer. The double-plane lens equations are given by \citep{Schneider:1993}
\begin{align}
    \bm{y}& = \bm{x} - m_1 \frac{\bm{x}}{|\bm{x}|^2} -m_2 \frac{\bm{x}^{(2)}-\bm{x}^{(2)}_{m_2}}{|\bm{x}^{(2)}-\bm{x}^{(2)}_{m_2}|^2} \, , \label{eq:double_plane_lens_eq1} \\
    \bm{x}^{(2)} &= \bm{x} - m_1 \beta \frac{\bm{x}}{|\bm{x}|^2}\, ,\label{eq:double_plane_lens_eq2}
\end{align}
where we define
\begin{align}
    m_1 = \frac{M_1}{M_1+\frac{D_{2S}D_L}{D_{\rm LS}D_{2}}M_2} \, , \quad
       m_2 &= \frac{M_2\frac{D_{2S}D_L}{D_{\rm LS}D_{2}}}{M_1+\frac{D_{2S}D_L}{D_{\rm LS}D_{2}}M_2}\, ,\label{eq:m12_def} 
\end{align}
$\beta = {D_{12}D_S}/{D_{\rm LS}D_2}\approx {D_{12}}/{D_{\rm LS}}$ for the source and lenses in the same galaxy, 
$D_{12}$, $D_{2S}$, $D_2$, and $D_S$ are respectively the distance between the two lens planes, between the source and lens-plane-2, between lens-plane-2 and the observer, and between lense-plane-1 and the observer, and $D_{\rm LS}=D_{2S}+D_{12}$, where we assume $D_{12}\ll D_{2S}\sim D_{\rm LS} \ll D_2 \sim D_{S}$. A light ray emitted at the source position $\bm{y}$ 
crosses
lens-plane-2 at $\bm{x}^{(2)}$ and then crosses lens-plane-1 at $\bm{x}$, which 
is the image position as seen by the observer.\footnote{Note that $\bm{y}$ and $\bm{x}$ are 2D vectors describing the direction to the physical source position and its apparent image, respectively, in the planes perpendicular to the line-of-sight to the observer.} Here $\beta \leq a/D_{\rm LS}\approx 10^{-5} M_{10}^{1/3}T_{\rm yr}^{2/3}D_{\rm kpc}^{-1}\ll 1$ 
since $D_{12}\leq a \ll D_{\rm LS}$, $a$ being the SMBHB separation. 
Hence the second lens equation Eq.~\eqref{eq:double_plane_lens_eq2} is approximately $\bm{x}^{(2)}\approx\bm{x}$, and the light ray crosses both lens planes at approximately the same position. The first lens equation in Eq.~\eqref{eq:double_plane_lens_eq1} is thus approximately equivalent to the single-plane lens equation in Eq.~\eqref{eq:leqn} in the main text. The SMBHB lensing system may be approximated with having a single lens plane because of the close proximity of the two SMBH components and because the source and SMBHB are in the same host galaxy far from the observer.

\subsection{Lens equations and magnification}\label{app:leqn}

As argued above, we can model the SMBHB as two point masses in the same plane. The lens equation of a general lens is given by 
\begin{equation}
    \label{eq:lens_eq_general}
    \bm{y} = \bm{x} - \bm{\nabla}\psi(\bm{x}),
\end{equation}
where $\psi(\bm{x})$ is the dimensionless gravitational lensing potential by projecting the three-dimensional Newtonian potential to the lens plane \citep{maneghetti:2022book}. Note that the lens equation can be obtained by setting $\bm{\nabla}_{\bm{x}} t_d =0$, where $t_d$ is the dimensionless time delay 
\begin{equation}
    \label{eq:delay_dimensionless}
    t_d =\frac 12 |\bm{x}-\bm{y}|^2 - \psi(\bm{x}),
\end{equation}
consistent with Fermat's principle \citep{Blandford:1986zz}.

In particular, for a binary lens, 
\begin{equation}\label{eq:psi}
\psi(\bm{x}) = m_1\log|\bm{x}-\bm{x}_{m_1}|+m_2\log|\bm{x}-\bm{x}_{m_2}|.
\end{equation}
Substituting this potential in Eq.~\eqref{eq:lens_eq_general} gives the lens equation of a binary lens in Eq.~\eqref{eq:leqn}. The magnification of a point source is given by 
\begin{equation}
\label{eq:mu_det}
    \mu =\frac{1}{\det A},
\end{equation}
where
\begin{equation}
    \label{eq:A_def}
    A = \frac{\partial \bm{y}}{\partial \bm{x}},
\end{equation}
{For an unresolved point source with multiple images, the total magnification is 
\begin{equation}\label{eq:mu_tot}
    \mu_{\rm tot}(\bm{y}) = \sum_{\bm{x}_i:\bm{y}(\bm{x}_i)=\bm{y}} \frac{1}{\left|\mathrm{det}A(\bm{x}_i)\right|}
\end{equation}
}
For an extended source with surface brightness intensity profile $I(\bm{y})$, the magnification is \citep{Schneider:1992}
\begin{equation}
    \label{eq:mu_e}
    \mu_{\rm ext} = \frac{\int d^2y\, I(\bm{y})\mu_{\rm tot}(\bm{y})}{\int d^2y\, I(\bm{y})},
\end{equation}
where $\mu_{\rm tot}(\bm{y})$ is the point-source magnification computed using Eq.~\eqref{eq:mu_tot}. 

In this work, we assume that star is a circular source with radius $r$ (in units of the Einstein radius) with uniform brightness and computed the etended-source magnification using two ways. (1) We put a grid of $200\times200$ over the source area and compute the average point-source magnification. (2) We compute the magnification 
using the code VBMicrolensing~\citep{Bozza2021} \footnote{\url{https://github.com/valboz/VBMicrolensing}}. The code evaluates the magnification of extended sources using a contour integral algorithm \citep{Bozza2010}. Both methods sum up the contribution of magnification from all parts of the sources and all images without considering the time delay. We will justify this assumption for our systems in Sec.~\ref{sec:delay}.

\subsection{The time delay}\label{sec:delay}

 When a source is close to the caustic curve on the interior side of it, the highly magnified images are close to the critical curves and close to each other. Consider a source near a fold, for example, the two most magnified images near the critical curve are located at $ x_1=\pm \sqrt{  x_2/3h}$
 \citep{Schneider:1992}, where we have dropped the tildes and transform the coordinates such that the $x_1$ and $x_2$ axes are aligned with the local critical curve. $x_1,x_2$ and $ y_1,y_2$ measure the distance of the image and source from the nearby critical or caustic curve. 
The time delay of a binary lens is given by
\begin{align}
T_d(\bm{y},\bm{x}) = &\frac{D_S \xi_0^{^2}}{cD_LD_{\rm LS}} \left[\frac 12 |\bm{x}-\bm{y}|^2 - \psi(\bm{x})\right],\label{eq:delay}
\end{align}
where the terms inside the bracket is the dimensionless time delay $t_d$ defined in Eq.~\eqref{eq:delay_dimensionless}. $t_d$ between the two brightest images near a fold caustic for a general lensing system is given by \citep{Ezquiaga2025}
\begin{align}
    \label{eq:delay_analytical}
    \Delta t_d &\approx \frac{4\sqrt{2}}{3}\frac{y_2^{3/2}}{T_{222}^{1/2}}
    \approx \frac{4\sqrt{2}}{3\sqrt{3}} \frac{y_2^{3/2}}{d\sqrt{\eta}},
\end{align}
where we have used Eq.~\eqref{eq:T_222_tight} in the second line, assuming that we are studying the primary caustic of a tight binary.

The time delay is estimated to be 
\begin{align}
    \label{eq:delay_physical}
    \Delta T_d = &\frac{D_S \xi_0^{^2}}{cD_LD_{\rm LS}} \Delta t_d \notag \\
    &= 0.06 \, {\rm s} \frac{1}{\sqrt{\eta}}\left(\frac{y}{100R_{\odot}}\right)^{3/2}M_{10}^{5/12}D_{\rm kpc}^{-1/4}T_{\rm yr}^{-2/3},
\end{align}

where $y$ is the distance of the source from the caustic curve. The time delay between the dominant images near caustic crossing is thus much shorter than the peak duration of both a \qpls{} magnification curve (see Eq.~\eqref{eq:caus_cross_time_detailed}) and a magnification curve with a static caustic (see Eq.~\eqref{eq:peak_T}).

For subdominant images during a caustic crossing, or images produced when the source is outside the caustic curve, they do not have as small separations or time delays as in Eq.~\eqref{eq:delay_analytical}. For the primary caustic of a tight binary, these images are located close to the Einstein ring which is centered at $\bm{c} =((m_1-m_2)d/2,0)$ (see Eq.~\eqref{eq:z_crit_a1}). The distance between images is $\sim1$ in units of the Einstein radius. Consider two images near the Einstein ring at positions $\bm{x}_{i} = r_{i}\bm{n}_{i}+\bm{c}$ for $i=1,2$, where $\bm{n}_{i} = (\cos \theta_{i},\sin \theta_{i})$ and $r_{i}\sim1$. The time delay between the images is 
\begin{align}
\Delta t_d
&= -(\bm{y}-\bm{c}) \cdot (\bm{n}_1-\bm{n}_2) \notag\\
&\quad + \frac{m_1m_2d^2}{2}\left(\cos 2\theta_1 -\cos 2\theta_2\right) \notag\\
&\quad + \mathcal{O}(d^3).
\label{eq:delay_general_two_images}
\end{align}
{The corresponding source position is near the primary
caustic, hence $|\bm{y}-\bm{c}|\sim Q$, where
$Q=\eta d^2$ is the dimensionless
binary quadrupole. Introducing
$\bm{\beta}=(\bm{y}-\bm{c})/Q$, we obtain
\begin{align}
\Delta t_{d,ij}
&=Q\mathcal{F}_{ij}+\mathcal{O}(d^3),\\
\mathcal{F}_{ij}
&=-\bm{\beta}\cdot(\bm{n}_i-\bm{n}_j)
+\frac{1}{2}(\cos2\theta_i-\cos2\theta_j).
\end{align}
The geometrical factor $\mathcal{F}_{ij}$ depends on
the source position and the image pair considered,
and is typically of order unity for widely separated
images near the primary caustic.}

The corresponding observer-frame time delay is
\begin{align}\label{eq:delay_physical_general}
|\Delta T_{ij}| &=\frac{D_S \xi_0^2}{cD_LD_{\rm LS}} |\Delta t_{d,ij}|
\simeq
\frac{4GM}{c^3}Q|\mathcal{F}_{ij}|
=\frac{4GM}{c^3} \eta d^2 |\mathcal{F}_{ij}|
\notag\\
&=11\,{\rm s}\,
|\mathcal{F}_{ij}|
\eta M_{10}^{2/3}D_{\rm kpc}^{-1}
T_{\rm yr}^{4/3}\frac{(1-e^2)^2}{(1+e\cos\nu)^2}.
\end{align}

For example, for a source at a cusp and the image
pair consisting of the merging cusp images and the
opposite image, $|\mathcal{F}_{ij}|=4$ at leading
order, and for a source at the center of a primary caustic, the maximum value of $|\mathcal{F}_{ij}|$ is 1 to leading order. 
For the source at the center, images are at the top, bottom, left, and right of the Einstein ring, the second term in $\mathcal{F}_{ij}$ vanishes for opposite images, and so $\mathcal{F}_{ij}$ is 1 for top vs. right but 0 for top vs. bottom.

These delays are generally much shorter than the peak duration of both a \qpls{} magnification curve (see Eq.~\ref{eq:caus_cross_time_detailed}) {by a factor of 
\begin{align}
    \frac{|\Delta T_{ij}|}{t_{\rm mag,tight}} =& 
    4\eta^{2} d^{4} r^{-1} \frac{R_{\rm g}}{R_{\rm sep}}\frac{v_{\perp}}{c}|\mathcal{F}_{ij}|
    \notag\\ 
=&
2.0\cdot 10^{-4}
(4\eta)^2
\left(\frac{|\mathcal F_{ij}|}{4}\right)
R_{10}^{-1}M_{10}^{5/6}T_{\rm yr}^{5/3}D_{\rm kpc}^{-3/2}
\notag\\
&\times\frac{(1-e^2)^{5/2}}
{(1+e\cos\nu)^2}.
\end{align}
where $v_{\perp}$ is the tangential orbital velocity. 
In comparison to the magnification time crossing an approximately static caustic (Eq.~\eqref{eq:peak_T}) is
\begin{align}
\frac{|\Delta T_{ij}|}{T_{\rm peak}}
&=
2\eta d^2\frac{R_g}{R_{\rm src}}
\frac{v_\star}{c}|\mathcal{F}_{ij}|
\notag\\
&=
\eta\left(\frac{|\mathcal{F}_{ij}|}{4}\right)
R_{10}^{-1}T_{\rm yr}^{4/3}
\frac{(1-e^2)^2}{(1+e\cos\nu)^2}
\notag\\
&\quad\times
\begin{cases}
4.8\cdot 10^{-4}
M_{10}^{7/6}D_{\rm kpc}^{-3/2},
& D_{\rm LS}\ll D_{\rm infl}/3,
\\[5pt]
1.8\cdot10^{-3}
M_{10}^{0.90}D_{\rm kpc}^{-1} Z^{-1},
& D_{\rm LS}\gg D_{\rm infl}/3.
\end{cases}
\end{align}
where in the last equality we substituted the $M$-$\sigma$ relation.
}

It is thus reasonable to sum the contribution from all images when estimating the magnification of the star at each time.

\section{The Fourier spectra of an inspiralling \qpls{} system} \label{app:spectra_insp}

\begin{figure*}
     \centering
    \includegraphics[width=1\linewidth]{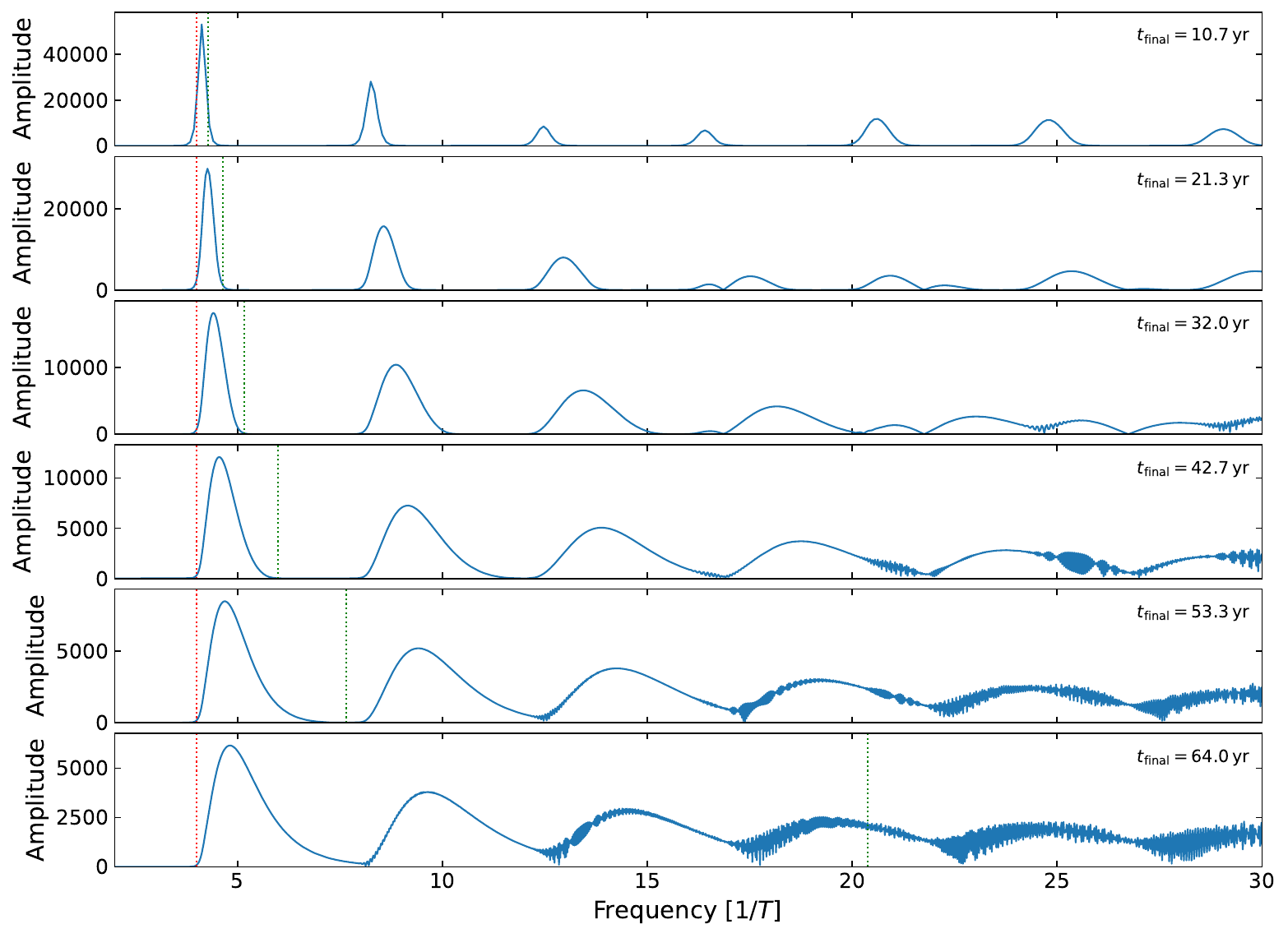}
    \caption{The Fourier spectra of the \qpls{} signal of an inspiralling binary truncated at different final times $t_{\rm final}$, which correspond to the top panel of Fig.~\ref{fig:FFT_inspiral}.  The red dotted line marks $4\times$ the initial orbital frequency of the binary and the green dotted line marks $4\times$ the orbital frequency of the binary at $T_{\rm final}$. }
    \label{fig:FFT_inspiral_byparts}
\end{figure*}

Fig.~\ref{fig:FFT_inspiral_byparts} shows the Fourier spectra of the \qpls{} signal by an inspiralling binary, presented in the top panel of Fig.~\ref{fig:FFT_inspiral} from Sec.~\ref{sec:insp}. Different panels show the spectra of the same \qpls{} signal truncated at different final time $t_{\rm final}$. The red dotted line marks $4\times$ the initial orbital frequency of the binary and the green dotted line marks $4\times$ the orbital frequency of the binary at $t_{\rm final}$. It can be seen that in the top five panels, the first peak is well bounded by the initial and final frequencies, as expected. In the bottom panel, the final frequency overlaps with the higher harmonics.

\bsp	
\label{lastpage}
\end{document}